\documentclass[12pt,tightenlines]{revtex4} 
\usepackage{color}
\usepackage{graphicx}
\usepackage[small,bf]{caption}
\usepackage{epsfig,wrapfig,sidecap}
\usepackage{fancyhdr,lastpage}
\usepackage[T1]{fontenc}
\usepackage[latin1]{inputenc}
\usepackage[english]{babel}
\usepackage{helvet}

\newcommand{\shorttitle}{ASCR/HEP Exascale Report} 
\newcommand{\projectnum}{}
\newcommand{\piname}{}
\newcommand{\smarial}[1]{\fontsize{10pt}{0pt} \em{#1}}

\begin{document}

\setcounter{page}{1}

\noindent\line(1,0){470}\\

\noindent{\bf \scshape{ASCR/HEP  Exascale Requirements Review Report}}

\bigskip

\bigskip

\noindent{\small{\bf \scshape{Lead Authors}}}\\
\noindent{\small{\bf \scshape{\bf HEP:}}}\\
\noindent Salman Habib$^1$, Robert Roser$^2$\\

\noindent{\small{\bf \scshape{ASCR:}}}\\
\noindent Richard Gerber$^3$, Katie Antypas$^3$, Katherine Riley$^1$, Tim
Williams$^1$, Jack Wells$^4$, \\Tjerk Straatsma$^4$ 

\medskip

{\em
\noindent $^1$~Argonne National Laboratory, 9700 S. Cass Ave., Lemont,
IL 60439\\
\noindent $^2$~Fermi National Accelerator Laboratory, Batavia, IL
60510\\
\noindent $^3$~Lawrence Berkeley National Laboratory, 1 Cyclotron Road,
Berkeley, CA 94720\\
\noindent $^4$~Oak Ridge National Laboratory, 1 Bethel Valley Road,
Oak Ridge, TN 37831\\}

\bigskip
 
\noindent{\small{\bf \scshape{Contributors (White Papers and Case
      Studies)}}}\\ 
A.~Almgren$^1$, J.~Amundson$^2$, S.~Bailey$^1$, D.~Bard$^1$,
K.~Bloom$^3$, B.~Bockelman$^3$, A.~Borgland$^4$, J.~Borrill$^1$,
R.~Boughezal$^5$, R.~Brower$^6$, B.~Cowan$^4$, H.~Finkel$^5$,
N.~Frontiere$^5$, S.~Fuess$^2$, L.~Ge$^4$, N.~Gnedin$^2$,
S.~Gottlieb$^7$, O.~Gutsche$^2$, S.~Habib$^5$, T.~Han$^8$,
K.~Heitmann$^5$, S.~Hoeche$^4$, K.~Ko$^4$, O.~Kononenko$^4$,
T.~LeCompte$^5$, Z.~Li$^4$, Z.~Luki\'c$^1$, W. Mori$^9$,
P.~Nugent$^1$, C.-K.~Ng$^4$, G.~Oleynik$^2$, B.~O'Shea$^{10}$,
N.~Padmanabhan$^{11}$, D.~Petravick$^{12}$, F.J.~Petriello$^5$,
J.~Power$^5$, J.~Qiang$^1$, L.~Reina$^{13}$, T.J.~Rizzo$^4$, R. Roser$^2$, 
R.~Ryne$^1$, M.~Schram$^{14}$, P.~Spentzouris$^2$,
D.~Toussaint$^{15}$, J.-L.~Vay$^1$, B.~Viren$^{16}$,
F.~Wurthwein$^{17}$, L.~Xiao$^4$

\medskip

{\em
\noindent $^1$~Lawrence Berkeley National Laboratory, 1 Cyclotron Road,
Berkeley, CA 94720\\
\noindent $^2$~Fermi National Accelerator Laboratory, Batavia, IL
60510\\
\noindent $^3$~University of Nebraska-Lincoln, NE 68588\\
\noindent $^4$~SLAC National Accelerator Laboratory, 2575 Sand Hill
Road, Menlo Park, CA 94025\\ 
\noindent $^5$~Argonne National Laboratory, 9700 S. Cass Ave., Lemont,
IL 60439\\
\noindent $^6$~Boston University, Boston, MA 02215\\
\noindent $^7$~Indiana University, Bloomington, IN 47405\\ 
\noindent $^8$~University of Pittsburgh, Pittsburgh, PA 15260\\
\noindent $^9$~University of California, Los Angeles, Los Angeles, CA
90095\\  
\noindent $^{10}$Michigan State University, East Lansing, MI 8824\\
\noindent $^{11}$Yale University, New Haven, CT 06520\\
\noindent $^{12}$National Center for Supercomputing Applications,
Urbana, IL 61801\\
\noindent $^{13}$Florida State University, Tallahassee, FL 32306\\
\noindent $^{14}$Pacific Northwest National Laboratory, Richland, WA
99354\\ 
\noindent $^{15}$University of Arizona, Tucson, AZ 85721\\
\noindent $^{16}$Brookhaven National Laboratory, Upton, NY 11973\\
\noindent $^{17}$University of California, San Diego, La Jolla, CA
92093\\ }

\bigskip

\newpage

\begin{center}
{\bf\scshape{Executive Summary}}
\end{center}

High energy physics is entering a challenging and exciting period over
the next decade with new insights into many of the fundamental
mysteries of the Universe moving tantalizingly within reach.  The
discovery of a Higgs boson in 2012 was just the first of many
anticipated discoveries as new Large Hadron Collider (LHC) runs are
performed at greater energy and luminosity, giving rise to myriad
questions. Are there many Higgs? Is supersymmetry correct?  Are there
new fundamental forces of nature yet to be discovered? Is there
physics beyond the Standard Model, and if so, what form does it take?
Next-generation experiments aim to solve the current mysteries of
neutrino physics: What is the origin of neutrino masses and how are
the masses ordered? Do neutrinos and antineutrinos oscillate
differently?  Are there additional neutrino types or interactions? Are
neutrinos their own antiparticles? As cosmological surveys and dark
matter experiments come online, a flood of new data will become
available to help answer fundamental questions: What is dark matter?
What is dark energy?  What is the origin of primordial fluctuations?

The high energy physics community -- both theoretical and experimental
researchers -- is actively seeking answers to these questions, and
there is a strong anticipation and sense of excitement that we are on
the threshold of paradigm-shifting discoveries. Success in these
endeavors, however, requires many tools of discovery. Forefront among
them, as in so many scientific fields today, is a need to perform
massive computations on an unprecedented scale and to collect, store,
and analyze complex datasets at never-before-seen rates and volumes.

To address these issues, the US Department of Energy (DOE) Office of
Science (SC) Offices of High Energy Physics (HEP) and Advanced
Scientific Computing Research (ASCR) convened a programmatic
requirements review that included leading researchers in high energy
physics, high-performance computing (HPC) experts from ASCR facilities
and scientific computing research areas, and DOE HEP and ASCR staff on
June $10-12$, 2015 in Bethesda, Maryland. The goal of the review was
to identify HPC needs and requirements to support HEP research
(computing, data, networking, and services) and seek ways for HEP and
ASCR to work together to provide the necessary facilities and
services.

The HEP community has historically leveraged HPC capabilities
primarily to address theory and modeling tasks. Energy
Frontier-related experimental efforts are now beginning to employ HPC
systems.  The estimated computational needs for these experiments far
exceed the expected HEP investment in computational hardware that will
be required to execute the envisioned HEP science program by
2025. Thus, partnering with ASCR is essential to the success of the
HEP science program.

The 2025 timeline will define a phase transition in terms of HEP
dataflows; three new facilities will come online and the landscape
will change dramatically. The high-luminosity LHC will be operating,
producing data samples at unprecedented scales.  Fermilab's flagship
Deep Underground Neutrino Experiment (DUNE) will be operating, as will
the next-generation dark energy survey experiment, the Large Synoptic
Survey Telescope (LSST). Data sizes produced each year could be 200 times
greater than what is being produced by today's operating
experiments. In addition, these new experiments will each require a
simulation program that dwarfs what we are doing today in order to
take advantage of the expected improvement in statistical precision.

One of the primary goals of computing within HEP will be to transition
to HPC capability when appropriate. This will require success at a
number of levels. Critical components of the HEP code base will have
to be refactored to take advantage of HPC architectures.  Equally
importantly, the HPC environment will need to adapt to HEP use cases,
including the adoption of ``edge'' services, networking modifications
and optimizations, available storage cache at the facilities to
optimize data transfer, and associated workflow management and
diagnostic tools. Taking these steps will be essential to addressing
escalating computing and data demands in the HEP community and, more
precisely, to informing HEP scientific programs about computing
opportunities and challenges in the future exascale environment.

This report summarizes and details the findings, results, and
recommendations derived from the June 2015 meeting. The high-level
findings and observations are as follows.

\begin{itemize}
\item Larger, more capable computing and data facilities are needed to
  support HEP science goals in all three frontiers: Energy, Intensity,
  and Cosmic. The expected scale of the demand at the 2025 timescale
  is at least two orders of magnitude -- and in some cases greater --
  than that available currently.
\item The growth rate of data produced by simulations is overwhelming
  the current ability, of both facilities and researchers, to store
  and analyze it. Additional resources and new techniques for data
  analysis are urgently needed.
\item Data rates and volumes from HEP experimental facilities are also
  straining the ability to store and analyze large and complex
  data volumes.  Appropriately configured leadership-class facilities can
  play a transformational role in enabling scientific discovery from
  these datasets.
\item A close integration of HPC simulation and data analysis will aid
  greatly in interpreting results from HEP experiments. Such an
  integration will minimize data movement and facilitate
  interdependent workflows. 
\item Long-range planning between HEP and ASCR will be required to
  meet HEP's research needs. To best use ASCR HPC resources the
  experimental HEP program needs 1) an established long-term plan for
  access to ASCR computational and data resources, 2) an ability to map
  workflows onto HPC resources, 3) the ability for ASCR facilities to
  accommodate workflows run by collaborations that can have thousands
  of individual members, 4)  to transition codes to the
  next-generation HPC platforms that will be available at ASCR
  facilities, 5) to build up and train a workforce capable of
  developing and using simulations and analysis to support HEP
  scientific research on next-generation systems.
\end{itemize}

\newpage

\begin{center}
{\bf\scshape{Introduction}}
\end{center}

A group of ASCR and HEP researchers met in Bethesda, Maryland, on June
$10-12$, 2015 to discuss the needs of the HEP scientific community in
the emerging exascale environment.  This was the first in a series of
DOE SC Exascale Science Requirements Reviews
conducted by DOE ASCR HPC facilities to
identify mission-critical computational science objectives in the
$2020-2025$ time frame. The aim of the reviews is to ensure that ASCR
facilities will be able to meet SC needs through the specified time
frame. Participants in the review included facilities staff, program
managers, and scientific and computational experts from the HEP and
ASCR communities. The review was co-organized by the HEP Forum for
Computational Excellence (HEP-FCE).

The review began with a series of talks discussing the HEP science
drivers, focusing on the scientific goals over the next decade and how
exascale computing would play a role in achieving them. Key topics
included a deeper understanding of the properties of the Higgs boson
and its implications for the fundamental laws of nature via collider
experiments (HEP's Energy Frontier); the construction of a new
scientific facility at Fermilab that will host an international effort
to understand the neutrino sector with unprecedented precision
(Intensity Frontier); and large-scale sky surveys to investigate the
nature of dark energy and dark matter, neutrino masses, and the origin
of primordial fluctuations (Cosmic Frontier).

Following the scientific overview, the computational use cases were
divided into two areas: (1) {\em compute-intensive}, referring to the
use of HPC resources in accelerator modeling, lattice quantum
chromodynamics (QCD), and computational cosmology; and (2) {\em
  data-intensive}, referring to the use of HPC, as well as other
computational resources (e.g., cloud computing), to tackle
computational tasks related to large-scale data-streams arising from
experiments, observations, and HPC simulations. The use of HPC
resources is well-established in the compute-intensive arena, whereas
the data-intensive set of use cases is an emerging application for HPC
systems. The evolution of the above computational streams, and their
interaction, will function as an important driver for a future
exascale environment that encompasses elements of computing, data
transfer, and data storage. For this reason, participants not only
discussed the computing requirements, but also articulated how the
entire computational fabric (e.g., networking, data movement and
storage, cybersecurity) needed to evolve in order to best meet HEP
requirements.  All of the HEP facilities plan enhancements that will
significantly increase the data volume, velocity, and data complexity,
as well as lead to an increased demand for much-improved modeling and
simulation of scientific processes. A common concern related to these
plans is whether the current scientific computing and data
infrastructure will be able to handle the impending demand for
computational resources.

Review participants later assembled into breakout sessions to discuss
issues associated with the demands imposed on compute-intensive and
data-intensive computing resources.  These discussions amplified the
points made in a series of white papers and case studies prepared by
the HEP community in advance of the meeting.  The review provided a
rare opportunity for experts from all of the ASCR facilities and R\&D
programs and HEP scientists to interact as a community and learn about
each other's expertise, the challenges faced, and the exciting
opportunities made possible by the exascale computing environment. The
review generated the following specific findings, which are detailed
in the body of this report.

\begin{itemize}
\item Compute-intensive applications in accelerator modeling, lattice
  QCD, and computational cosmology continue to remain a critical need
  for HEP and their requirements are expected to escalate. These groups
  have the scientific needs and the expertise to use next-generation
  HPC systems at extreme scale.
\item HEP is expanding its range of HPC expertise to data-intensive
  use cases -- driven by large data streams from simulations and from
  experiments. The HEP experimental software stack, however, is
  currently not well-suited to run on HPC systems and will need to be
  significantly modified in order to use these resources more
  effectively.
\item Strong partnership between ASCR and HEP is necessary to optimize
  systems design for the exascale environment taking into account the
  range of requirements set by the compute-intensive as well as the
  data-intensive applications.
\item Long-range planning of computational resources is of paramount
  importance to HEP programs. Stability of ASCR systems and services
  (which are typically allocated annually) must align with HEP
  timescales for proper planning to take place, and to optimize the
  design and use of ASCR resources for HEP science campaigns.
\item In many ways, the current ASCR and HEP computing facilities are
  complementary: ASCR resources are more focused toward
  compute-intensive applications, while HEP facilities are more
  oriented toward data storage, high-throughput computing (HTC) (the
  grid), and complex scientific workflows. As the two interact more in
  a future driven by mission needs, it is important that they evolve
  in a coordinated and mutually beneficial fashion.
\item Effective uniformity of the computing environment across the
  ASCR facilities would be very helpful to the user community and
  important to simplifying engagement with HEP experimental
  workflows. A federated authentication/cybersecurity model would be
  very desirable.
\item As part of establishing useful partnerships, it was considered
  desirable to have ASCR staff partially embedded in science teams.
\item To enhance opportunities for establishing long-term partnerships
  across the ASCR and HEP programs, a pilot ASCR/HEP Institute was
  suggested to foster long-term collaborations that develop
  capabilities benefiting both programs.
\end{itemize}

As a broader message, there was a general consensus that a
mission-oriented ASCR/HEP partnership for long-range activities should
be initiated.

Prior to the meeting, a number of white papers and case studies were
collected that describe, in broad terms, the current computational
activities across the HEP science spectrum (white papers) and then
detail specific examples (case studies) to provide concrete data in a
number of representative applications. The case studies cover both
compute-intensive and data-intensive examples and vary in size from
``mid-scale'' to ``extreme-scale'' applications, where by this
terminology we mean the distinction between small (mid-scale) and
large (extreme-scale) computational footprints on facility-scale
computing platforms. Both the white papers and the case studies
project needs in their areas on the $2020-2025$ timescale.

\newpage

\noindent
{\bf \scshape Table of Contents}\\

\begin{tabular}{lr}

\noindent {\scshape
  {1~High Energy Physics: Vision and Grand Challenges}}.................................&
1\\
\noindent {\scshape {2~Research Directions and Computing Needs/Requirements}}......................& 4\\
\hspace{0.7cm}{\em 2.1~Energy Frontier}.................................................................................................& 4\\
\hspace{0.7cm}{\em 2.2~Intensity Frontier}..............................................................................................& 5\\
\hspace{0.7cm}{\em 2.3~Cosmic
  Frontier}................................................................................................& 6\\
\hspace{0.7cm}{\em 2.4~HPC
  Applications}.............................................................................................& 6\\
\hspace{0.7cm}{\em 2.5~HTC
  Applications}.............................................................................................& 7\\ 
\hspace{0.7cm}{\em 2.6~Future Architectures and
  Portability}.................................................................& 7\\ 
\hspace{0.7cm}{\em 2.7~Evolution of HPC
  Architectures}........................................................................& 8\\  
\hspace{0.7cm}{\em 2.8~Resource
  Management}.......................................................................................& 9\\  
\noindent {\scshape {3~Path Forward}}........................................................................................................&10\\
\hspace{0.7cm}{\em 3.1~Compute-Intensive
  Applications}........................................................................& 10\\  
\hspace{0.7cm}{\em 3.2~Data-Intensive
  Applications}..............................................................................&
11\\  
\hspace{0.7cm}{\em 3.3~Evolution of the ASCR/HEP Partnership}........................................................&
13\\ 
\noindent {\scshape {4~Summary of HEP Requirements}}.......................................................................& 14\\
\noindent {\scshape {5~References}}.............................................................................................................& 17\\
\noindent {\scshape {Appendix 1: White
    Papers}}.....................................................................................&
18\\ 
\hspace{0.7cm}{\em A1.1~Exascale Accelerator
  Simulation}.....................................................................& 19\\
\hspace{0.7cm}{\em A1.2~Computational Cosmology at The Exascale}....................................................& 23\\
\hspace{0.7cm}{\em A1.3~Lattice
  QCD}...................................................................................................& 26\\
\hspace{0.7cm}{\em A1.4~HPC in HEP
  Theory}......................................................................................& 29\\
\hspace{0.7cm}{\em A1.5~Energy Frontier
  Experiments}..........................................................................& 31\\
\hspace{0.7cm}{\em A1.6~HEP Experiments: Data Movement and 
  Storage}...........................................& 34\\ 
\hspace{0.7cm}{\em A1.7~Cosmic Frontier
  Experiments}.........................................................................& 38\\ 
\hspace{0.7cm}{\em A1.8~Intensity Frontier
  Experiments}......................................................................& 41\\ 
\hspace{0.7cm}{\em A1.9~Evolution of HEP Faciltities: Compute Engines}............................................& 44\\ 
\noindent {\scshape {Appendix 2: Case Studies}}......................................................................................& 47\\ 
\hspace{0.7cm}{\em A2.1~Advanced Modeling of Accelerator Systems 
  (ACE3P)}....................................& 48\\  
\hspace{0.7cm}{\em A2.2~Luminosity Optimization in High Energy
  Colliders}\\
$~~~~~~~~~~~~$ {\em using High Performance Computers
  (BeamBeam3D)}....................................& 51\\ 
\hspace{0.7cm}{\em A2.3 Computational Cosmology
  (HACC)}................................................................& 54\\
\hspace{0.7cm}{\em A2.4~Cosmic Reionization
  (ART)}...........................................................................& 57\\
\hspace{0.7cm}{\em A2.5~Dark Energy Survey in Hindsight (DES)}........................................................& 60\\
\hspace{0.7cm}{\em A2.6~Lattice QCD 
  (MILC/USQCD)}.......................................................................& 63\\ 
\hspace{0.7cm}{\em A2.7~Event Generation 
  (Sherpa)}.............................................................................& 66\\  
\hspace{0.7cm}{\em A2.8~Energy Frontier Experiment
  (ATLAS)}...........................................................& 69\\   
\hspace{0.7cm}{\em A2.9~Cosmic Microwave Background
  (TOAST)}.....................................................& 72\\   
\noindent {\scshape {Appendix 3: Acronym
    Index}}..................................................................................& 75\\  
\noindent {\scshape {Appendix 4:
    Acknowledgments}}............................................................................& 77\\

\end{tabular}

\newpage


\setcounter{page}{1}

\begin{center}
{\bf\scshape{1~High Energy Physics: Vision and Grand Challenges}}
\end{center}

Particle physics is the study of both the very small and the very
large, with the underlying goal to understand matter and energy at a
fundamental level. In the realm of the very small -- at the highest
energies probed by accelerator experiments -- the current particle
physics Standard Model does a remarkable job of describing the
physical world we live in. But because it is not complete, it also
points to new directions in which exciting progress can be made with
experiments operating at new energy, precision, and data collection
thresholds.

On the largest scales, the expanding universe connects particle
physics to cosmology, allowing fundamental physics to be probed by
using the universe as a laboratory, with telescopes as the analogs of
detectors in accelerator experiments.  By the very nature and scale of
these activities, most of the exploration of nature in high energy
physics is, by definition, ``Big Science.''  The tools required to
explore both the quarks and the cosmos are, at the same time, massive,
yet exquisite in the precision achieved by the eventual measurements.

Over the past few years, the entire US HEP community came together to
embark on an extensive planning exercise, ``to develop an updated
strategic plan for US high energy physics that can be executed over a
10-year timescale, in the context of a 20-year global vision for
the field''~\cite{p5}.

Through this process, a great number of exciting scientific
opportunities were identified. Five intertwined science drivers for
the field categorize the essence of these opportunities:
\begin{enumerate}
\item Use the Higgs boson as a new tool for discovery
\item Pursue the physics associated with neutrino mass
\item Identify the new physics of dark matter
\item Understand cosmic acceleration: dark energy and inflation
\item Explore the unknown: new particles, interactions, and physical
  principles
\end{enumerate}

A report organized around these drivers -- the P5 report -- was
produced detailing the path forward for HEP in the United States and
internationally, based on a realistic funding scenario~\cite{p5}. (See
also the white papers from the Snowmass 2013 Study~\cite{snowmass},
which provided important input into the P5 process.) Together, the HEP
program and community pursue the science drivers identified in the P5
report -- which are broadly categorized by the HEP categories of the
Energy, Intensity, and Cosmic Frontiers -- via experimental paths that
are closely linked to technological advances, and to progress in
theory, modeling, and simulations; computing plays an essential role
in all of these aspects. Over the next decade, several new projects
will be brought online that are potential game changers for HEP. These
include the Cosmic Frontier projects LSST, Cosmic Microwave
Background-Stage 4 (CMB-S4), and Dark Energy Spectroscopic Instrument
(DESI); Intensity Frontier projects DUNE and Muon-to-Electron
Conversion Experiment (Mu2e); and Energy Frontier projects including
the High-Luminosity LHC, among others. Results from any of these could
profoundly alter our understanding of the world we live in.

Computing has always been -- and continues to be -- an integral and
essential part of all activities in high energy physics, which is a
data- and compute-intensive science. As the field of particle physics
advances, computing must continue to evolve in order to satisfy the
ever-increasing appetite of its researchers. For example, the volume
of physics data from LHC experiments already stresses both the
computing infrastructure and related computational expertise, and LHC
operations in the next decade will likely result in order-of-magnitude
increases in data volume and analysis complexity. The data needs
associated with experiments exploring the cosmos will greatly expand
as vast new surveys and high-throughput instruments come
online. Cosmological computing is making significant progress in
connecting fundamental physics with the structure and evolution of the
Universe at the necessary level of detail. Lattice QCD provides the
most precise values of heavy quark masses and the strong coupling
constant, important for using Higgs boson decays at the LHC to test
the Standard Model and probe for new physics. Precision perturbative
QCD and electroweak calculations are also beginning to use large-scale
computational resources. The use of high-performance computing is
advancing full 3-D simulations of nearly all types of
accelerators. The aim is to enable ``virtual prototyping'' of
accelerator components on a much larger scale than is currently
possible, potentially alleviating the need for costly hardware
prototyping. All of these use cases will see significant enhancements
and improvements in the future.

The HEP community fully recognizes the significance of exascale
computing and is counting on its availability as an essential resource
for the next set of major HEP projects, early in the next
decade. Broadly speaking, there are two types of HEP use cases
relevant to the exascale computing environment. The first are {\em
  compute-intensive}, referring to the use of HPC systems in HEP
applications such as lattice QCD, computational cosmology, and
accelerator modeling. The second are {\em data-intensive}, referring
to the use of HPC and other computational resources (e.g., analysis
clusters, cloud resources, HTC systems) to managing and analyzing
data streams arising from experiments, observations, and HPC
simulations. As the quality of simulation and modeling improves, and
the associated datasets become more complex, a number of applications
are emerging that merge the two categories by being simultaneously
compute- and data-intensive.

Historically, the HEP theoretical community has been a long-standing
customer of ASCR HPC resources, a relationship that has been further
nurtured by partnerships initiated and continued by the Scientific
Discovery through Advanced Computing (SciDAC) program. The situation
with HEP experiments has been somewhat different; for the most part
experiments in all three HEP frontiers have been responsible for
handling their own computational needs, which have been mostly
satisfied using distributed high throughput computing (`the grid').

The data from HEP experiments is growing at a rapid rate in terms of
volume and throughput. It is now doubtful that experimental HEP will,
as a field, own all of the computing resources it needs moving
forward. Provisioning will be, at best, capable of meeting
steady-state demands -- the HEP community will need to find creative
ways to meet peak demands. At the same time, to extract the best
possible results from the observations, joint analysis with
sophisticated theoretical predictions is needed; such analysis comes
with its own set of large-scale data and computational needs.

The emerging exascale computing landscape offers many opportunities
for HEP both in terms of compute-intensive HPC use cases, as well as
the data-intensive use cases -- for example, using HPC machinery (or
what it evolves into) to perform data reconstruction and simulation
tasks that have historically fallen within the domain of HTC systems
in Energy Frontier experiments.  The data-intensive community is
increasingly becoming aware of the value of the HPC resources and
expertise available.  These use cases bring new challenges in the form
of data movement and storage, data persistence, and integrating these
technologies into the scientific workflows of the experiments.

A number of white papers and case studies are presented in this report
in Appendices 1 and 2. The white papers provide broad coverage of HEP
science and describe the science cases and modes of computational
usage. The case studies present individual examples that provide more
details about the computational methodologies and requirements.

The aim of this Exascale Requirements Review is to understand the
requirements that must be satisfied for ASCR facilities to be able to
meet HEP needs through the $2020-2025$ timeframe. The approach taken
in this review is broad, covering the entire range of HEP activities
and taking into account not just ASCR computing systems, but the
broader environment of data storage and transfer, software evolution,
and the desired characteristics of the exascale environment in the
face of HEP's data-intensive needs. In this broader context, the
specific examples just discussed above are connected to a number of
important issues that are the central to the review. These are:
\begin{itemize}   
\item {\em Simulations:} The ability to effectively perform simulations is
  paramount for successful HEP science.  As experimental and
  observational datasets grow -- with ever-higher-resolution detectors
  -- the demand for finer-grained, higher-precision simulation will
  continue to increase, requiring an increase in computational
  resources and the overall efficiency with which they are used.
\item {\em Architectures:} In order to satisfy increasing
  computational demands, and keep abreast of technological changes,
  the field needs to make better use of next-generation computing
  architectures. With increased concurrency at the nodal level, and
  more complex memory and communication hierarchies, the complexity of
  software and systems will continue to increase, and such systems
  will need to be better exploited and managed.
\item {\em Data Access:} Distributed storage access and network
  reliability are essential for data-intensive distributed computing, a
  hallmark of HEP experimental practice and a growing issue for other
  applications. Emerging network capabilities and data access
  technologies improve researchers' ability to optimally use resources
  independent of location. Treating networks as a resource that needs
  to be managed and planned for is an important area of future ASCR
  and HEP interaction. This will require close collaboration of HEP
  scientists with networking experts from the Energy Sciences Network
  (ESnet). 
\item {\em Training:} As a field, HEP, across all its frontiers and
  applications, must continue to develop and maintain expertise and
  re-engineer frameworks, libraries, and physics codes to adapt to the
  emerging hardware landscape, as mentioned above. There is a large
  code base that needs to be re-factored and re-engineered, and there
  is a shortage of trained experts to do it. Training the next
  generation of HEP computational scientists and software developers
  needs urgent attention.
\end{itemize}

\newpage

\begin{center}
{\bf\scshape{2~Research Directions and Computing Needs/Requirements}} 
\end{center}

It is convenient to subdivide the field of particle physics into three
broad categories of scientific pursuit, namely the Energy Frontier,
Intensity Frontier, and Cosmic Frontier. Taken as an ensemble, these
three fields together address the five science drivers outlined in
the P5 program and listed in the Introduction. The computational needs
do not divide as neatly as the scientific ones, however, and there are
significant areas of overlap, as well as unique needs, in
computational use cases and requirements across the three frontiers.

The Energy Frontier focuses on studying the fundamental constituents
of matter and conditions in the very early universe by accelerating
charged particles to very high energies
\begin{wrapfigure}[23]{l}{2.5in}
\centering \includegraphics[width=2.5in]{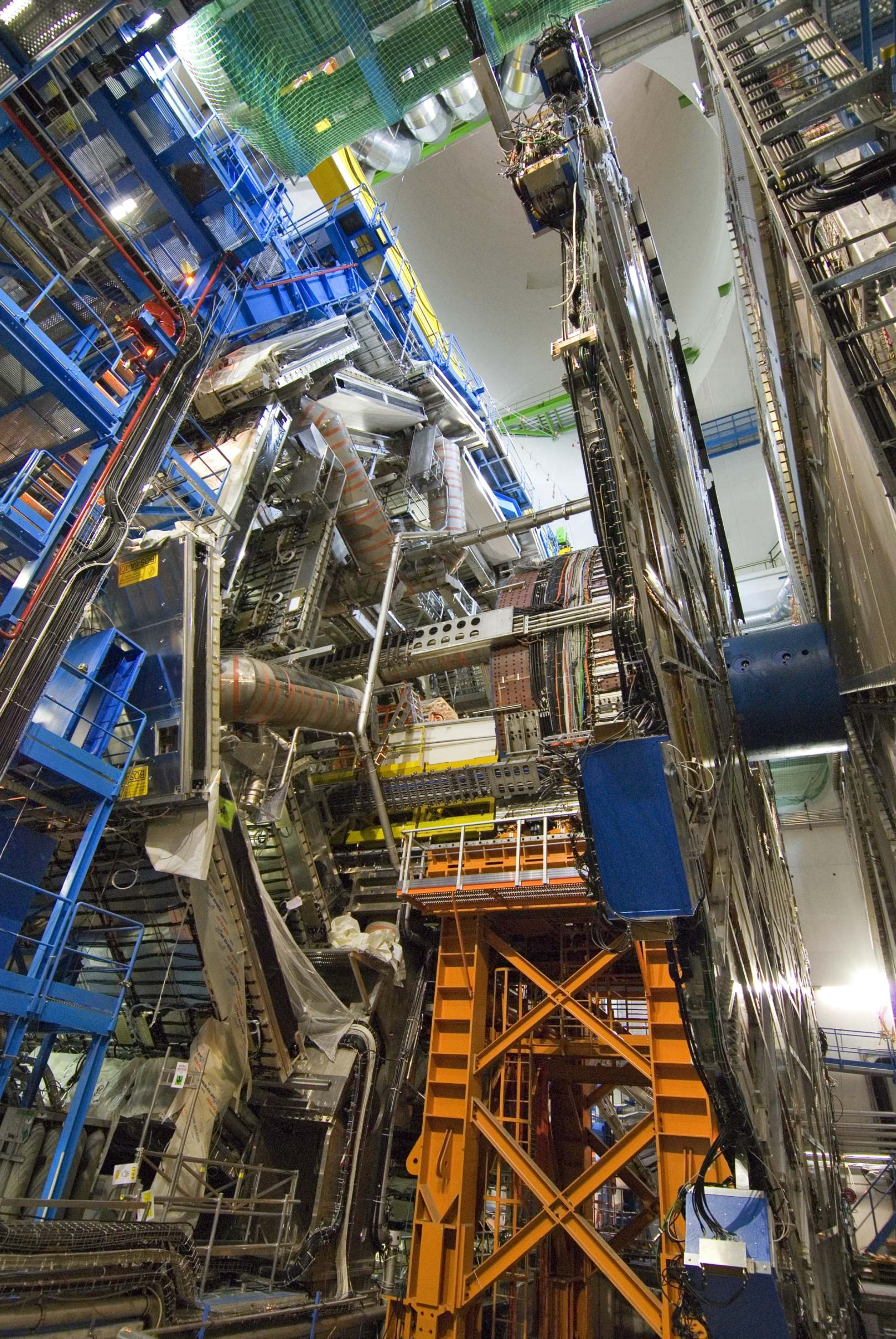}

\vspace{-0.2cm}

\caption{\small{\em LHC's ATLAS detector under construction at
    CERN in 2007.}}
\label{atlas}
\end{wrapfigure}
\noindent
in particle colliders; massive detectors are used to study the
resulting collision events. The primary science direction in the
Energy Frontier is the search for hints of physics beyond the Standard
Model of particle physics and the investigation of the properties of
the recently discovered Higgs boson~\cite{higgs_atlas,
  higgs_cms}. Intensity Frontier experiments use intense particle
beams to generate rare events searched for by sensitive detectors,
characterized by well-understood backgrounds and background rejection
methods. The primary arena of investigation in the Intensity Frontier
is the neutrino sector; in partcular, the neutrino masses and the
neutrino mixing matrix~\cite{neut}. Cosmic Frontier experiments detect
particles from space to investigate the nature of cosmic acceleration,
primordial fluctuations, and neutrino masses (via optical surveys and
CMB measurements)~\cite{weinberg_rev} and search for dark matter
candidates using direct and indirect detection
strategies~\cite{dm_review}.

In addition to the three Frontiers, the HEP program also has an
important enabling technology component in the areas of particle
accelerators and advanced detectors and instrumentation. These areas
also have significant computational requirements.

\medskip

\noindent{\bf\scshape{2.1~Energy Frontier}}  

Located at the LHC in Geneva, Switzerland, are two general-purpose
detectors that are designed to investigate a broad program of physics
opportunities at the Energy Frontier.  The two detectors are CMS
(Compact Muon Solenoid)~\cite{cms} and ATLAS (A Toroidal LHC
Apparatus)~\cite{atlas} (Figure~\ref{atlas}). These detectors are
designed to record the ``interesting'' high-energy proton-proton
collisions and keep track of the position, momentum, energy, and
charge of each of the resulting particles from the collision event.
Utilizing this information combined with simple conservation laws,
scientists attempt to reconstruct what particles were generated by the
collision and the physical processes that created it.  By looking at
ensembles of events, the hope is to gain a deeper insight into the
fundamental laws of nature that operate at high energies and,
therefore, to also gain improved understanding of the physics of the
early universe.

LHC scientists pursue two different types of physics analysis: in the
first, they collect precise measurements to improve the Standard Model
parameters and identify small deviations, and in the second, they
search for the unexpected, which could lead to completely new physical
models and theories.

Over the next decade, LHC scientists will characterize the newly found
Higgs boson to see whether it behaves as expected. There may be more
hidden surprises, as a new energy regime is explored.  Signatures of
supersymmetry, a well-developed theory that provides the ``unification
mechanism'' required by the Standard Model, have not been observed so
far~\cite{craig_susy}.  Scientists on the LHC will be looking for the
myriad super-particles that are predicted.  Of course, at these higher
energies, there may be surprises that have not yet been envisioned.

White papers in this report relevent to the Energy Frontier include
A1.3 {\em Lattice QCD}, A1.4 {\em HPC in HEP Theory}
(compute-intensive), A1.5 {\em Energy Frontier Experiments}, and A1.6
{\em HEP Experiments: Data Movement and Storage} (data-intensive). The
case studies are A2.6 {\em Lattice QCD (MILC/USQCD)}, A2.7 {\em Event
  Generation (Sherpa)}, and A2.8 {\em Energy Frontier Experiment
  (ATLAS)}.

\medskip

\noindent{\bf\scshape{2.2~Intensity Frontier}}  

The Intensity Frontier comprises the set of experiments that require
intense particle beams and/or highly sensitive detectors to study rare
processes with ever-greater precision and sensitivity.  The neutrino
sector is a primary area of interest in the Intensity Frontier.
\begin{wrapfigure}[13]{l}{4.5in}
\centering \includegraphics[width=4.5in]{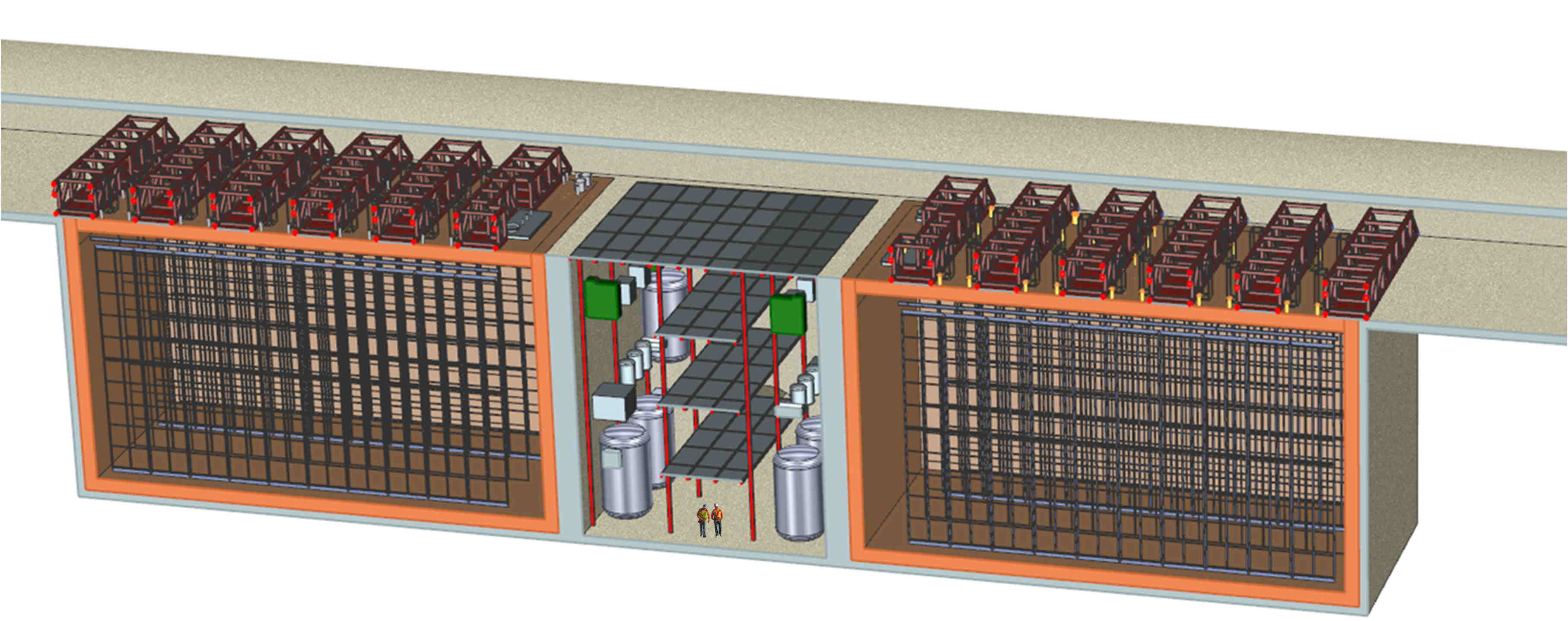}

\vspace{-0.2cm}

\caption{\small{\em Engineering design study of 10-kiloton cryogenic
    detector modules for DUNE.}}
\end{wrapfigure}
\noindent 
We now know that neutrinos exist in three types (lepton flavors) and
that they undergo quantum oscillations, i.e., they change flavor as
they propagate in space and time. The observed oscillations imply that
neutrinos have masses (neutrinos of given flavors are different
superpositions of the mass eigenstates). Many aspects of neutrino
physics are puzzling, and the experimental picture is
incomplete. Powerful new facilities are needed to move forward,
addressing the following questions: What is the origin of neutrino
mass? How are the masses ordered (referred to as the mass hierarchy
problem)? What are the masses? Do neutrinos and antineutrinos
oscillate differently?  Are there additional neutrino types or
interactions? Are neutrinos their own antiparticles?  The answers to
these questions can have profound consequences for understanding the
current make-up of the universe and its evolution. A special facility
at Fermilab~\cite{pip2_fnal} is being built to provide a high flux
beam necessary to perform this science.  The flagship future program
in the Intensity Frontier is DUNE, designed to study neutrino
oscillations~\cite{dune_fnal}. Although Intensity Frontier experiments
have historically not been at the forefront of computation,
computationally intensive lattice QCD calculations have been important
for these experiments.

Other experiments in the Intensity Frontier include 1) Muon
g-2~\cite{g-2}, which tracks the precession of muons in an external
magentic field in order to precisely measure the anomalous magnetic
moment of the muon and look for evidence of new physics, 2)
Mu2e~\cite{m2e}, which will investigate the very rare -- currently
unobserved -- process of muon-to-electron conversion, 3) Belle
II~\cite{belle2}, which will investigate sources of CP violation and
provide precision tests of the Standard Model at Japan's SuperKEKB
high-luminosity collider.

Intensity Frontier white papers are A1.3 {\em Lattice QCD}
(compute-intensive) and A1.8 {\em Intensity Frontier Experiments}
(data-intensive). The associated case study is A2.6 {\em Lattice QCD
  (MILC/USQCD)}.

\medskip

\noindent{\bf\scshape{2.3~Cosmic Frontier}}  

Unlike the other two frontiers that are predominantly
accelerator-based and study very small length scales, the Cosmic
Frontier program focuses on the detection and mapping of galactic and
extra-galactic sources of radiation utilizing a variety of
well-instrumented telescopes, both ground- and satellite-based, to
better understand the fundamental nature
\begin{wrapfigure}[18]{l}{3.5in}
\centering \includegraphics[width=3.5in]{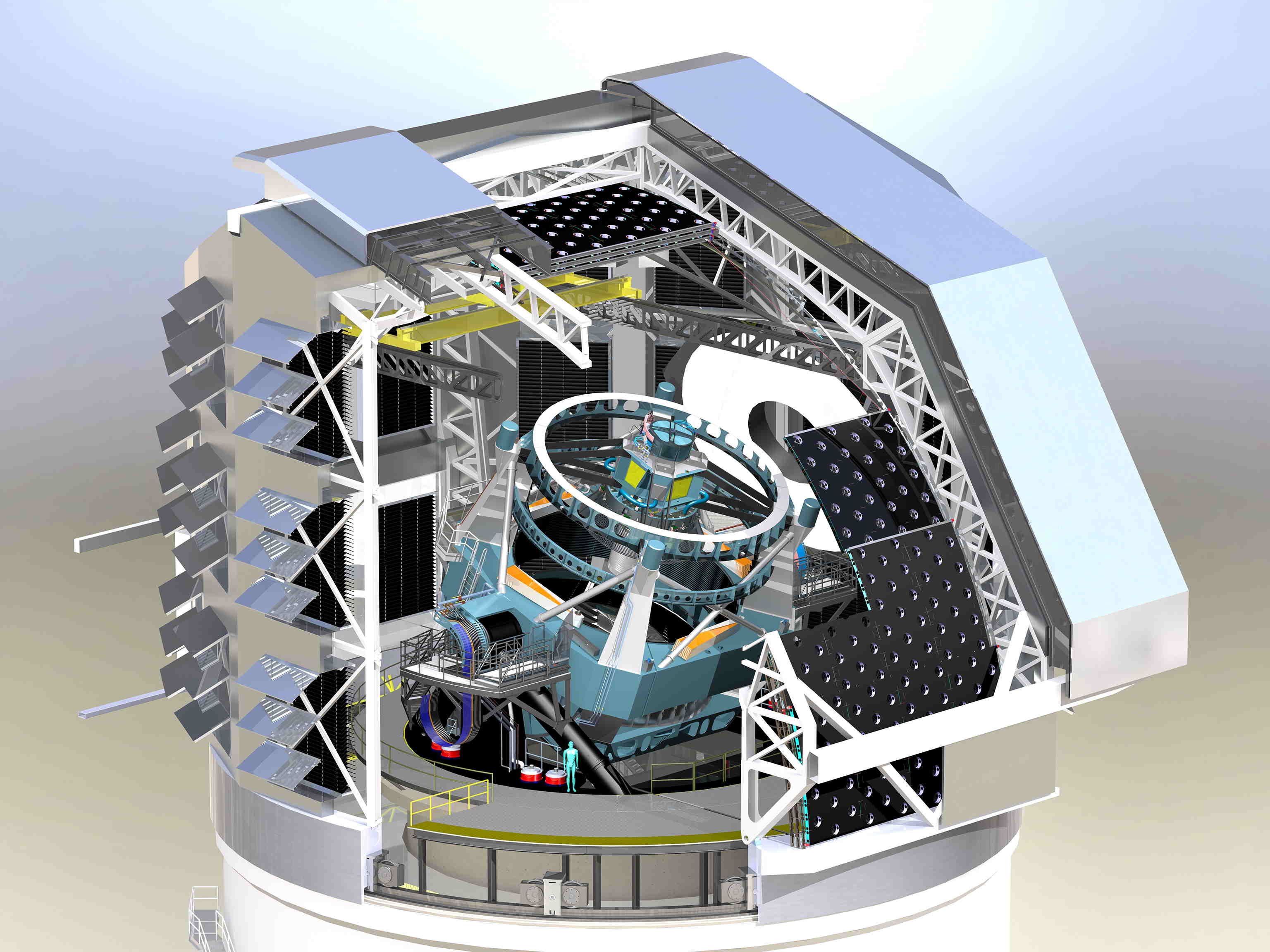}

\vspace{-0.2cm}

\caption{\small{\em LSST: cutaway of the dome showing the telescope
    within.}}   
\end{wrapfigure} 
\noindent
of the dynamics and constituents of the Universe. The primary science
thrusts within this frontier are understanding the nature of cosmic
acceleration (investigating dark energy); discovering the origin and
physics of dark matter, the dominant matter component in the universe;
and investigating the nature of primordial fluctuations, which is also
a test of the theory of inflation. A number of sky surveys in multiple
wavebands are now scanning the sky to shed light on these
questions. Near-future observations will be carried out by the
DESI~\cite{desi} and LSST~\cite{lsst} surveys in the optical, and by
the CMB-S4 survey~\cite{cmbs4} in the microwave band. These surveys
will generate extremely large datasets in the hundreds of petabytes
(PB). Very large radio surveys, such as the Square Kilometer Array
(SKA)~\cite{ska}, are also in planning stages.

Dark matter detection experiments fall under this Frontier. These
include direct detection experiments using cryogenic detectors (e.g.,
LZ~\cite{lz} and SuperCDMS~\cite{s_cdms}) and indirect detection using
high-energy particles from space (e.g., Fermi~\cite{fermi_exp} and the
High-Altitude Water Cherenkov Observatory
(HAWC)~\cite{hawc_exp}). Computational requirements in this sector are
small- to medium-scale and do not reach the extreme requirements of
large-scale sky surveys.

Cosmic Frontier white papers are A1.2 {\em Computational Cosmology at the
  Exascale} (compute-intensive) and A1.7 {\em Cosmic Frontier Experiments}
(data-intensive). Case studies are A2.3 {\em Computational Cosmology
  (HACC)}, A2.4 {\em Cosmic Reionization (ART)}, A2.5 {\em Dark Energy Survey in
  Hindsight}, and A2.9 {\em Cosmic Microwave Background (TOAST)}. 

\medskip 

\noindent{\bf\scshape{2.4~HPC Applications}}

HEP computing can be roughly divided into two broad categories --
compute- and data-intensive -- as discussed in the previous
sections. Historically, the two have been dealt with in very different
ways, using different architectures and facilities.

The category of compute-intensive tasks in HEP belongs primarily to
the HPC realm. The principal use cases of HPC within the HEP
community include groups working on
\begin{wrapfigure}[15]{r}{2.4in}
\centering \includegraphics[width=2.4in]{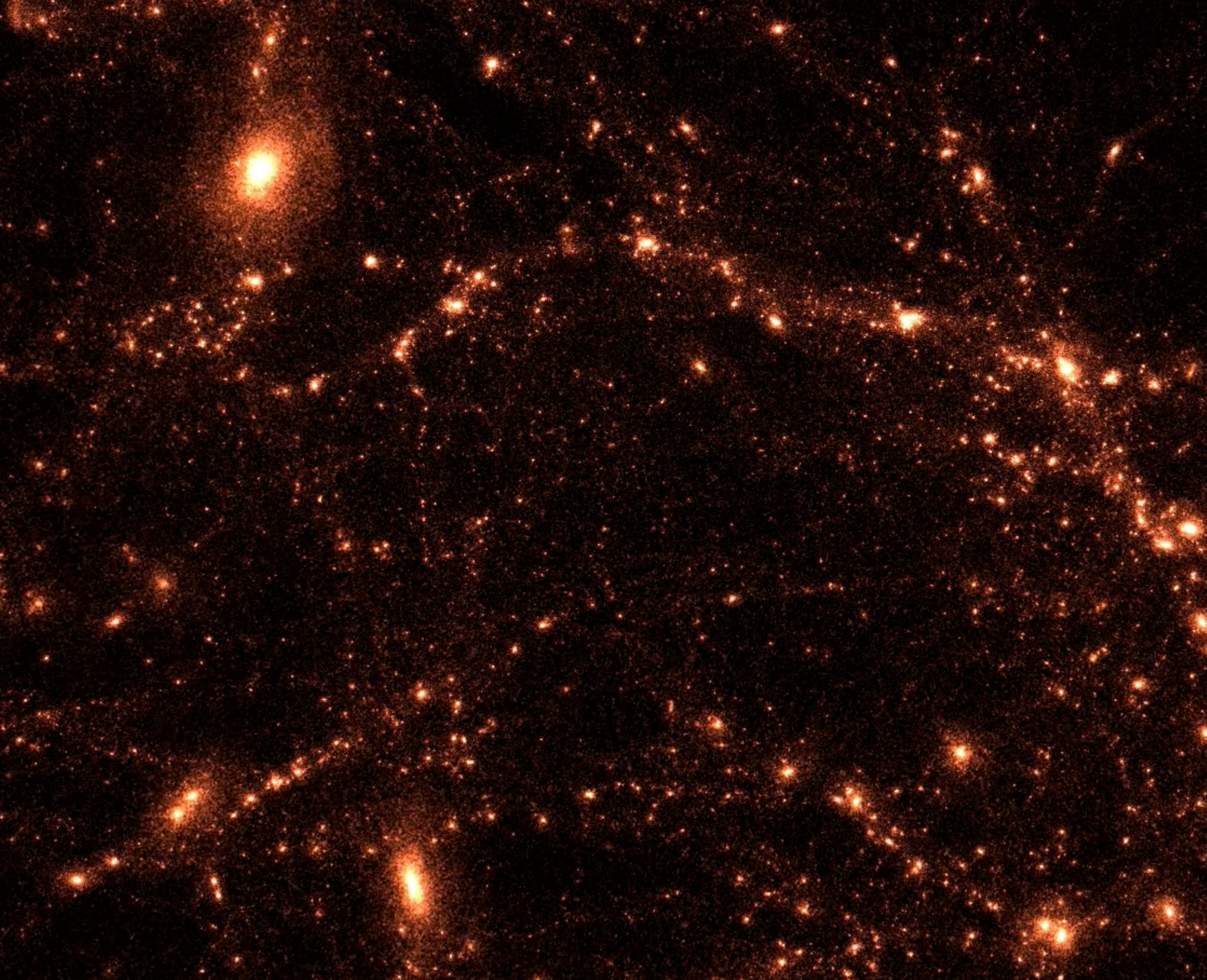}

\vspace{-0.2cm}

\caption{\small{\em Zoomed-in image of a region of the universe in the
    trillion-particle `Outer Rim' simulation.}}
\label{outer_rim}  
\end{wrapfigure} 
\noindent
accelerator modeling, cosmology simulations (Figure~\ref{outer_rim}),
and lattice QCD; all have been long-time partners with the ASCR
community.  These users possess considerable computational
sophistication and expertise; they leverage significant
state-of-the-art computing resources to solve, at any one time, a
single, complex, computationally demanding problem. 

Very large datasets can be generated by the above HPC applications --
potentially much larger than those from any experiment. Analyzing
these datasets can prove to be as difficult a problem as running the
original simulation. Thus, data-intensive tasks currently go hand in
hand with many HPC applications that were originally viewed as being
only compute-intensive.

\medskip

\noindent{\bf\scshape{2.5~HTC Applications}}

In contrast to HPC applications, the data-intensive use cases in HEP
experiments 
\begin{wrapfigure}[14]{r}{2in}
\centering \includegraphics[width=2in]{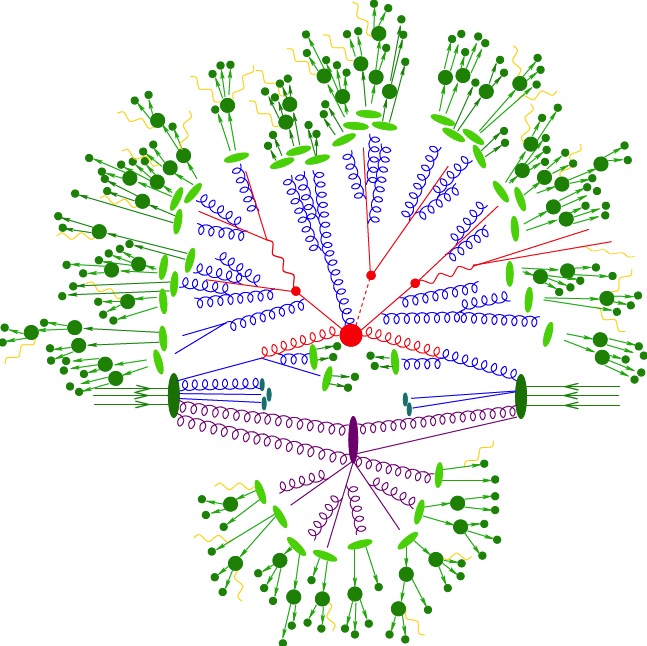}

\vspace{-0.2cm}

\caption{\small{\em Collision event generated with Sherpa.}} 
\label{sherpa_event}  
\end{wrapfigure} 
\noindent 
exploit HTC systems to take advantage of the inherent parallelism of
event-centric datasets. HEP teams have built large computing grids
based on commodity hardware in which each computing node handles a
computing problem from start to end.  A typical example of an HTC
application is event simulation (Figure~\ref{sherpa_event}) and
reconstruction in Energy and Intensity Frontier experiments.

The HEP community that has primarily focused on HTC applications is
actively considering the use of HPC resources. Sherpa, for example,
has been ported to HPC systems, and ATLAS and CMS have ongoing efforts
that use HPC resources for event simulation. It is fair to state,
however, that the community is at an early phase of learning how to
leverage HPC resources efficiently to address their computing
challenges.

\medskip

\noindent{\bf\scshape{2.6~Future Architectures and Portability}}

\begin{figure}\centering \includegraphics[width=6.5in]{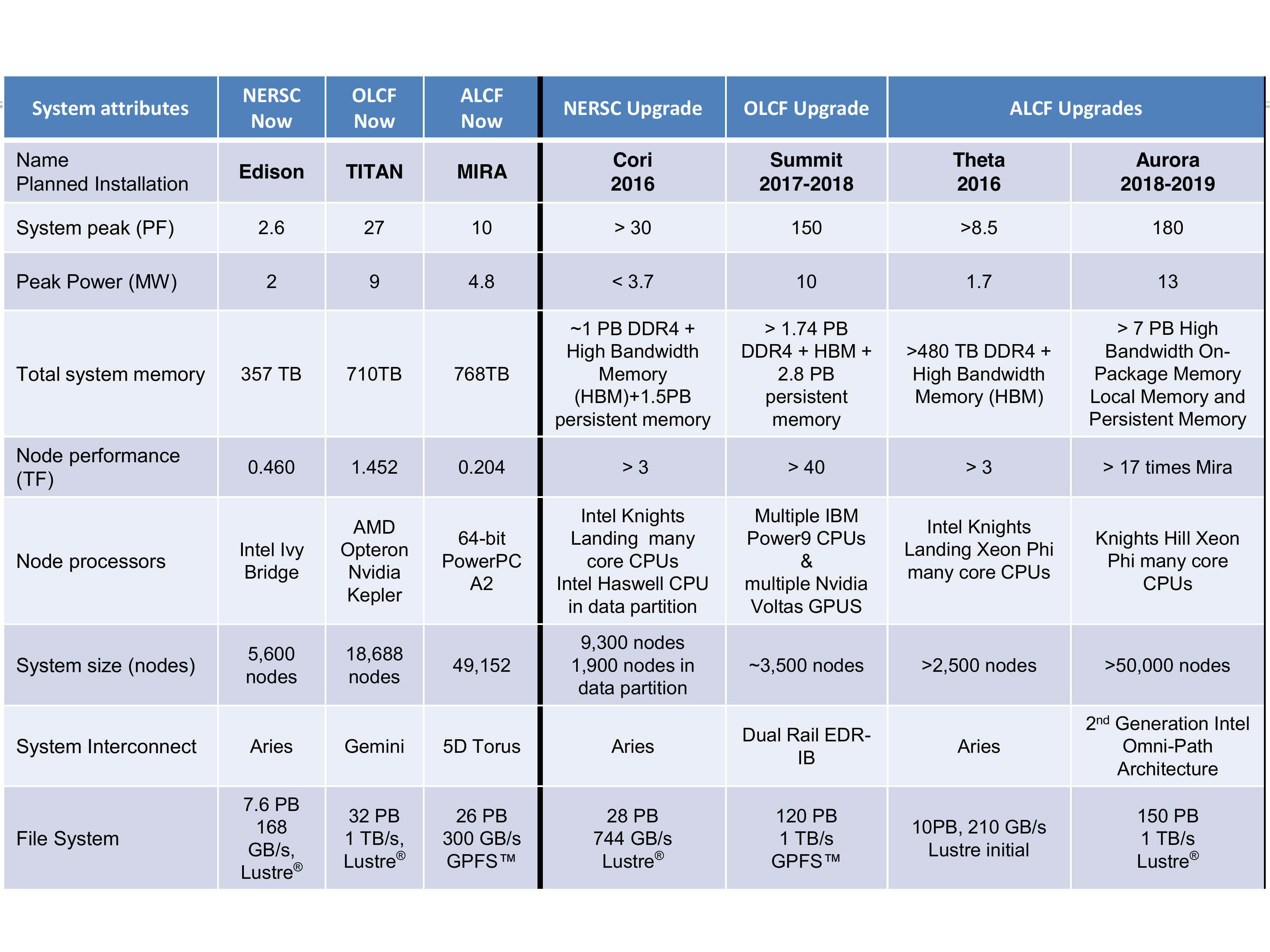}

\vspace{-.8cm}

\caption{Evolution of the main computational systems at the ASCR HPC
  facilities from the current petascale systems to near-future
  pre-exascale systems, continuing the two swim lanes of many-core and
  CPU/GPU nodes. Note the increase in peak performance and the
  reduction in the ratio of system memory to performance, a trend
  shared by the ratio of inter-node bandwidth to the peak flop
  rate. These trends are part of the challenge posed by
  next-generation systems.}
\label{ascr_systems}
\end{figure}

A common problem that both the compute- and data-intensive communities
within HEP face is the possible proliferation of ``swim lanes'' in
future computational architectures and the difficulty with writing
portable code for these systems. Currently, and in the next generation
of large ASCR systems (`pre-exascale'; see Figure~\ref{ascr_systems}),
there are only two types of computational architectures available --
CPU/GPU (accelerated) and many-core (non-accelerated). While HPC users
can imagine running codes on these two types of architectures -- and
even this is limited to only a few teams -- data-intensive users have
a much more difficult planning decision to make. Disruptive changes
cannot be made often to the HEP experiment software stack, and even
then, only with considerable difficulty. This means that it is very
likely that future evolution of this software will follow conservative
trends, which for now, appear to lead down the many-core
path. Although we cannot predict the detailed future of the exascale
environment with precision, from what is currently known, this
strategy would appear to make sense. The above argument suggests that
a parallel effort in developing portable programming models for the
exascale would be particularly beneficial for data-intensive
applications. These are not typically characterized by a few
highly-tuned kernels, but by a number of chained subprograms that may
not be individually tuned for performance (nor is there usually an
attempt to apply global optimization). Therefore, in this case,
portability may not necessarily be accompanied by an unavoidable loss
in performance, as is the case for the vast majority of HPC
applications.

\medskip

\noindent{\bf\scshape{2.7~Evolution of HPC Architectures}}

The evolution of HPC architectures is driven by a number of
technological, financial, and power-related constraints. Although
system peak performance is continuing to follow a Moore's Law
scenario~(Figure~\ref{ascr_systems}), future architectures are
focusing performance within relatively loosely coupled nodes, with
high local concurrency and less RAM per core than in current
systems. Optimization of computational efficiency will likely drive
the use of simpler data structures and communication-avoiding
algorithms in HPC applications.

At the same time, there is a general trend to investigate the notion
-- if not to immediately adopt it -- that future HPC platforms should
be able to perform some subset of scientifically important,
data-intensive tasks. This trend is attributable to the need for more
computational power from data-intensive applications, but also to the
substantial advantages of having data analysis performed on the same
systems where the underlying detailed theoretical modeling and
simulations are also being run.

In order to optimize HPC architectures for HEP data-intensive
applications, it would be desirable for the future HPC environment to
provide more balanced I/O on local and global scales and to make more
memory available per core, or to present faster access to remote
memory. Availability of node-local NVRAM can not only provide
I/O-related benefits (``burst buffers'') but will also allow a
write-once, read-many (WORM) capability that can be exploited by
data-intensive applications (e.g., for holding detector
description/status databases).  In addition, there are issues related
to the global infrastructure: How is data moved to the facility,
transferred to the compute cores, and then transferred off again in a
timely fashion?  Furthermore, addressing the need for automated
work/data flows is a key requirement (partly connected to
cybersecurity and partly to interactions with schedulers), which adds
a requirement for elasticity, which HPC systems do not currently
implement, not so much because of technological limitations, but
because of implementation of usage policies.  It is not uncommon today
for HEP data-intensive users running scientific workflows -- from a
single console -- to launch thousands of jobs each day on facilities
located all over the world without logging into each facility
separately.

As a further complication to data-intensive computing, many
experimental software stacks are very complex; enabling some of them
to run on HPC systems will likely require use of software containers
as essential resources. The expertise to rewrite the sofware may not
be available on the required timescale. Fortunately, rapid progress is
being made in this area.

\medskip

\noindent{\bf\scshape{2.8~Resource Management}}

It is important to keep in mind that the scale of HEP science is very
large; there are almost no ``single-investigator'' activities in areas
of HEP science relevant to this review, including research in theory
and modeling. HEP experiments typically plan $5-10$ years out, for
example, in terms of accelerator luminosity, but also in terms of data
and computing resources.  Competing yearly for major allocations, the
procedure currently used by ASCR facilities, creates an uncertainty
that is close to unacceptable for large HEP-supported projects or
science campaigns in all three frontiers; large resource fluctuations
can seriously jeopardize attainment of the HEP community's science
goals. There needs to be a mechanism in place to provide the necessary
resource stability.

An exascale HPC environment possessing a significant data-intensive
capability would be extremely attractive to the HEP community.  As an
example of synergy, HEP computing facilities could leverage the
associated technology and provision their own data centers in
accordance with the larger ASCR facilities.  Not only would this be
cost effective, it would also allow the design of a uniform HEP code
base for the system architectures adopted by both the HEP and ASCR
facilities.  This strategy would simplify code maintenance, expand the
pool of experts who can leverage these resources effectively, and
enable HEP users to execute their scientific workflows based solely on
availability and not on the machine architecture.

\newpage

\begin{center}
{\bf\scshape{3~Path Forward}}
\end{center}

In this section we cover specific technical, organizational, and other
challenges facing the HEP community as it moves forward to take
advantage of the next generation of computational systems and data
storage and transfer technologies. Exascale users will face a number
of difficulties, such as stability in hardware and software, system
resilience issues, evolving programming models, complex system
hierarchies, etc. -- some of which are merely extensions of current
problems and some of which will be new to the exascale
environment. Here, we focus on the issues that are particularly
relevant to HEP applications.

\medskip

\noindent{\bf\scshape{3.1~Compute-Intensive Applications}}

The computational challenges ahead for the HEP community have been
broadly outlined in the white papers -- this material covers, in
principle, most of HEP's HPC-oriented computational practice. The
compute-intensive areas are covered in the following white papers A1.1
{\em Exascale Accelerator Simulation}, A1.2 {\em Computational
  Cosmology at the Exascale}, A1.3 {\em Lattice QCD}, and A1.4 {\em
  HPC in HEP Theory}. These white papers discuss current science
topics and computational approaches, as well as future plans.

From the point of view of computational needs, given the science
targets, the raw demand in terms of core-hours (from all of the white
papers, individually and summed) seems to be very reasonable and
completely consistent with the expected availability of computational
resources at the Leadership Computing Facilities (LCFs) and NERSC on
the $2020-2025$ timescale -- the HEP requirements, in terms of
core-hours, amount to roughly 10\% of the total that is likely to be
available. It is important to note, however, that the different
applications make demands on HPC systems in different ways --
computational cosmology and lattice QCD have the highest computational
density, and computational cosmology codes are also memory-limited (as
are a few examples in accelerator simulation). To run the problems at
the required scales, even next-generation systems are on the smaller
side of some of the requirements. The HEP theory use case is more of a
mid-scale computing example, and makes relatively light demands on the
HPC system network.

The HEP compute-intensive community is abreast of the current
computational state-of-the-art, although new algorithm development
continues for a variety of tasks. Other work includes performance
optimization and adoption of better software engineering
practices. The evolution to next-generation systems is less clear,
however. Certain teams within the lattice QCD and computational
cosmology groups already have extensive experience with GPUs and
many-core hardware, but these two options should be viewed as a
disruptive set of technologies for most applications. The SciDAC
program and other opportunities, such as the Early Science Projects at
the LCFs and the NERSC Exascale Science Applications Program, will
help to accelerate the adoption of new programming paradigms.

The consensus in this area regarding what is needed is a set of common
tools across architectures -- efficient parallel primitives -- that
are separately optimized for each. These tools should possess flexible
data layouts so codes can adapt to different memory hierarchies.
Examples include operators such as shift, scan, reduce, transform,
scatter-gather, geometry remapping, etc. Moreover, it should be
possible to easily replace primitives with hand-tuned code, when
needed. It would be useful to compile a list of desired primitives and
to define appropriate interfaces. It is important to note here that
there is no desire for heavy-duty frameworks and libraries that would
be hard to develop, maintain, and modify. The emphasis is primarily on
lightweight tools, especially because the future computing environment
is expected to evolve considerably. A possible suggestion is that a
few application codes should function as testbeds for developing the
envisioned toolkits. Additionally, emphasis should be placed on the
benefits of having better network fabrics on the HPC systems and on
increased memory/node. At a lower level, there is some concern
directed at the problems of node-level memory management. A common
interface for managing data placement and movement across different
levels of the memory hierarchy would be very desirable.

As noted in two of the white papers (A1.2 and A1.3), data storage and
transfer will become significant issues. In fact, it can be argued
that for the compute-intensive field of computational cosmology, the
point where current needs are not being met has already been crossed.

Finally, as mentioned in Section~2.4, compute-intensive applications
can generate very significant amounts of data -- potentially
significantly larger than the data volume from the experiments -- that
has then must be processed in a number of different ways. This issue
is of particular concern to the lattice and computational cosmology
groups, who either have their own resources for off-line analysis and
storage (lattice) or are also leveraging other approaches, such as
in-situ analysis and use of software containers and virtualization
(cosmology). This general area is currently characterized by a number
of ad hoc approaches; achieving production-level status in the next
two years is an important goal, so as to keep in sync with the arrival
of next-generation systems at the LCFs and NERSC.

\medskip

\noindent{\bf\scshape{3.2~Data-Intensive Applications}}

Up to this point, the data-intensive HEP community has been very
successful in terms of its ability to leverage computing technology in
order to carry out its science. Yet, there are significant challenges
ahead that must be overcome before HEP groups can effectively leverage
large-scale HPC resources for many of their applications. The relevant
white papers here are A1.5 {\em Energy Frontier Experiments}, A1.6
{\em HEP Experiments: Data Movement and Storage}, A1.7 {\em Cosmic
  Frontier Experiments}, and A1.8 {\em Intensity Frontier
  Experiments}. Currently, the Cosmic Frontier and Intensity Frontier
experiments generate relatively modest amounts of data, and the
primary use cases are the two LHC experiments. In the future, the
other Frontiers will be characterized by significantly larger data
volumes, but they are unlikely to stress the available resources in a
way comparable to the LHC needs. Except for a few cases in the Cosmic
Frontier, all of HEP's data-intensive computing uses the grid model;
forecasts of future computing needs versus what the HEP grid can
provide are making it very clear (see white papers A1.5 and A1.7) that
access to alternative computational resources will be necessary.

With some very rare exceptions, it is not an easy task to run the HEP
experiment software stack ``out of the box'' in an efficient way on
HPC systems.  The current code base is not well-suited to run on HPC
machines (designed for HTC and not for HPC resources, lack of
vectorization, threading, unsuitable code organization, heavy I/O with
small files), and it will need to be significantly modified and
appropriately refactored in order to use these resources efficiently.

This is a significant problem, and the HEP workforce as it stands does
not possess the skill set to tackle it on the required timescale.
Training -- through summer schools, boot camps, and mentors -- will
therefore become increasingly important. ASCR expertise can play a
significant role in helping to modernize the HEP workforce in this
area. An effort is now underway to develop a set of
``mini-apps''~\cite{miniapp} through the HEP-FCE -- for both data- and
compute-intensive applications -- to help HEP researchers understand
the new computing environment and ensure that HEP codes can scale
appropriately at reasonable performance levels.  The mini-apps can
also be used as testbeds for prototype exascale hardware to ensure
that HEP applications and workflows will perform adequately.

More generally, however, the problem is that it is simply not possible
to train thousands of HEP researchers to become HPC experts, nor is it
desirable. The goals of such a training program would be to develop a
core of expertise within HEP, which would allow structured refactoring
of the current code over time, and to construct usable frameworks that
would allow thousands of scientists to run large-scale analysis or
other data-intensive tasks without having to become HPC experts.

The complexity of the current HEP experiment software base (many
millions of lines), a substantial fraction of which needs to be
available at the level of the compute nodes, and the fact that it
changes on relatively short time scales, is a significant problem for
HPC systems. Fortunately, this problem can be substantially mitigated
by the use of software containers, such as Docker. Although HPC
systems currently do not offer this capability, the hope is that they
soon will.

One of the significant issues facing data-intensive applications in
the exascale environment involves incorporating HPC machines into the
experiments' workflows. One difficulty is that Energy Frontier
applications require real-time remote access to resources such as
databases while the associated jobs are running. Because of the
fine-grained nature of the computational tasks, the associated I/O
bandwidth requirements are a potentially significant issue and need to
be properly characterized, not only for the systems being used, but
also at the facility scale. A second difficulty is that HPC systems
are scheduled for maximal utilization in a way that is problematic for
large-scale complex workflows, especially those that require true
real-time access. The problem is that a given workflow may require
widely varying resources as it runs, in ways that may not be easily
predictable beforehand. This mode of operation is essentially
orthogonal to current batch scheduling implementations on HPC
systems. Moreover, truly addressing the elasticity requirement implies
a substantial degree of overprovisioning, which is hard to imagine
implementing at current HPC facilities. Note that having containers
available will not solve the problem of elasticity, although it will
help by potentially providing the ability to quickly spin up and spin
down machine partitions dedicated to a particular workflow.

A significant barrier at the moment is that each of the ASCR
facilities is configured differently.  There is little uniformity with
respect to scheduling jobs, cybersecurity, staging data, etc.  It
would be significantly easier if all HPC systems could be accessed via
a uniform fabric.  ``Edge Servers'' needed to stage data outside of
the facility firewall will play an important role in data-intensive
applications, including potentially addressing access control issues
on HPC systems.  A uniform implementation of access protocols would
make interfacing to the facilities much simpler.  Lastly, uniformity
in cybersecurity protocols across the facilities would be very
desirable.  A single, federated model that works at all three centers
would be ideal.

The volume of data produced by experiments is rising rapidly. The
current LHC output is on the order of 10PB/year and in the HL-LHC era
(2025), this number is expected to rise to 150PB/year. No experiment
in the Cosmic or Intensity Frontier will come close to this data rate
or total volume (EB scales). For the most part, one can expect that if
the volume is sufficiently large, the projects will budget for and
purchase the required storage. Below a certain threshold, the ASCR HPC
centers provide disk storage in parallel file systems and much larger
archival tape storage using HPSS. The timescale over which this
storage continues to exist is subject to success or failure in the
ASCR Leadership Computing Challenge (ALCC) and Innovative and Novel
Computational Impact on Theory and Experiment (INCITE) process (and to
a much lesser extent at NERSC). In any case, if the HPC systems are
used as a data-intensive computing resource, much more attention will
have to be paid to storage resources in terms of capacity, latency,
and bandwidth (as is also the case for compute-intensive applications
that generate large amounts of data, as mentioned in Sections~2.4 and
3.1). In particular, high-performance, reliable, and predictable data
transfer across multiple storage endpoints would be an essential
requirement.

\medskip

\noindent{\bf\scshape{3.3~Evolution of the ASCR/HEP Partnership}}

In the context of a future exascale environment -- or even in the near
term -- a number of needs for the HEP computing program have been
identified thus far, as described above. These needs include
programming on new architectures and associated algorithm development,
data storage and transfer, refactoring of a potentially significant
fraction of the code base (over ten million lines of code),
co-evolution of ASCR and HEP facilities, and training of HEP
personnel. To address these needs, the HEP community is very
interested in identifying and exploring new ways to partner with
ASCR. The exascale environment opens up exciting new scientific
opportunities; the knowledge and expertise that ASCR has are
invaluable to the HEP community in its pursuit of new scientific
opportunities.

One difficulty is that current partnership opportunities are limited
under today's organizational structure.  Outside of the SciDAC  
programs, and informal contacts, there are limited established
mechanisms to develop long-term partnerships between the two
communities. This is an area that calls for serious attention.

A possibility discussed at the review is to enhance or re-direct
partnerships where some entity -- like a joint ASCR/HEP institute (and
similar entities for other sciences if appropriate) -- would provide
multi-faceted partnerships with ASCR research and facilities
divisions.  These new entities would have broad-ranging computational
expertise and be dedicated to the scientific success of the community
they are supporting, forming valuable additions and expansions to the
successful SciDAC Program. As an important example, the fact that a
complex HEP software suite needs to run ``everywhere'' requires an
approach where a joint ASCR/HEP group of experts (``embedded power
teams'') work together to solve the associated portability problems
within the larger computational context of a given project. In this
mode, scientists across different domains would get to know each
other, establish a common language, and have enough long-term
commitments to tackle difficult problems. In other words, the broader
computational support model should involve working with large
collaborations, as well as with small teams.

Another issue involves long-range planning to ensure availability of
resources; such planning is essential for HEP experiments because of
their extended lifetimes of five years or more (Section~2.8.) As
discussed in the white paper A1.5 {\em Energy Frontier Experiments},
``it is possible that by 2025, most processing resources will be
supplied through dynamic infrastructures that could be accessed
opportunistically or through commercial providers.'' If this is really
to come to pass, close interaction will be needed not only between the
ASCR computing facilities and HEP counterparts to support programmatic
needs, but also between HEP and ESnet to make sure that networking
resources are adequately provisioned. Joint design of ``Edge Servers''
(Section~3.2) and co-evolution of HEP facilities (Section~2.8) would
also be facilitated by such an interaction. To summarize, a
mission-oriented ASCR/HEP partnership for long-range activities needs
to be initiated.

\newpage

\begin{center}
{\bf\scshape{4~Summary of HEP Requirements}}
\end{center}

The purpose of this section is to encapsulate the quantitative
estimates of computing, storage, and networking that are presented in
the white papers. The white papers covered 1) the individual
compute-intensive HEP applications (accelerator modeling,
computational cosmology, lattice QCD, and HEP theory), 2)
data-intensive computing requirements from experiments at the three
frontiers, and 3) a separate white paper on Energy Frontier experiment
data movement and storage and one on the evolution of HEP
computational facilities, as driven by experimental requirements and
the external computing environment. The authors of the white papers
were asked to collate information from all the major use cases in
their domains, and, in particular, to provide an assessment of the
science activities and computational approaches on the $2020-2025$
timescale, as well as estimates for the computing, data, and services
needs over the same period. The case studies presented in Appendix~2
are meant to provide more in-depth examples of specific use cases,
which can range from medium-scale computing to full-machine exascale
simulations.

It is not easy to assess requirements roughly a decade into the future
for a diverse set of scientific applications because new breakthroughs
and unexpected obstacles are, by definition, difficult to predict. The
difficulty is compounded by the fact that two architecture changes are
expected over this timeframe -- the change from multi-core to
many-core (GPUs included in this classification) at all the major ASCR
computing facilities (pre-exascale) followed by another architecture
change at the exascale/beyond-exascale level, which could be
significantly disruptive. In addition, computational needs in
different areas do not have equal priority. Much depends on which
computations are considered more relevant at a certain time for HEP
science, and which can be pushed into the future. Computational needs
driven by ongoing projects and those planned or under construction
have been taken into account in the white papers. The future, however,
may turn out differently if the associated timelines change in ways
that lead to lack of coordination with the evolution of the
computational resources.

Given these caveats, it is nevertheless encouraging to note that the
compute-intensive HEP applications have kept pace with hardware
evolution -- including key pathfinder roles in several cases (for
example, lattice QCD for the IBM Blue Gene systems and computational
cosmology for Roadrunner, the world's first petaflop system). In
addition, as noted in the white papers, the science drivers show no
sign of letting up in their hunger for computational resources. The
situation in the case of experiments is more difficult to assess. In
the case of the Cosmic Frontier, it is becoming clear that the future
will lie in using ASCR facilities (NERSC is the host for DESI and also
the likely host for LSST Dark Energy Science Collaboration (LSST-DESC)
computing; the LCFs will provide additional resources) as the dominant
source of computational resources. HEP Energy Frontier experiments
have historically not used HPC sites for computing, although this is
rapidly changing. Intensity Frontier experiments have not been major
consumers of computational resources, but this situation is also
changing, even on the current timescale. How quickly all of these
areas will be able to take full advantage of ASCR resources depends on
the pace with which the relevant components of the HEP production and
collaboration software are refactored as well as how the ASCR
facilities evolve -- in their turn -- to address the HEP use cases.

We now consider the computational, data, and networking requirements
that can be extracted from the white papers. Unless explicitly stated,
the numbers listed in this section, in core-hours (for a standard 2015
X86 core), refer only to the computational requirements associated
with HEP use of ASCR facilities. The disk storage requirements are
given as an aggregate number, but it is conceivable that this storage
can be split between ASCR and HEP facilities depending on the
particular use case. Wide area network (WAN) bandwidth requirements
are given in a few cases where it is expected that the requirement
will potentially stretch ESnet capabilities. Local bandwidth within
the facilities is assumed to be at least as good as the wide area
requirement.

Accelerator modeling can be divided into 1) electromagnetics and beam
dynamics simulations for current and near-future technology machines
and 2) dedicated simulations to help in the design of future
accelerators. Currently, both electromagnetics and beam dynamics
consume on the order of 10M~core-hours annually each and these numbers
are expected to scale up to $10 - 100$ billion (G)~core-hours by
2025, depending on the use cases being run. Large-scale simulations
for future accelerators (machines that may be built on the $2030+$
timescale) focus on plasma-based acceleration schemes. While these
simulations are currently at the 10M~core-hours level, they can scale
up to an annual requirement of $1\rm{G} - 100$G~core-hours (or more)
by 2025, but there is significant uncertainty regarding the upper
value. Storage (or networking) has historically not been a major issue
for accelerator modeling, and this is unlikely to change in the
future.

Computational cosmology will have to support a large number of Cosmic
Frontier projects, some of which are quickly reaching the scale of HEP
experiments in terms of collaboration size. Current annual simulation
usage is at the $100\rm{M} - 1$G~core-hours scale and is expected to
increase to $100\rm{G} - 1000$G~core-hours by 2025. In addition,
large-scale cosmology simulations are already memory-limited and they
are likely to saturate the system memory of machines in the exascale
era. Storage requirements are likely to be large; currently they are
already at the level of 10PB of disk and they are likely to easily
exceed 100PB by 2025. Furthermore, because large-scale distributed
analysis will be needed by the collaborations, there will be
significant networking requirements. Currently, a pilot project with
ESnet is aiming to establish a 1PB/week production transfer rate for
moving simulation data. By 2025, the burst requirements will be
approximately 300Gigabits/s (Gb/s), which is roughly the same scale
as that required by Energy Frontier experiments.

Lattice QCD has a long history of efficient use of supercomputing
resources and this trend will continue into the foreseeable future. Current
annual usage is in the 1G-core-hour class and is expected to
increase to $100\rm{G} - 1000$G~core-hours by 2025. Memory
requirements (unlike computational cosmology) are nominal. Disk
storage, which has historically not been a major requirement, will
only increase slowly, from $\sim$1PB currently to $\sim$10PB by
2025. Theory requirements (event generation, perturbative QCD) are at
$1\rm{M} - 10$M~core-hours currently and these will likely
increase to $100\rm{M} - 1$G~core-hours by 2025. Memory and storage
needs for this effort will likely remain at easily satisfiable levels.

Cosmic Frontier experiments are currently running at the $10\rm{M} -
100$M~core-hours scale on HPC resources; on the 2025 timescale, this
is likely to increase to $1\rm{G} - 10$G~core hours. Disk storage
requirements are currently at roughly the $\sim$PB scale and are
likely to increase to $10 - 100$PB by 2025. Network requirements are
unlikely to stress future capabilities at the same level as Energy
Frontier experiments.

Energy Frontier experiments have begun using HPC systems relatively
recently, primarily for event simulation tasks. However, annual usage
has already reached the 100M~core-hour level on ASCR resources. This
usage level should be compared to the total US contribution to LHC
computing, which is on the order of $100\rm{M} - 1$G~core-hours
annually on HTC systems. By 2025, the requirement on HPC resources
could reach $10\rm{G} - 100$G~core hours, with extensive storage
needs, exceeding 100PB of disk space (the total global storage
requirement will reach the exabyte scale). The network requirements
will also be high, at the level of 300Gb/s (more in a continuous mode,
rather than burst operation).

Intensity Frontier experiments are at the $\sim$10M~core-hour level of
annual usage. This number is expected to increase to $100\rm{M} -
1$G~core hours annually. Storage requirements are roughly at the
$\sim$PB level currently and are expected to increase to $10 - 100$PB
by 2025.

The information discussed above is summarized in
Table~\ref{tab:req}. The sum of HEP requirements, although difficult
to pin down precisely because of the uncertainties discussed above, is
projected to be $\sim$10\% of the expected ASCR facility resources at
ALCF, NERSC, and OLCF by 2025.

\begin{table}
\captionsetup{justification=centerlast}
{\footnotesize
\begin{tabular}{|c|c|c|c|c|c|}
\hline
Computational & Current & 2025 &
Current & 2025 & 2025 Network \\ 
Task & Usage & Usage &
Storage (Disk) & Storage (Disk) & Requirements (WAN)\\ \hline\hline

Accelerator & $\sim 10\rm{M} - 100\rm{M}$ & $\sim 10\rm{G} - 100$G &
  &  &  \\ 
Modeling & core-hrs/yr & core-hrs/yr &
 & &  \\ \hline

Computational & $\sim 100\rm{M} - 1$G & $\sim 100\rm{G} - 1000$G &
$\sim$10PB  & $>$100PB &  300Gb/s\\ 
Cosmology & core-hrs/yr & core-hrs/yr &
 & & (burst) \\ \hline

Lattice & $\sim$1G & $\sim 100\rm{G} - 1000$G &
$\sim$1PB  & $>$10PB &  \\ 
QCD & core-hrs/yr & core-hrs/yr &
 & & \\ \hline

Theory & $\sim 1\rm{M} - 10$M & $\sim 100\rm{M} - 1$G &
  &  &  \\ 
 & core-hrs/yr & core-hrs/yr &
 & &  \\ \hline

Cosmic Frontier & $\sim 10\rm{M} - 100$M & $\sim 1\rm{G} - 10$G &
$\sim$1PB  & $10 - 100$PB & \\ 
Experiments & core-hrs/yr & core-hrs/yr &
 & &  \\ \hline

Energy Frontier & $\sim 100\rm{M}$ & $\sim 10\rm{G} - 100$G &
 $\sim$1PB & $>$100PB &  300Gb/s\\ 
Experiments & core-hrs/yr & core-hrs/yr &
 & &  \\ \hline

Intensity Frontier & $\sim 10\rm{M}$ & $\sim 100\rm{M} - 1$G &
 $\sim$1PB & $10 - 100$PB &  300Gb/s\\ 
Experiments & core-hrs/yr & core-hrs/yr &
 & &  \\ \hline

\end{tabular}
}
\caption{\label{tab:req} Approximate exascale requirements for major HEP
  computational activities. All quantitative estimates are for HEP
  requirements at ASCR facilities, i.e., no HEP facility estimates are
  included. Note that the quantities given in this Table only represent
  order of magnitude estimates. If no numbers are provided, the usage is considered to be
  small enough that it should be well within the facility
  specifications. The nature of the uncertainties in the estimates is
  discussed in the text.}
\end{table}

\newpage

\begin{center}
{\bf\scshape{Appendix 1: White Papers}}
\end{center}

The HEP computational landscape is diverse but can be sensibly
categorized by the type of computing (compute-intensive or
data-intensive) and by the individual application areas. The bulk of
the HPC-relevant computing in HEP is carried out in the areas of
accelerator modeling, computational cosmology, and lattice field
theory. All of these topics, as well as a smaller, but growing effort
in HEP theory, are represented by individual white papers. The bulk of
the computing in experimental HEP areas is dominated by data-intensive
applications, in which simulations also play an important role. The
three HEP frontiers -- Cosmic, Energy, Intensity -- have their own
white papers, and there is a separate one to address data movement and
storage in Energy Frontier experiments, as this is a very significant
component of HEP's computational program.

The white papers provide compact introductions to each area's current
science activities and computational approaches, and their expected
evolution on the $2020-2025$ timescale. Based on the relevant
requirements, estimates are presented for the compute, data, and
services needs in each area on the $2020-2025$ timescale. We emphasize
that each white paper attempts to represent a broad activity
comprising many researchers, scientific problems, and varied
computational tools. For this reason, the information in the white
papers is necessarily incomplete. References are provided separately
with each white paper to improve the coverage and include access to
more detailed information. The second Appendix on case studies gives
more specific examples of HEP science use cases and goes into detail
on the computational needs for these selected examples.

\newpage

\noindent{\bf\scshape{1.1~Exascale Accelerator Simulation}}  

\medskip

\noindent{\bf Authors:} J. Amundson, R. Ryne (leads), B. Cowan,
W. Mori, C.-K. Ng, J. Power, J. Qiang, and J.-L. Vay 

\medskip

\noindent{\bf{\em 1.1.1~Current Science Activities and Computational
    Approaches}}

Current research in accelerator simulation~\cite{accsim} can be
divided into three main categories: beam dynamics~\cite{beams},
electromagnetics and advanced accelerators. In beam dynamics, the
problems involve single- and multi-pass transport in externally
applied fields, self-fields, and fields generated by the
beam-environment interaction. In electromagnetics, simulations are
performed in order to obtain detailed predictions of fields in
accelerator components, primarily RF structures. Advanced accelerator
research focuses on future-generation concepts such as laser-plasma
accelerators (LPA or LWFA), plasma wakefield accelerators (PWFA), and
dielectric wakefield accelerators (DWA). All the above are heavy
consumers of contemporary high-performance computing resources.

Along with these three areas, another important topic involves
simulating the interaction of beams with materials~\cite{mars}. Such
simulations are important for designing shielding systems and for
predicting radioactivation from particle loss. This also includes the
simulation of particle production in high power targets and simulation
of machine-detector interface phenomena.

Nearly all the major accelerator codes are multi-physics codes. Beam
dynamics codes involve nonlinear optics, space charge, wakefield, and
other effects~\cite{impact, synergia}. Additional phenomena of
importance are beam-beam effects~\cite{beambeam} (for collider
studies), spin dynamics, and secondary particle emission (for
electron-cloud and other studies). Virtual prototyping involves
multi-physics analysis including electromagnetic, thermal and
mechanical effects~\cite{materials}. Such simulations are used, e.g.,
to control microphonics in superconducting structures. Codes used for
advanced accelerator modeling~\cite{adv1, adv2, adv3, adv4} model the
interaction of beams, lasers, and plasmas, but also require the
inclusion of ionization, radiation reaction, single quantum events,
spin polarized beams, and recombination. Modeling is essential to
explore and gain insight into the complex phenomena of laser- and
plasma-based concepts.

The techniques most widely used for simulations involving beams,
lasers, and/or plasmas are variations on the particle-in-cell (PIC)
approach, including electrostatic, quasi-static, and fully
electromagnetic PIC. LPA simulations also use laser-envelope
models. Electromagnetic simulations involving complicated geometries
generally use finite element techniques or cut cell techniques.

Until recently, the dominant computing model in the field has been a
pure MPI approach. However, in recent years many accelerator
applications have moved to a hybrid MPI-OpenMP approach. The most
recent advances include ports to multiple GPU- and Intel Phi-based
architectures, including hybrid MPI-OpenMP/CUDA approaches. While a
substantial amount of progress has been made in this area, most
mainstream applications are using pure MPI or MPI+OpenMP hybrids.

\medskip

\noindent{\bf{\em 1.1.2 Evolution of Science Activities and
    Computational Approaches on the 2020/2025 Timescale}}

In light of the P5 report~\cite{p5}, the computational needs for
2020/2025 fall into three categories: simulations in support of (1)
high-priority domestic facilities at Fermilab, (2) high-priority
off-shore facilities (i.e., LHC upgrades), and (3) R\&D in advanced
technologies (i.e., laser-/beam-plasma and dielectric accelerators).

\begin{enumerate}
\item {\em Simulations in support of high-priority domestic
    facilities:} The evolution of these activities is dominated by the
  shifted role of Fermilab to the Intensity Frontier. This includes
  PIP-II, DUNE, Mu2e, g-2, and PIP-III R\&D. Successful execution of
  PIP-II depends on current activities including electromagnetics
  simulations for the PIP-II linac and beam dynamics simulations of
  all the various components of the Fermilab accelerator
  complex. Accelerator simulation efforts for PIP-III will require
  greater fidelity simulations with better statistics than ever
  before, due to the extremely small loss requirements that will
  follow from extreme high-power running.
\item {\em Simulations in support of LHC upgrades:} The evolution of
  these activities is dominated by the High Luminosity LHC Upgrade
  (HL-LHC). Similar to the previous paragraph, the emphasis on higher
  intensity in the injector chain will require large-scale,
  multi-physics simulations in order to predict losses, explore
  mitigation strategies, and optimize performance. In addition, the
  increased intensity and number of bunches will present new
  challenges to the beam-beam simulations. Such simulations are needed
  to maximize luminosity and hence discovery potential.
\item Simulations in support of R\&D in Advanced Concepts: This area
  focuses on creating the next generation of accelerating structures
  (either plasma or dielectric), which can be excited by short,
  intense beams or laser pulses to sustain extremely high accelerating
  gradients. LPA and PWFA activities currently emphasize $\geq10$ GeV
  stages, improved emittance and energy spread control, and staging.
  DWA activities are focused on beam breakup control of the drive beam
  in the presence of large wakefields, and structure heating. Major
  activities are the BELLA Center~\cite{bella}, FACET\cite{facet},
  AWA~\cite{awa}, and potential upgrades on the 2020/2025 time scale.
\end{enumerate}

Structure design and optimization cuts across all these areas. Virtual
prototyping has the potential to dramatically reduce the time and cost
to develop structures, to optimize their performance, and to reduce
risk. End-to-end integrated simulation of the PIP-III linac consisting
of multiple cryomodules will require 2-3 orders of more computing
resources than current simulations of a single cryomodule.

For all the above, exascale modeling will involve scaling to $>1$M
cores, new data structures for GPUs and Xeon Phis, new programming
approaches, and new algorithms.  Along with these priorities, DOE HEP
supports other accelerator projects that will benefit from future HPC
resources. Examples include ASTA, ATF, IOTA, UMER, R\&D in high-field
magnets, and R\&D to understand effects that limit gradients in room
temperature and superconducting cavities. Exascale activities will, in
general, be carried out through collaborations involving national labs
and universities.

In the 2020/2025 time scale we will have to accelerate three current
trends: the move to deeper multi-level parallelism, the increased use
of multithreaded shared memory programming techniques, and the
improved utilization of instruction-level vectorization. The latter
will probably require both advances in compilers as well as
modifications of programming practices. Use of multithreaded
techniques currently exists in OpenMP-based and CUDA-based GPU
methods. While current efforts typically target scalability to a few
threads or small number of GPUs, future efforts will require
scalability that encompasses many threads and/or GPUs. Multi-level
parallelism has its current realization primarily in multi-level MPI,
hybrid MPI-OpenMP, MPI-CUDA, and MPI-OpenMP-CUDA. Some instances are
application-specific, as with bunch-bunch train splitting. Multi-level
parallelism is also used for parallel optimization. The architectures
of the exascale era will will push us to 4- or 5-level parallelism,
and involve scaling to millions of cores.

To make use of millions of cores, we will need software whose
performance has been optimized for the exascale. This includes
parallel FFTs, linear solvers scalable in memory and on multi-core
computer architectures, parallel AMR capabilities, and technologies
for load balancing such as parallel particle- and field-managers, all
capable of running at exascale.

We expect even greater impetus for multi-physics modeling. Simulations
that combine beam dynamics, electromagnetics, lasers, plasmas, and
beam-material interactions would open a new era in accelerator
design. It would enable, e.g., self-consistent modeling of
radioactivation from dark current, from halos and ultra-low losses in
high-intensity beams, and complete advanced concepts-based collider
designs. The architectural complexity of exascale systems will present
a challenge to efficiently componentize and share computational
capabilities and components.

\medskip

\noindent{\bf{\em 1.1.3 Compute, Data, and Services Needs on the 
    2020/2025 Timescale}}
 
Raw CPU power is the driver for the majority of the needs of
accelerator modeling. Storage needs are typically small compared with
other areas of computational HEP. As an example of the CPU
requirements, consider PIP-III. A 2015 INCITE project that included
PIP-II~\cite{pip2} estimated that 8-bunch simulations over 4500 turns
of the Main Injector and Recycler would require 10M core-hours on a
current BG/Q system. Scaling for better statistics (10$\times$),
longer simulations (5$\times$) and full machines (from 8 bunches to
588 bunches) require nearly 4,000 times the CPU. This represents only
a fraction of the total effort required for a campaign of end-to-end
simulations. Another example is provided by beam-beam modeling for
HL-LHC~\cite{hl_lhc}: Runs with 4000 coupled bunches, with 4 working
points optimization (not even including the 8 crab cavity multipoles),
with 3-level parallelization (with modest 100$\times$100$\times$512
cores), would easily use a fraction of an exascale system. In regard
to structure design, estimates indicate an expected need for 50M
core-hrs/yr~\cite{struct_req}. Finally, modeling of advanced concepts
demands exascale resources~\cite{adv_exa}. This is especially when
simulating staged systems, and for parametric studies for tolerance to
non-ideal effects. Such simulations can use 100 million CPU-hours per
run or more, although reduced models can lower requirements to be in
accord with available resources.

\newpage

\noindent{\bf\scshape{1.2~Computational Cosmology at the Exascale}}  

\medskip

\noindent{\bf Authors:} A. Almgren, K. Heitmann (leads), N. Frontiere,
S. Habib, Z. Lukic, P. Nugent, B. O'Shea

\medskip

\noindent{\bf{\em 1.2.1~Current Science Activities and Computational 
    Approaches}}

Large-scale simulations play key roles in cosmology today, including:
1) exploring fundamental physics and probes thereof, e.g., dark
energy, dark matter, primordial fluctuations, and neutrino masses, 2)
providing precision predictions for a range of cosmological models,
important for data analysis, 3) generating detailed sky maps in
different wavebands to test and validate analysis pipelines, 4)
understanding astrophysical systematics, e.g., baryonic effects, and
5) providing covariance estimates.

Cosmology simulation codes fit into two main categories: gravity-only
(``N-body'') solvers and ``hydrodynamics'' codes that also include gas
physics and associated subgrid modeling (e.g., cooling, astrophysical
feedback, star formation).  The tasks above require four kinds of
simulations using a mix of these capabilities, depending on the
specific application: 1) gravity-only and hydrodynamics simulations
over a range of cosmological models to address the first and second
tasks, 2) very high resolution large volume gravity-only simulations
(for, e.g., large galaxy catalogs) and medium resolution large volume
hydrodynamics simulations (for, e.g., thermal Sunyaev-Zel'dovich maps
and Lyman-alpha investigations) to address the third task, 3) very
high resolution hydrodynamics simulations including treatment of
feedback effects to address the fourth task, and 4) a very large
number (well beyond thousands) of moderately accurate gravity-only
simulations to address the fifth task.

Approaches used today for the gravity-only solvers include
particle-mesh plus short-range solvers (particle-particle or tree
methods)~\cite{springel, habib}, pure tree methods~\cite{hot, stadel},
and pure grid methods.  Codes that include hydrodynamics coupled with
an N-body representation of dark matter include grid-based hydro
methods, typically using Adaptive Mesh Refinement (AMR)~\cite{bryan,
  nyx, ramses, art}, Smooth Particle Hydrodynamics
(SPH)~\cite{springel2}, and Moving Mesh Methods~\cite{hopkins}.

Analysis of the data generated in the simulations is fundamental to
addressing the research goals.  Currently most of this occurs in
post-processing, but the large amount of data from future simulations
and increased computational expense due to more complex analyses will
result in more reliance on in-situ approaches.

\medskip

\begin{wrapfigure}[16]{l}{4in}
\centering \includegraphics[width=4in]{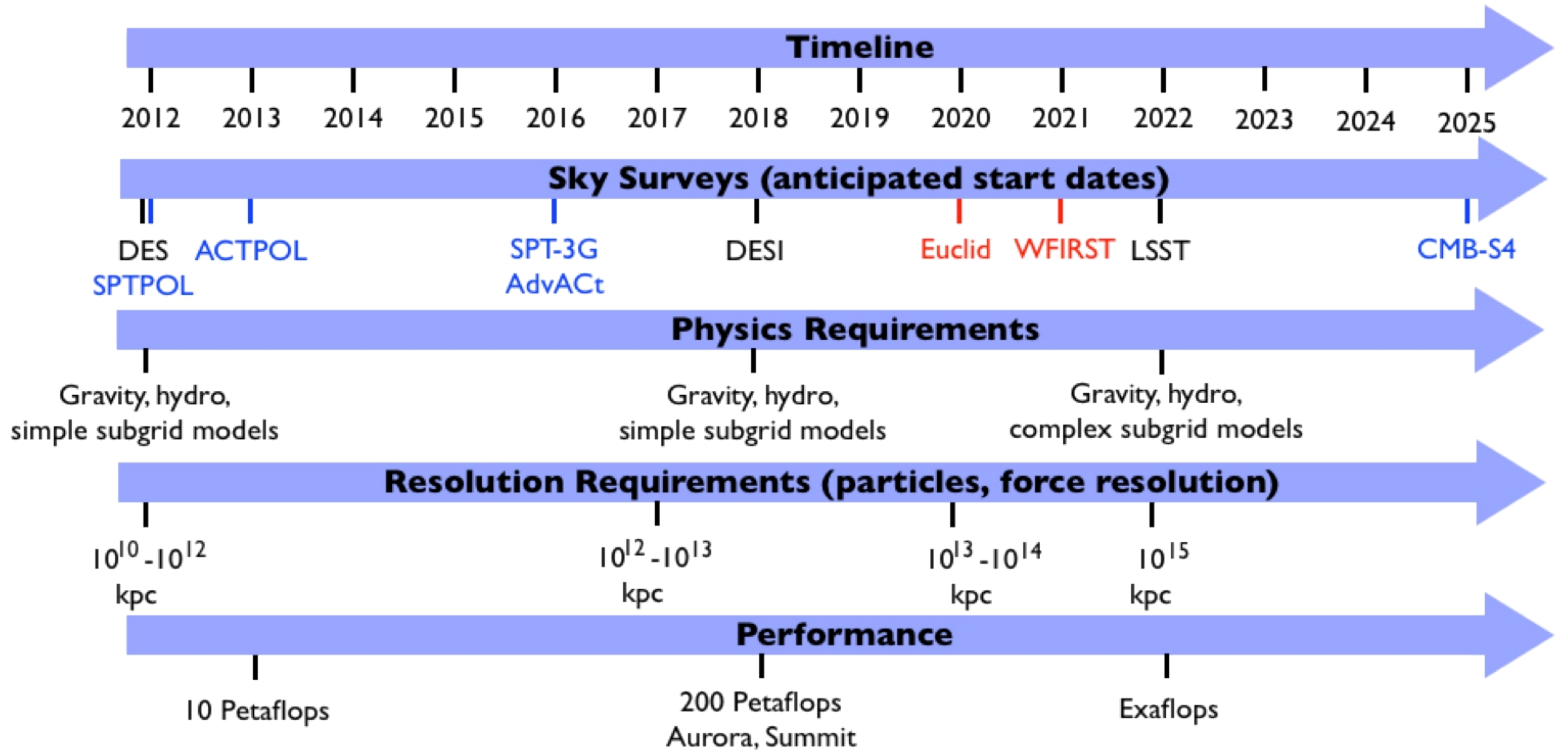}

\vspace{-0.2cm}

\caption{\small{\em Timelines for cosmological surveys (blue: CMB,
    black: optical, ground-based, red: optical/NIR satellite)
    and supercomputing resources as well as simulation
    requirements. DOE HEP plays a major role in the ground based
    optical surveys and in some of the CMB surveys.}}
\end{wrapfigure}

\noindent{\bf{\em 1.2.2~Evolution of Science Activities and 
    Computational Approaches on the 2020/2025 Timescale}}

Figure 1 is an overview of the cosmological surveys that will dictate
our science activities until 2025. Future activities will be similar
to those today, but the simulations will have to keep up with
observational improvements. Requirements are most stringent for
optical surveys since they probe much smaller scales than CMB
surveys. With increasing depth, fainter galaxies at larger distances
need to be resolved, requiring larger simulations (larger volumes and
more particles) with more detailed physics implemented in the
hydrodynamics codes.

For gravity-only simulations, the ultimate goal is to have enough mass
resolution to resolve dark matter halos that host dwarf galaxies in a
cosmological volume. These simulations will be needed for surveys such
as LSST. In the next decade, we want to cover volumes of several Gpc
and achieve a mass resolution of ~10$^6$-10$^7$M$_{\odot}$. This means
that we have to simulate up to a hundred trillion to a quadrillion
particles, leading to memory requirements of $\sim$4-40 PB. In
addition, the requirement to capture small-scale structure in
large-volume simulations will demand finer force resolution, leading
to very compute-intensive runs. Although the general approach to
gravity-only simulations will likely not change much in the next
decade, two main challenges exist: 1) global load balance and
efficient communication to enable use of large parts of supercomputers
(e.g., FFTs), 2) local load-balancing to follow the formation of
complex sub-structure. Despite these challenges, no other major
roadblocks are anticipated in the future.

In the case of grid-based hydrodynamics, scientific targets include
modeling the Lyman-alpha forest at scales relevant to baryon acoustic
oscillation measurements (with box-sizes of $\sim$1Gpc) while
maintaining resolution to resolve density fluctuations responsible for
the forest ($\sim$10kpc).  This leads to memory requirement in the
range of 4-64 PB.  Another challenge in modeling the forest arises
from the fact that with future precision requirements the ionizing
background can no longer be treated as uniform; radiation transport --
and probably multigroup radiation transport -- is going to become the
norm.  This will be computationally costly and will present enormous
scaling challenges, depending on the method used. Similar requirements
arise in other areas of cosmological hydrodynamic studies, e.g., in the
study of clusters of galaxies, or the evolution of galaxies.  The
resolution needed to build physically reliable subgrid models for star
formation and feedback, as well as AGNs, is much more stringent and of
the order 100pc, bringing again the total memory requirements into the
PB range.

Improvements in grid-based hydrodynamics codes will focus both on
on-node performance and load balancing.  We will need to make
effective use of all the cores on the new many-core nodes subject to
low-memory per core and on-chip NUMA effects.  This will require new
approaches for working on domain-decomposed or block-structured AMR
blocks of data using finer granularity.  ``Logical tiling'', used to
control the working size of the block of data being operated on, can
improve performance due both to improving the use of cache and
allowing threading over blocks rather than loops. ``Regional tiling''
alters the layout of the data on a node by optimizing for the on-node
NUMA effects.  Both of these strategies fit well into the AMR paradigm
in which the cost of metadata is minimized by keeping the size of
individual AMR grids large.

SPH codes provide a different approach to the hydrodynamics
problem. They have the advantage of computational efficiency compared
to AMR codes but have suffered from problems such as lack of mixing in
the past. Significant progress has been made recently with regard to a
number of the accuracy concerns and new SPH techniques show great
promise. In terms of implementations on future architectures, these
improved SPH methods will be an attractive option. More work is needed
in assessing the accuracy issues and also in the implementation of
these new approaches on future architectures.

In addition to the performance-improving paths identified above, a
major development component will be in sub-grid modeling. Here,
uncertainties are currently very large and ensembles of runs will have
to be carried out to understand the implications of different modeling
assumptions in detail.

Finally, as the amount of data generated by large simulations
increases, we will move from primarily using post-processing for
diagnostics to a mode of in-situ data analysis.  This addresses the
data storage issue but requires additional run-time optimization of
the diagnostic routines as they will either compete with the
simulation code for resources or require additional data movement to
cores that will not compete with the simulation itself.

\medskip

\noindent{\bf{\em 1.2.3 Compute, Data, and Services Needs on the 
    2020/2025 Timescale}}

In general, all cosmological simulations are memory-limited. The
memory requirements on next-generation supercomputers will be in the
tens of PB for gravity-only simulations. Each time step would produce
tens of PB of data, usually around 100 snapshots are needed for a
complete analysis. This amount of data will probably go beyond the
available resources in 2020/2025, therefore efficient in-situ analysis
frameworks will be essential. In addition to a handful of very large
simulations, we will also need to carry out suites of simulations of
medium size.

For the hydrodynamic simulations, the memory requirements per particle
or grid element are much higher than for the gravity-only solver
because of the additional fields being evolved. For both the
gravity-only and hydrodynamics runs, the size of the largest runs will
be dictated by the total memory available on the supercomputer. As
mentioned above, ensembles of runs will be carried out to explore the
effects of different sub-grid models.

The current usage for cosmological simulations is roughly 400M
core-hours at the LCFs and NERSC. Simulation requirements for surveys
are still being processed by the community but are expected to be very
substantial. The demand will scale faster than the available compute
because of the need for multiple runs. The hydrodynamics runs will add
a multiplicative factor of 20 beyond the N-body requirements. Finally,
while some of the science can be accomplished using run-time
diagnostics that do not require the storage of the solution, a large
community would like to use the results for a variety of additional
science tasks. This means that the storage of the output from at least
several key runs of different types is highly desirable. Storing and
moving such large amounts of data is very difficult and it is not
clear 1) how to efficiently serve the data, and 2) how to provide
analysis capabilities that can deal with very large datasets. The
community has started to address these questions, but with increasing
amounts of data a more rigorous and coordinated plan over the next
decade is needed to address both of these issues.

\newpage

\noindent{\bf\scshape{1.3~Lattice QCD}}  

\medskip

\noindent{\bf Authors:} R. Brower, S. Gottlieb (leads), D. Toussaint 

\medskip

\noindent{\bf{\em 1.3.1~Current Science Activities and Computational 
    Approaches}}

QCD is the component of the standard model of sub-atomic physics that
describes the strong interactions.  It is a strong coupling theory,
and many of its most important predictions can only be obtained
through large scale numerical simulations within the framework of
lattice gauge theory.

These simulations are needed to obtain a quantitative understanding of
the physical phenomena controlled by the strong interactions, to
calculate the masses and decay properties of strongly interacting
particles or hadrons, to determine a number of the basic parameters of
the standard model, to make precise tests of the standard model, and
to search for physical phenomena that require physical ideas which go
beyond the standard model for their understanding.

Lattice field theory calculations are essential for interpretation of
many experiments done in high-energy and nuclear physics, both in the
US and abroad. Among the important experiments that have recently been
completed, or are in the final stages of their data analysis are,
BaBar at SLAC, CLEO-c at Cornell, CDF and D0 at Fermilab, and Belle at
KEK, Japan. New data is beginning to arrive from the LHCb experiment
at the LHC, and BESIII in Beijing. In the longer term Belle II will
provide very high precision data. In many cases, lattice QCD
calculations of the effects of strong interactions on weak interaction
processes (weak matrix elements) are needed to realize the full return
on the large investments made in the experiments. The uncertainties in
the lattice QCD calculations limit the precision of the standard model
tests in many cases, but in some, the experimental errors are larger
than those from theory. Our objective is to improve the precision of
the theoretical calculations so that they are not the limiting
factor. A number of calculations are now reaching sub percent level.

Having precise values of heavy quark masses and the strong coupling
constant are important for using Higgs boson decays at the LHC to test
the standard model and probe for new physics. The most precise values
come from lattice QCD and will be improved upon in the future. Lattice
QCD is also important for better understanding of neutrino production
via the axial vector current. The dominant domestic US high energy
experimental program will be in neutrino physics for the foreseeable
future.

The anomalous magnetic moment of the muon is an important experiment
that was carried out at Brookhaven National Lab and was moved to
Fermilab envisioning a factor of four improvement in the
precision. Currently, there is a 3-4 standard deviation difference
between theory and experiment. Most of the theoretical error is due to
QCD effects, and there is a major effort within USQCD (and abroad) to
improve lattice QCD calculation of these effects and thus reduce the
theoretical error. If the theoretical error is not reduced on a time
scale commensurate with the new experiment, the experimental effort
will not result in a critical test of the standard model.

Beyond lattice QCD, lattice field theory calculations are in use to
explore models for dynamical symmetry breaking (in contrast to the
simple Higgs mechanism) and nonperturbative aspects of
supersymmetry. These calculations are likely to grow in importance and
computational demand in the 2020/2025 time period.

The United States lattice gauge theory community, organized as the
USQCD collaboration, operates computing clusters at DOE labs that
currently provide around 450 million conventional core-hours per year,
as well as eight million GPU hours. In addition, about 300 million
hours of time from the INCITE program at DOE computing centers is
distributed among the US lattice gauge theory projects. Several
lattice gauge theory groups have allocations at local centers or at
XSEDE centers. For example, the MILC collaboration has a 2015
allocation of about 8.6 M core hours. On Blue Waters, the USQCD
collaboration has 31.5 M node-hours split between high energy and
nuclear physics calculations. (Note the distinction between core-hours
and node-hours.) For 2015, about 110 M units of NERSC mpp time are
devoted to lattice QCD under the office of High Energy Physics and a
comparable amount is devoted to Nuclear Physics research.

\medskip

\noindent{\bf{\em 1.3.2~Evolution of Science Activities and 
    Computational Approaches on the 2020/2025 Timescale}}

A number of white papers and reports have addressed the needs and
challenges of computing for lattice field theory~\cite{usqcd1, usqcd2,
  snow1, snow2, nersc_req, ex_2008}.  In the space available here, we
can provide only a brief summary of the scientific goals and
computational needs for the 2020/2025 time period. Broadly, the field
needs to provide the theoretical input required to interpret current
and planned particle physics experiments. The experiments are designed
to study the properties of the standard model of elementary particle
and nuclear physics, and more excitingly, 
\begin{wrapfigure}[23]{l}{3in}
\centering \includegraphics[width=3in]{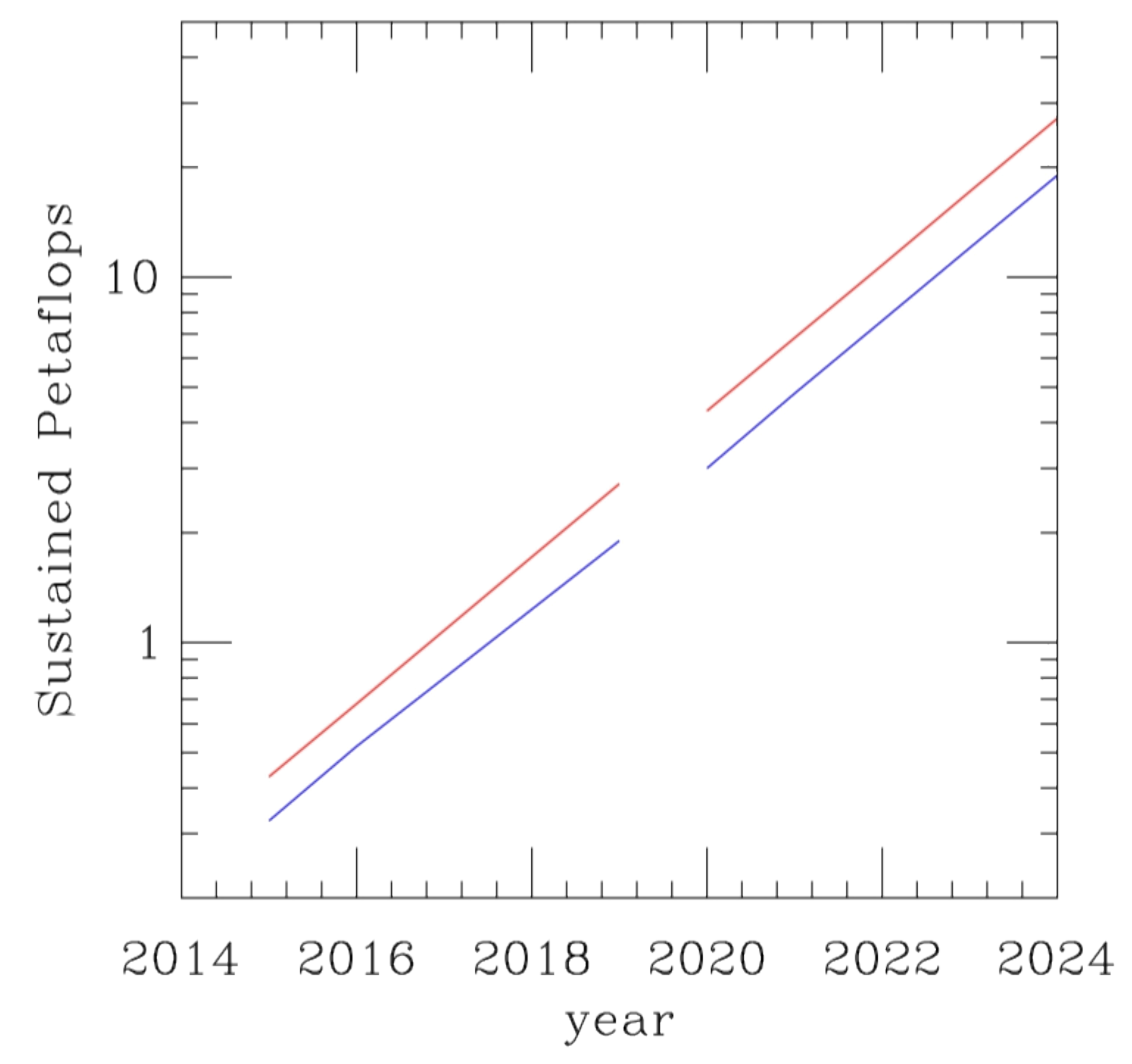}

\vspace{-0.2cm}

\caption{\small{\em Sustained total performance needed on leadership
    (red) and capacity class (blue) systems as a function of
    time. Projections up to 2019 based on Snowmass Lattice Field
    Theory report. Future projects assume continued exponential
    growth.}}
\end{wrapfigure}
\noindent
to find evidence for new
forces or new particles. Many of the parameters of the standard model
such as the quark masses, strong coupling and elements of the CKM
mixing matrix require input from lattice QCD to extract the parameter
from experimental measurements. In many cases, more than one physical
process can be used to determine a particular element of the CKM
matrix. If these processes are studied and imply different values for
the matrix element, that is evidence that there are additional
interactions not included in the standard model. However, this
requires high precision in both the measurement and the theoretical
calculation. As one example, there currently is a small tension
between the CKM matrix element $|V_{us}|$ determined in leptonic
and semileptonic decays of the kaon. Improvements to both theoretical
and experimental precision would be needed to find evidence for new
physics. For kaon semileptonic decay, the error in $|V_{us}|$ from
theory is about 50\% larger than the experimental error.

To improve future calculations, we will reduce the spacing between
grid points, increase the volume of the system, calculate with up and
down quark masses at their physical values (including isospin
breaking) and add electromagnetic effects. Taking into account the
first three improvements, we would like to use a lattice spacing of
0.03 fm with a physical size of 7.5 fm. So, this could be a $256^3
\times 512$ grid compared to the $144^3 \times 288$ grid in production
today. For the NERSC requirements study, it was estimated that we
would require 160 billion (Hopper) core hours for the physics we want
to do on this project, and we would be able to run on the order of 1
million cores. Including electromagnetism and isospin breaking would
increase the cost very roughly by a factor of two. Using different
formulations for the lattice quarks, which helps to control systematic
errors, could add up to a factor of 10, depending on the formulation.

\medskip

\noindent{\bf{\em 1.3.3 Compute, Data, and Services Needs on the 
    2020/2025 Timescale}}

An alternative planning approach is based on actual code floating
point requirements. This was used in the Snowmass report which covers
the period 2015 to 2019. Requirements are in terms of delivered
Tflop/sec-years are presented in Table 1. The sustained speeds are not
required for a single job, but for all jobs running concurrently. A
greater number of jobs can be expected to run on the capacity hardware
than on the leadership class computers.

The requirements are exponentially rising and in broad terms, it would
be reasonable to expect that to continue beyond 2019 (see Figure 1). Of
course, we hope for algorithmic improvements which have come several
times in the past. However, these improvements have been used to
improve the quality of results with current resources, not to reduce
the requested 
\begin{wraptable}[15]{l}{2.7in}\footnotesize
  \begin{tabular}{|c|c|c|} \hline \hline
    {Year}&Leadership Class &Dedicated Clusters\\
    \hfill & (TFlop/sec-yrs) & (TFlop/sec-yrs)\\
    \hline
    2015 & 430 & 325 \\
    2016 & 680 & 520 \\
    2017 & 1080 & 800 \\
    2018 & 1715 & 1275 \\
    2019 & 2720 & 1900 \\
    \hline\hline \end{tabular}

\vspace{-0.2cm}

  \caption{\label{tabdat}{\em Resources assumed for the planned
      program. Conversion factor for sustained Tflop/sec-years,
      assuming 8000 hours/yr, is 1 Tflop/sec-yr = 3.0M core-hr on IBM
      BG/Q hardware.}}
\end{wraptable}
\noindent  
resource. Perhaps that will change in the future. In any
case, it is worth emphasizing that research in algorithms and
investment in people who can both code efficiently and develop new
algorithms are very important.

Traditionally, lattice QCD has not had very large storage
requirements. This is changing to some extent because of new
preconditioning techniques that require storage of many eigenvectors
on each configuration.  For the year starting July, 2015, members of
USQCD requested 1.3 PB of disk storage and 2 PB of tape.  That is
split between nuclear and particle physics. For lattice field theory
calculations, the required number of floating point operations grows
as a higher power of the inverse lattice spacing than the number of
grid points. On that basis alone, we expect slower growth of storage
needs.

\newpage

\noindent{\bf\scshape{1.4~HPC in HEP Theory}}  

\medskip

\noindent{\bf Authors:} S. Hoeche (lead), R. Boughezal, T. Han,
F. J. Petriello, L. Reina, T. G. Rizzo

\medskip

\noindent{\bf{\em 1.4.1~Current Science Activities and Computational 
    Approaches}}

HPC computing in HEP theory can be broadly classified into activities
relating to new physics searches, and efforts to make precise
predictions for Standard Model (SM) reactions.

Among existing new physics models, supersymmetric extensions of the SM
are promising candidates for discovery at the Large Hadron Collider
(LHC) or at dark matter experiments. Even in its most minimal form
(the MSSM), supersymmetry has over 100 free parameters, and models in
different regions of parameter space can therefore have very different
experimental signatures. Converting measurements into limits on the
parameters of the theory requires a detailed simulation of a plethora
of parameter points. This can be achieved, for example, in the
framework of the phenomenological MSSM, or by using effective theories
and simplified models. Several Tools have been developed to perform
the subsequent limit setting procedure in a fully automated fashion,
including Atom, FastLim and CheckMate. All approaches have in common
that they require large computational resources, due to a wealth of
experimental data.

Scans of new physics parameter spaces typically involve generating
sets of $\sim$250K-500K models, and simulation of particle physics
events at the LHC for each one. For a single set of models at 25/fb,
the simulation data takes up about 1-2TB and involves about 1-2M
CPU-hours of computing time. Simulation data are archived. The
software typically consists of single-core executables. Because of the
Monte-Carlo approach used in the simulation, a trivial parallelization
is easy to implement, and memory limitations on new architectures are
not restrictive at present.

The computation of SM reactions at high precision has reached a high
degree of automation, with next-to-leading order (NLO) QCD
perturbation theory being the standard means of obtaining cross
sections at collider experiments like the LHC. These calculations have
been instrumental in extracting properties of the Higgs boson, and
they will continue to play a dominant role in the future. The field is
rapidly moving towards full automation of NLO electroweak
corrections. Dedicated next-to-next-to leading order (NNLO) QCD
calculations exist for important reactions such as Higgs-boson
production, Drell-Yan lepton pair production, di-photon and
vector-boson pair production, top-quark pair and single-top
production, both at the inclusive and at the fully differential
level. Many of those have been completed in the past years. First
results for Higgs-boson and vector-boson plus jet calculations have
just been obtained, and the first complete three-loop result for
inclusive Higgs-boson production at a hadron collider was also just
presented. These ultra-precise computations will become mandatory as
future data leads to decreased experimental errors and theory
uncertainties begin to limit our understanding of Nature. The high
precision of fixed-order calculations is matched by a correspondingly
high precision in resummation, obtained either through more
traditional approaches or though soft-collinear effective theory
(SCET). The matching of fixed-order calculations to parton shower
Monte-Carlo event generators used to make particle-level predictions
for experiments has been fully automated at the NLO in the past years,
and first solutions at NNLO exist. The need for precise theory input
to experiments requires multiple predictions from either of these
calculations, leading to a strain on existing computational resources.

Typical NLO QCD calculations currently require between 50K and 500K
CPU-hours each, with storage needs between 0.1 and 1.5 TB. The results
can be re-analyzed to obtain predictions for different SM
parameters. NNLO differential calculations require between 50K and 1M
CPU-hours each. Parton-shower matched predictions require of the order
of 100K CPU hours, depending on the complexity of the
final-state. Programs have been parallelized using both MPI and
OpenMP, as well as POSIX threads. First studies exist for scalable
applications on accelerators, and one of the generators running on
Xeon Phi is ready for physics analysis. In a recent study, NNLO
computations of vector/Higgs boson plus a jet have demonstrated strong
scaling up to 10$^6$ threads using a hybrid MPI+OpenMP approach. Ideas
on how to extend beyond this level are being studied.

\medskip

\noindent{\bf{\em 1.4.2~Evolution of Science Activities and 
    Computational Approaches on the 2020/2025 Timescale}}

Scans of new physics parameter spaces will likely be done in a manner
similar to today, and they will involve a similar sample size during
the course of LHC Run II. For a single set of models at the full
luminosity, the simulation data will consume $\sim$100M CPU-hr of
computing time and take up $\sim$100TB of storage space. If possible,
simulation data will be archived, possibly on tape. No significant
evolution is expected on the software side, though it can be expected
that accelerators with i386 instruction set can be utilized.

Precision calculations in the SM will likely move to a stage where
NNLO sets the standard in the same way that NLO sets the standard
today. Matching these calculations to resummed predictions and
particle-level simulations, and improving the resummation implemented
in Monte Carlo programs itself will be a key task of the community
over the coming years. At the same time the full exploitation of
higher-order calculations to extract parton densities from
experimental data will play a crucial role to reduce theoretical
uncertainties stemming from the parametrization of QCD dynamics at
large distances.

Computational approaches are likely to rely more and more on
multi-core architectures and HPC. Current cutting edge calculations
have demonstrated that HPC facilities allow to increase the
final-state multiplicity in NLO calculations by at least
one. Algorithmic advances will push the limits further. When using NLO
calculations as an input to NNLO calculations by means of qT
subtraction or jettiness subtraction, the fast evaluation of NLO
results using HPC will be crucial. Similar considerations apply to
NNLO calculations making use of antenna and sector improved residue
subtraction.

\medskip

\noindent{\bf{\em 1.4.3 Compute, Data, and Services Needs on the 
    2020/2025 Timescale}}

Both new physics searches and precision SM calculations will require
considerable support from supercomputing centers and data centers, if
the pace of development shall be kept. An estimate may be extracted
from the scaling of NLO calculations, performed during the past years:
The calculation of W/Z/Higgs production requires an increase in
computing power by a factor 3-4 for every additional jet in the final
state. Memory requirements increase in a similar manner. For NNLO
calculations the projection is more difficult. Complete NNLO results
are available only for electroweak objects with zero or one jets in
the final state, and different techniques are used to obtain these
various results. The available calculations point towards an increase
in computing time by about a factor 10 for each additional object in
the final state.

\newpage

\noindent{\bf\scshape{1.5~Energy Frontier Experiments}}  

\medskip

\noindent{\bf Authors:} K. Bloom and O. Gutsche

\medskip

\noindent{\bf{\em 1.5.1~Current Science Activities and Computational 
    Approaches}}

The Energy Frontier thrust of High Energy Physics will be focused on
the Large Hadron Collider (LHC) through 2025 and probably for many
years beyond.  The LHC, operated by CERN, the European particle
physics laboratory, collides protons (and sometime ions) at the
highest energies ever achieved in the laboratory.  In 2015, the LHC is
resuming operations after a two-year shutdown at a center-of-mass
energy of 13 TeV, much greater than the 8 TeV that was achieved in
2012.  The particle collision rate (instantaneous luminosity) will
increase by a factor of two.  These collisions are recorded by four
independent multi-purpose detectors, each of which is operated by
scientific collaborations of hundreds to thousands of physicists.  The
largest U.S. involvement is in the ATLAS and CMS experiments.  In this
paper we will use CMS as an exemplar; similar arguments about the
evolution of the scale of resource needs would apply to ATLAS.

The 2010-12 LHC run (Run 1) had a prodigious scientific output, and
the coming run could be even more remarkable, if Nature is on our
side.  The most significant result from the Run 1 was the discovery of
the Higgs boson~\cite{higgs}, announced in 2012 to worldwide acclaim
and recognized in the awarding of the Nobel Prize in 2013.  But this
was just one physics result -- the CMS Collaboration has submitted
nearly 400 papers so far on measurements from the 2010-12
data~\cite{cms_exp}.  Searches for new phenomena such as supersymmetric
and/or dark matter particles will take center stage in the 2015-18 run
(Run 2), as the increase in LHC collision energy will open
opportunities to observe new heavy particles, should they exist.

CMS records in a typical Run 2 year about 5 PB of raw data per year;
another 12 PB of raw simulations are also produced, making the Energy
Frontier experiments especially data-intensive within the context of
HEP.  These data must be processed and then made available for
analysis by thousands of physicists.  The computing model for the LHC
experiments has been one based around high-throughput computing (HTC)
rather than high-performance computing (HPC), as the computing problem
is embarrassingly parallel.  CMS runs computing jobs at more than
independent facilities spread around the world.  In general, input
datasets are pre-placed at the facilities, and computing jobs are
routed to the correct facility.  In the coming run, there will be
greater use of a worldwide data federation that makes any CMS data
available to any CMS computing site with low latency~\cite{cmscomp};
this will give greater flexibility in the use of resources.  The
computers at the facilities are commodity machines, with x86
processors that run Linux.  Jobs typically require up to 2 GB of
memory.  CMS is now able to run in a multi-threaded mode, with
multiple jobs sharing the memory available within a processor.  The
resources are all available through the Worldwide LHC Computing
Grid~\cite{wlhccg} and are provisioned through
glideinWMS~\cite{giwms}, a pilot system with a global job queue.
There have been efforts to use facilities beyond those dedicated to
CMS, such as NERSC, SDSC and cloud systems, with increasing success,
but only Linux-based resources have been used.  Strong demands on
computing resources for the current run require us to increase usage
at these centers, and in particular to make greater use of the HPC
resources available there.

\medskip

\noindent{\bf{\em 1.5.2~Evolution of Science Activities and 
    Computational Approaches on the 2020/2025 Timescale}}

2020 will mark the start of Run 3 of the LHC~\cite{run3}.  The
accelerator performance will only be slightly modified compared to Run
2, with instantaneous luminosities increasing only by about 15\% over
the expected values in 2018.  Thus no significant changes are planned
for the CMS computing model at that time.  Modifications will be
evolutionary, to scale the Run 2 tools to handle increases in the data
volume.  x86 will remain the primary computational architecture, but
we will expand the use of multi-core capabilities to improve memory
usage.  As ever-larger datasets will require expanded computing
resources, there will be greater efforts to use resources beyond those
owned by the experiments, such as those in academic and commercial
clouds and all architectures available at ASCR computational
facilities.  We expect that there will be a greater reliance on data
federations rather than local data access as network bandwidth
improves.

2025 could lead to much more revolutionary changes.  It will mark the
start of the High Luminosity LHC (HL-LHC), with instantaneous
luminosities a factor of 2.5 greater than those of Run 3.  The
experimental collaborations are planning on major upgrades to their
detectors to accommodate the greater density of particles that will be
produced in the collisions~\cite{upgrades}.  See the table below for
the expected increases in trigger rate, data volume and processing
time. If we were to only take advantage of the expected growth in CPU
power for fixed cost at 25\% per year and in storage at 20\% per year,
we would expect deficits of a factor of four or twelve in CPU and a
factor of three in storage under the assumption of fixed budgets for
computing resources, even after accounting for potential algorithmic
improvements.  Clearly changes to the current paradigm must be
considered.  For instance, LHC computing will need to make use of
advanced computing architectures, such as HPC systems, GPUs,
specialized co-processors and low-power mobile platforms. Research and
development efforts on the software environment for these
architectures are in progress.  More dynamic provisioning systems will
be needed, so that experiments can control costs by purchasing
resources for average usage levels and then renting additional
resources needed for peak usage.  It is possible that by 2025, most
processing resources will be supplied through dynamic infrastructures
that could be accessed opportunistically or through commercial
providers.  Studies on how to create an elastic virtual facility have
begun.  Such facilities would need to access the data that is stored
on systems owned by the experiment.  These systems would probably need
to be served by a content delivery network that took advantage of
dynamic replication, predictive placement and network awareness.
Reliable high-bandwidth networks will be needed to transport the data
into the clouds.  A center with tens of thousands of cores would need
to have multiple 100 Gb/s links to serve the input data.  Fortunately
such bandwidth is expected to become more commonplace between now and
2025.

\medskip

\noindent{\bf{\em 1.5.3 Compute, Data, and Services Needs on the 
    2020/2025 Timescale}}

We extrapolate the resource needs of CMS for three sets of running
conditions: the start of LHC Run 3 in 2020, and the HL-LHC in two
scenarios starting at 2025, one low-luminosity and one high-luminosity
scenario. The extrapolations are given in the table.  The reference is
the performance of LHC Run 2 at 30 interactions per beam crossing
assuming two reconstruction passes over data and MC per
year~\cite{spdsheet}.

We restrict ourselves to quoting CPU needs only for central workflows
and also do not include any improvements in the software into the
extrapolation.  From experience, a factor of two improvement in
reconstruction time should be feasible; in the last four years, CMS
managed a speedup of factor four.  Analysis activities would also have
to be added to determine a complete picture of resource needs.

\begin{figure}
\centering \includegraphics[width=6in]{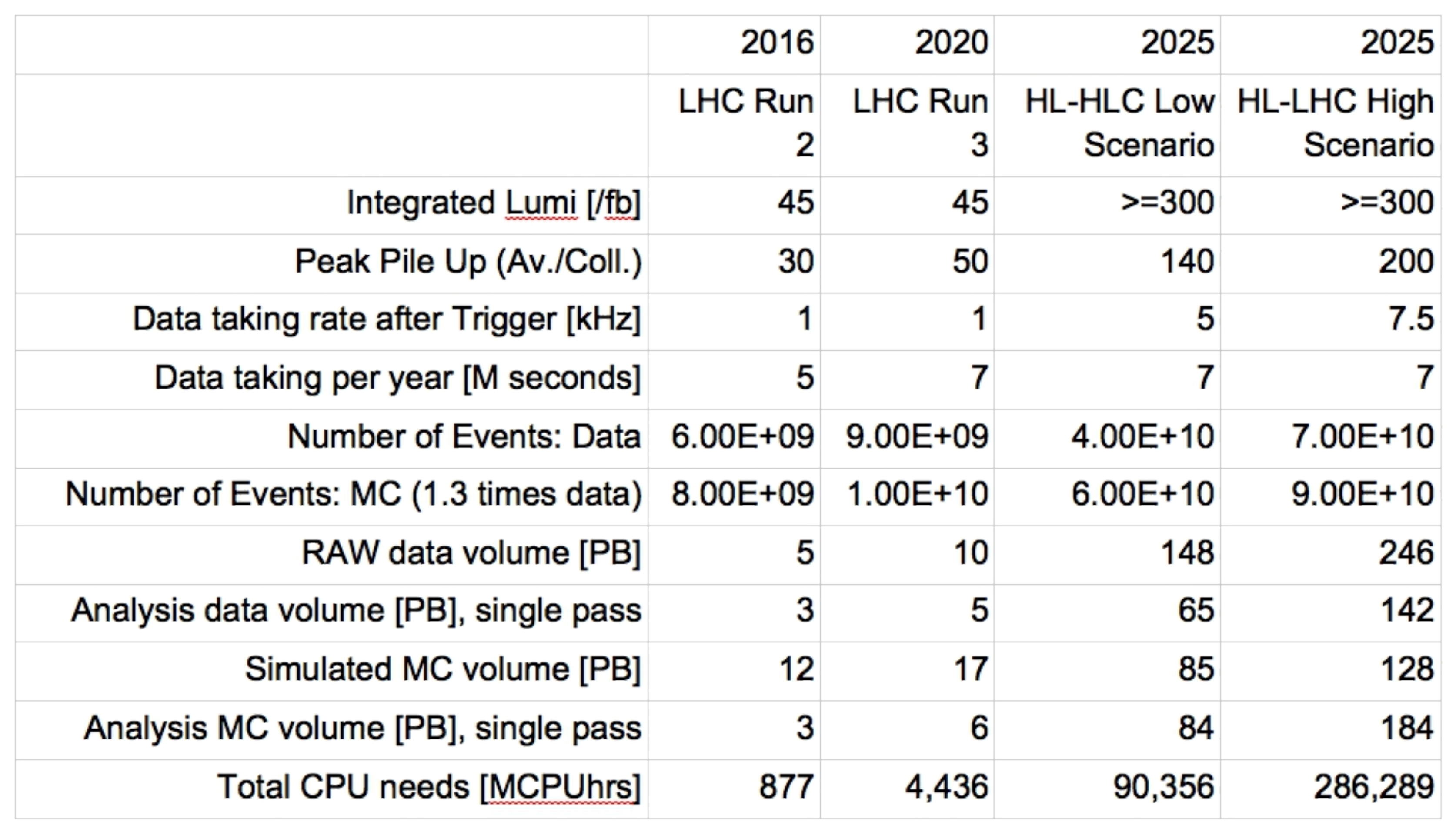}

\vspace{-0.2cm}

\caption{\small{\em Estimated resource needs for CMS.}}
\end{figure}

\newpage

\noindent{\bf\scshape{1.6~HEP Experiments: Data Movement and Storage}}  

\medskip

\noindent{\bf Authors:} F. Wurthwein, G. Oleynik, B. Bockelman

\medskip

\noindent{\bf{\em 1.6.1~Current Science Activities and Computational 
    Approaches}}

The landscape of experimental HEP is strikingly diverse. In addition
to the large LHC experiments, there are a number of Intensity and
Cosmic Frontier experiments at all sizes. While the LHC will continue
to be the largest data producer in 2020-25, experiments like DUNE and
LSST present their own challenges, and so do smaller experiments.

The ATLAS and CMS experiments produced a few tens of PB of data during
LHC's Run~1, 2010-12. Roughly a third of this is from the detector,
the rest is simulation. The $\sim$2:1 relationship is likely to be
stable; the output rate from the trigger system is thus a rough guide
to the increase in data volume over time. Both experiments started
with a data taking rate of $\sim$150Hz that increased to
$\sim$300-600Hz at the end of Run 1. For Run 2 (2015-2017) the initial
rate is 1kHz and is expected to reach $\sim$10kHz by 2025. In the ten
years of HL-LHC running (roughly 2025-35) each experiment will
transition from O(100) petabytes to O(1) exabyte of data.

Two copies of the RAW data are archived to tape. One is conceptualized
as a ``backup'' copy at CERN, while the other is distributed as an
active copy across the Tier-1 centers worldwide for each
experiment. Only a single copy of the processed data and simulations
is archived. The RAW data from the detector is understood to be the
most precious, while all else can be reproduced, if corrupted or
lost. Tape is the preferred archival technology because of its
durability and lower cost.

It is useful to divide the path from raw data to science results into
two steps. First, each collaboration centrally produces official
datasets from raw data and simulation. Second, small groups of
collaborators produce private custom datasets from (subsets of) the
official data, and analyze them to derive publishable results. The
first step is consuming ever increasing amounts of CPU power, to
produce an output format optimized for maximal flexibility to support
the full diversity of the LHC program, allowing for continued
improvements in physics object definitions and selections. The second
step typically begins with each small group producing ``slimmed'' and
``filtered'' custom datasets, resulting in smaller event sizes, fewer
events, and substantially faster to process data formats; further
analysis tends to be I/O limited. The custom datasets comprised
$\sim$4-400TB per individual group during Run 1, and are expected to
grow at a similar rate as the official data.

LHC computing was a huge success during Run 1. The speed and
robustness with which results were produced exceeds most prior large
scale experimental HEP programs, despite the significant increase in
scale in data volumes and computing needs. Thus, any proposed change
in how data is stored, transferred, processed, and managed, should
result in cost savings without loss of speed and robustness in
producing physics results. It is worth noting that the LHC program
globally involves O(10k) people. At a cost of \$100k per FTE, this
amounts to \$1 Billion in global personnel costs, dwarfing the annual
LHC computing budgets worldwide. Therefore, a ``cost-effective''
solution must maximize the effectiveness of the global human capital,
or it will invariably lead to loss of efficiency in producing science
results.

\medskip

\noindent{\bf{\em 1.6.2~Evolution of Science Activities and 
    Computational Approaches on the 2020/2025 Timescale}}

When considering the evolution of scientific and computational
activities, we distinguish ``technical'' and ``sociological''
opportunities with the aim of identifying more cost-effective
solutions as just discussed above.

Among the technical drivers, we identify three high level
concepts. First, the trend towards vectorization and parallelization
driven by a larger number of simpler cores on commodity/HPC
hardware. Second, the advent of Big Data technologies, and third the
advent of highly elastic computing. The three combined with the
divergence of CPU and I/O needs for the two steps discussed above are
likely to drive the need for integrated workflow management across a
diverse set of resource types.

In 2025, physics and detector simulation, and raw data processing may
be done on three different hardware platforms. Disk buffers in front
of tape archives may be minimal in size, as tape retrieval is tightly
integrated into the data processing workflow. Such a workflow might be
scheduling disk buffers at a remote processing center in addition to
disk buffers in front of the tape archive, and the wide area network
between the two disk buffers. By 2025, the majority of disk space in
ATLAS and CMS may thus be in the analysis systems. These systems may
be heavily I/O optimized using disk scheduling ideas from
Hadoop/MapReduce, in combination with an I/O layer optimized for
partial file reads that is already standard today in ROOT.

Finally, all of the above must be highly elastic. ATLAS and CMS take
months to produce releases validated for large-scale simulation and
data processing. Once the release is validated, the time for
simulation and processing would be shrunk as much as possible, leading
to large spikes in desired resource consumption. Since commercial
Cloud providers already operate distributed exascale systems, it is
natural that ATLAS and CMS will want to use such systems by 2025.

The most important opportunity for cost savings is in placing the
divide between the (centrally produced) official data, and custom data
produced by the scientists. It may be beneficial to centrally produce
data formats that are much less CPU intensive to process, trading
flexibility against reduced size per event. To make up for the lost
flexibility, such formats might be produced more often, leading to
more versions in use at a given time. Whether this will lead to more
human inefficiency than gains in computational efficiency needs to be
explored.

The common theme that emerges is that future computing approaches will
need to be much more agile, dynamic, and elastic in order to support a
more diverse set of hardware resources at scales that frequently
change throughout the year.

\medskip

\noindent{\bf{\em 1.6.3 Compute, Data, and Services Needs on the 
    2020/2025 Timescale}}

Data Storage and Movement Services needed to meet HEP needs in the
2020 timescale already exist for the LHC and other HEP experiments,
but there are challenges in scaling up by a factor of 10-18 in I/O
bandwidth and data volume~\cite{info2}, and to apply these services to
collaborations, which though smaller scale in terms of data volume,
can benefit from existing infrastructure and architectures. In
addition, services are in the process of becoming more agile, but that
process is still far from complete.  Common facility Services include:\\
\noindent{$\bullet$} Long term archival of data (tape storage)\\
\noindent{$\bullet$} Dataset services (e.g., data location aware staging service)\\
\noindent{$\bullet$} Federated storage access, local posix and WAN data access
  protocols\\
\noindent{$\bullet$} Network infrastructure\\
\noindent{$\bullet$} High throughput, low latency access storage for analysis
  computation\\
\noindent{$\bullet$} High throughput, low latency storage for production computation\\
\noindent{$\bullet$} Tape backed low latency disk cache\\ 
\noindent{$\bullet$} Global catalog (or mappings to local catalogs)\\
\noindent{$\bullet$} Management and monitoring infrastructure

Opportunities and challenges in these service areas are:\\
\noindent{$\bullet$} Delivering data is still a major challenge;
storage systems do not perform well with random access patterns from
multiple users. Services and storage architectures need to convert
inefficient patterns into efficient requests to the underlying storage
hardware.  An example of this is SAM~\cite{sam}, which understands the
physical location of files on tape and attempts to optimally stage
experiment defined datasets from tape to disk. A related challenge
will be to effectively utilize the anticipated 2
GB/s bandwidth of tape drives.

\smallskip

\noindent{$\bullet$} Smaller HEP collaborations are often limited, not
by resource restrictions, but in organizing data to efficiently
deliver it to computational workflows. Providing tools and
architectures to aid with this could be a great benefit.

\smallskip

\noindent{$\bullet$} The ``storage element'' (storage organized as a
POSIX-like filesystem with an access interface such as GridFTP) tends
to be a too-low-level abstraction, especially when multiple storage
systems are involved.  The human overhead of maintaining filesystem
consistency is high.

\smallskip

\noindent{$\bullet$} The largest data management systems (CMS's
PhEDEx, ATLAS's Rucio) have failed to gain widescale adoption outside
their respective experiments.  A successful reuse example is SAMGrid,
adopted by several HEP experiments.  This area has had a poor track
record of moving computer science innovations from R\&D to production.
Future experiments would benefit if data management
services could be generalized and made production ready.

\smallskip

\noindent{$\bullet$} The field has rallied around ROOT as a common I/O
layer, allowing investments in a single software package to benefit
the entire community. However, the core ROOT IO community is too
small; we have the opportunity for significant improvements but not
the scale of effort to achieve them.

\smallskip

\noindent{$\bullet$} Standardized cloud based storage interfaces (S3,
CDMI, WebDAV) have not been taken advantage of. Work is needed to
assess if they can meet production processing requirements.

\smallskip

\noindent{$\bullet$} Smaller scale HEP collaborations not directly
affiliated with National Labs are often on their own for providing an
active data archive and the expertise to manage it. Work is underway
to provide storage services to these experiments on a case by case
basis. A cohesive set of services for storing and retrieving data at a
set of National Labs would be a significant benefit.

Facility based long-term tape storage will be the most affordable
option in the 2020 timeframe, less than \$20/TB, and capacity scales
better than needed for the HL-LHC run. Tape form factors will likely
be the same and libraries will likely still be sized at around 10,000
slots.  Tape drive bandwidth is expected to increase by about a factor
of 8$\times$ to 2 GB/s~\cite{tape, info1}.  Providing data to these drives at full
rate will be a challenge. The takeaway for storage is that tape
storage and network expenditures will likely be lower, while CPU, disk
and tape drive costs will likely be higher than current expenditures.
Tape libraries will likely need to be refreshed prior to HL-LHC
luminosity running. For more information, see Refs.~\cite{info2,
  info1}.

\newpage

\noindent{\bf\scshape{1.7~Cosmic Frontier Experiments}}  

\medskip

\noindent{\bf Authors:} A. Borgland, N. Padmanabhan (leads),
S. Bailey, D. Bard, J. Borrill, P. Nugent

\medskip

\noindent{\bf{\em 1.7.1~Current Science Activities and Computational 
    Approaches}}

The current experiments supported (directly and indirectly) by DOE HEP
facilities are cosmic microwave background experiments and low
redshift dark energy experiments. The CMB experiments include both
space-based experiments (Planck) as well as ground-based experiments;
the key science goals are inflation (including a detection of the
gravitational waves from inflation), dark matter and dark energy. The
dark energy experiments aim to use imaging and spectroscopic surveys
to constrain the expansion and growth history of the Universe through
weak gravitational lensing and observations of the galaxy distribution
(including correlation functions, number counts etc). Exemplars of
this are the recently completed Baryon Oscillation Spectroscopic
Survey (BOSS), the extended BOSS (eBOSS) survey (ongoing) and the Dark
Energy Survey.

Exact analyses of these data are computationally intractable, and
therefore one must often rely on Monte Carlo methods to characterize
(a) the instrument responses and biases, (b) observational effects
from incomplete data/cuts, (c) statistical uncertainties and (d)
astrophysical systematics. The generation and analysis of these mock
catalogs is often the limiting step in every analysis.  For example,
to reach 1\% statistical uncertainties, the Planck group at NERSC
considered 10$^4$ realizations, each with O(10$^3$) maps.  Low
redshift galaxy surveys rely on large numbers of N-body simulations to
model the nonlinear formation of structure.

Traditional astronomical data sets are often 10s of terabytes, but are
often broken down into very large numbers ~O($10^6$) files, making
them unwieldy for storage. Tools to efficiently access these data are
either often missing or have yet to see widespread usage. Traditional
HPC models are often poorly suited for the analysis of such data
sets. An associated challenge is the distribution of these data (and
related simulations) amongst large and geographically diverse
collaborations.

\medskip

A third class of experiments supported by DOE HEP facilities are dark
matter experiments. These can be divided into three classes --
collider production (covered in the Energy Frontier), indirect
detection (Fermi Gamma-Ray Space Telescope, not part of this timeline)
and direct detection. Direct detection experiments use WIMP-nucleon
elastic scattering to put constraints on the WIMP mass. The current
Generation-2 program has two main experiments: LUX/LZ (Xenon) and
SuperCDMS SNOLAB (Ge, Si) both of which will be located underground to
shield them from cosmic rays. They are expected to start operating
around 2018-2019.

\medskip

The analysis of direct detection data closely follows the particle
physics model in that there are particle reactions in a detector with
associated detector information read out. Because of the low expected
WIMP signal rate, a thorough understanding of backgrounds is the
critical part of direct detection experiments. Monte Carlo simulations
along with dedicated calibration events are used to estimate
backgrounds from the detector and associated shielding.  Up until now,
the data sets from these experiments have been small and computing has
not been a priority.

\medskip

\noindent{\bf{\em 1.7.2~Evolution of Science Activities and 
    Computational Approaches on the 2020/2025 Timescale}}

The next decade will see an order of magnitude increase (or larger) in
data volume and science reach from the next generation of
experiments. Each of the three major areas described above have next
generation experiments in the planning/construction phase -- the CMB
experimental community is working towards a Stage IV CMB experiment
(CMB-S4) in the early 2020's; the BOSS and eBOSS surveys will be
succeeded by the Dark Energy Spectroscopic Instrument (DESI)
($\sim$2019-2024), and the Large Synoptic Survey Telescope (LSST) will
be the premier imaging survey throughout the next decade
($\sim$2020-2030).  The science reach of these surveys will require
significant increases in computational/storage needs, both for the
analysis of the raw data and its cosmological interpretation. We
discuss the needs of these surveys individually below. (An important
assumption in all of this is that we will be able to maintain
computational efficiency on the next generations of hardware,
including the expected heterogeneous processing architectures,
bandwidth to memory and storage, etc.)

The LSST survey will survey half the sky in multiple bands to
unprecedented depths; in additions, it will provide a time-domain view
of the sky by surveying it every few days. The primary goals of the
HEP LSST effort are to probe the nature of dark matter and dark energy
through weak gravitational lensing and galaxy clustering. The weak
gravitational lensing signal is very subtle and easily swamped by
systematic effects from the atmosphere and detector, as well as
imperfections in the analysis algorithms. To quantify/mitigate these
systematic effects, the LSST project is undertaking a very detailed
program to simulate all aspects of this measurement. The estimated
compute cost for this process is $\sim 10^7$ compute hours and 100 TB
of storage. The other dominant cost for the LSST analysis are the
simulations necessary for quantifying the uncertainties in the
measurements. The LSST DESC collaboration estimates requiring
$\sim$2500 simulations, each with a cost of $\sim$1M CPU hours, for a
total of O($10^9$) CPU hours and ~100 TB of storage. The other
analysis tasks are expected to be subdominant to this.

The DESI survey aims to obtain redshifts to $\sim$25M galaxies over
$\sim$14,000 sq. deg. of the sky. This represents an order of
magnitude increase over current galaxy surveys. The 
\begin{wrapfigure}[19]{l}{3.2in}
\centering \includegraphics[width=3.2in]{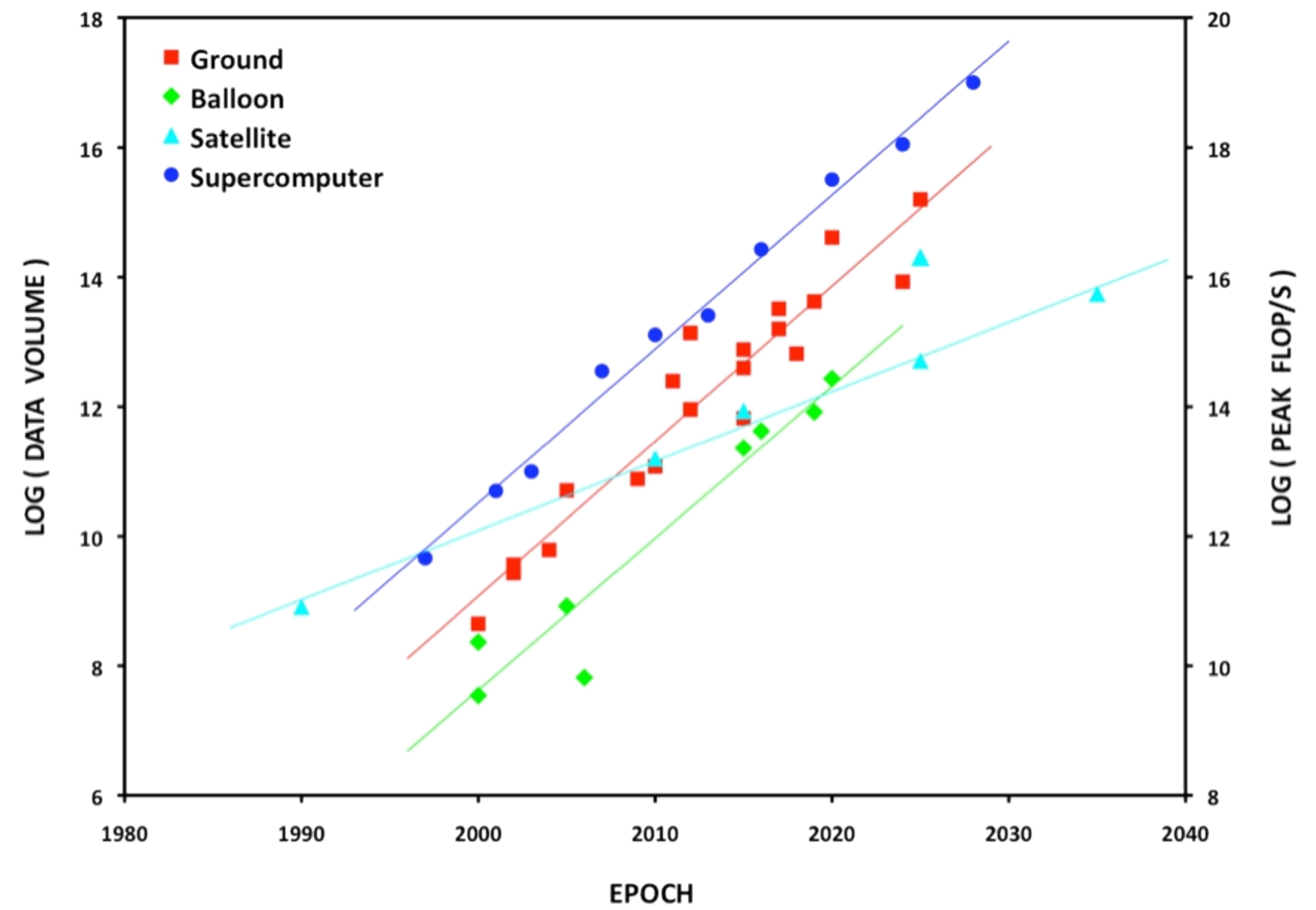}

\vspace{-0.2cm}

\caption{\small{\em Exponential growth over the past and coming 20
    years of CMB data volumes gathered by ground-based, balloon-borne
    and satellite missions, and supercomputer peak performance using
    the NERSC flagship system as a proxy.}} 
\end{wrapfigure} 
\noindent
estimated computational and storage requirements for the survey are
$\sim$100 M CPU hours and O(1) PB of data storage. As with the other
projects described here, the other dominant cost will be the
simulation requirements for error quantification (although these
should be less demanding than the LSST case).

The CMB-S4 experiments aim, amongst other goals, at constraining both
the CMB polarization caused by gravitational waves from the
inflationary epoch as well as constraining the total neutrino mass at
unprecedented precision. These will provide astrophysical windows into
scales impossible (or very hard) to reach by traditional
particle-physics experiments. Reaching these goals will require a 1000
fold increase in the data volume compared to the Planck satellite, and
will require improvements in analysis algorithms as well as their
implementations to make best use of the architectures of the next
generation of supercomputers. We can estimate both storage and compute
needs by scaling from experience with the Planck satellite. The Planck
satellite used $10^8$ CPU-hours for its analysis, a CMB-S4 experiment
will require $10^{11}$ CPU hours. Planck used O(250) TB in 2014, but
this constrained us to storing O($10^6$) maps instead of the full
O($10^7$); CMB-S4 will need O(10) PB in 2020.

An important set of considerations for these next generations of
surveys are the large and geographically diverse collaborations and
serving different views of the data to them. Furthermore, the
traditional data models and file systems do not scale well to these
data sets, since they often result in O($10^9$) files making
accesses/queries time consuming. The collaborations will therefore
benefit from tooling to simplify these tasks. Furthermore, different
analysis tasks often have very different scales of parallelism, making
them hard to chain together under traditional MPI/OpenMP
models. Finally, issues of openness and reproducibility are becoming
very important considerations for these collaborations.

Data volumes and processing needs for the G2 direct detection
experiments, while larger than current experiments, are expected to be
small. LUX/LZ expects 1.2 PB of data/year, with 250 cores needed for
prompt processing and 3k cores needed to reprocess one year of data in
three months. Including Monte Carlo simulations the LUX/LZ plan
includes support for 10 PB of disk storage and 10k cores during
operations.

SuperCDMS, which is concentrating on lower WIMP masses, will have data
volumes of the order of 100 TB/year, with a very small number of cores
needed for prompt processing. The plan is to adopt a very distributed
computing model to be able to access as much CPU as possible at
collaboration sites to minimize the time needed for data reprocessing
needs and Monte Carlo production. The overall CPU needs will be
similar to LUX/LZ.

\medskip

\noindent{\bf{\em 1.7.3 Compute, Data, and Services Needs on the 
    2020/2025 Timescale}}

The previous section discusses the needs of the three major cosmic
frontier surveys individually. We summarize these below:
\begin{itemize}
\item Compute: O(10$^{11}$) CPU hours. This is dominated by the anticipated
  requirements of the CMB-S4 experiment, with the other two
  experiments, estimating about two orders of magnitude less
  computation. Therefore, this should be able to accomodate an
  expanded compute requirement from these surveys. 
\item Data: O(10) PB. This is again dominated by CMB-S4, with the
  other surveys requiring an order of magnitude less data. 
\item Serving the raw and reduced data to large, geographically
  diverse collaborations. 
\item The ability to support a broad range of analysis/programming
  paradigms including (but not limited to) traditional HPC usage,
  massively (but trivially) parallel analysis tasks, to exploratory
  analyses. 
\item Developing the infrastructure to enable
  reproducibility/openness.
\end{itemize}

\newpage

\noindent{\bf\scshape{1.8~Intensity Frontier Experiments}}  

\medskip

\noindent{\bf Authors:} B. Viren and M. Schram

\medskip

\noindent{\bf{\em 1.8.1~Current Science Activities and Computational 
    Approaches}}

\noindent {\em Neutrino Experiments:} Past and existing neutrino
experiments such as MINOS, T2K, Daya Bay and Nova produce a relatively
modest amount of data in comparison to LHC experiments. Data transfer,
storage and production processing can be managed by a few experts
exploiting a few identified institutional clusters and mass storage
systems.

The reactor experiments require only fairly simple reconstruction
algorithms due to the physics of their interactions and by careful
design of their detectors.  More important is a careful understanding
of background, energy resolution and energy scale systematics.  This
emphasizes the need to produce large MC simulation samples.  A
simulation or real data processing campaign requires order 100s-1000s
of CPU-months and makes use of commodity institution clusters.

Accelerator-based neutrino experiments tend to require more
sophisticated reconstruction software and produce larger data volumes.
T2K uses the Super-Kamiokande water Cherenkov detector is read out by
about 14,000 PMTs.  MINOS and Nova are scintillation detectors; Nova
reads out approximately 300,000 channels.  Large MC simulation samples
equivalent to 10-100$\times$ the size of real data can be required. These
detectors employ a reconstruction technique that involves fitting
events to pre-generated detector response patterns.  These detectors
require approximately an order of magnitude more CPU than reactor
experiments and their experiments tend to employ Grid resources to
satisfy their needs.

\noindent {\em Belle II Experiment:} The Belle II Collaboration
includes more than 600 scientists from 100 institutions in 23
countries. The U.S. membership on Belle II is now 13\% of the
collaboration from 14 institutions. Belle II has adopted a distributed
computing model based on the grid. The Belle II computing Memorandum
of Understanding states that ``Each Belle II member institute
provides, either by itself or in collaboration with other institutes
at a regional or national level, the computing resources to produce,
store, and make available for analysis a fraction of the total dataset
for Belle II and to carry out a fraction of the overall physics
analyses. The fraction corresponds to the fraction of PhDs in the
international Belle II collaboration''.

\medskip

\noindent{\bf{\em 1.8.2~Evolution of Science Activities and
    Computational Approaches on the 2020/2025 Timescale}}

\noindent{\em Neutrino Experiments:} The next generation Deep
Underground Neutrino Experiment (DUNE) employs a 40kton liquid argon
time-proportional chamber (LArTPC) far detector with 1.5M channels
reading wires spaced every 5mm and acquiring waveforms at 2MHz for
about 5ms.  This produces approximately 10-20 GB per ``event''.  In
order to be sensitive to supernova bursts the readout must be capable
to sustaining 10s of seconds of data collection.  Such full-stream
readout can produce 100s of Exabyte per year.  However, most of the
ADC samples will be noise and can be discarded by using ``zero
suppression'' technique in which low-threshold portions of the
waveform are discarded.  This can reduce the raw data rates to the
TB/year.

The DUNE LArTPC is incredibly fine-grained compared to other neutrino
detectors (except bubble chambers).  Traditional reconstruction
techniques will, at best, scale linearly with the number of channels
and may scale as worse as $N^3$ and novel techniques exploiting the
unique characteristics of LArTPC are envisioned.  At the very least,
it is expected that production processing must exploit Grid resources.

\noindent{\em Belle II Experiment:} The Belle II distributed computing
system must handle $\sim$85 PB data volume for each year when the
SuperKEKB accelerator is operating at design luminosity. The Belle II
computing model includes several tasks such as raw data processing,
Monte Carlo event production, physics analysis, and data archiving.
Belle II has adopted the DIRAC framework for their Grid system. DIRAC
provides both a workload and data management system along with other
systems, such as data creation/manipulation workflow system, metadata
catalog system, etc.

The Belle II software/computing workflow has numerous elements. At the
core is the standalone Belle II software framework (BASF2) and
external dependencies (Geant4, ROOT, etc.). The code is currently
distributed on the grid using CVMFS servers. Grid sites are deployed
with commodity machines (x86 processors that run Linux) and require
queuing software (HTCondor, SLURM, etc.) and Grid middleware (gridftp,
voms, etc.). Currently Belle II jobs are submitted to the Grid as
single core jobs, however, Belle II is developing/testing a multicore
solution. Multicore jobs will reduce processing time and RAM per
core. This will allow Belle II jobs to more efficiently use
opportunistic resources such as Amazon EC2 and HPC clusters. Belle II
has started to test jobs on HPC resources and identified some
challenges when submitting jobs as backfill; these challenges are
expected to be partially resolved with multicore jobs. However, most
of the Belle II code and external library do not take advantage of the
HPC hardware and are not compiled optimally. Moreover, only one binary
is currently distributed on the Grid.

The Wide Area Network requirements for Belle II are similar to that of
the LHC experiments and the needs are expected to be satisfied by the
NRENs.
 
Belle II is currently developing extensions to the DIRAC framework to
enhance the distributed computing model. Central to this effort are
the ``Fabrication System'' and ``Data Management System''.

\medskip

\noindent{\bf{\em 1.8.3~Compute, Data, and Services Needs on the
    2020/2025 Timescale}}

\noindent{\em Neutrino Experiments:} In neutrino physics in the US,
DUNE will be the driving experiment in terms of computing and expects
to lean heavily on the trails blazed by MicroBooNE, the DUNE prototype
detectors and other LArTPC work.  If one contemplates scaling up to
the DUNE FD it is expected that the traditional approach of framing
out data files to a cluster of CPUs will no longer scale.  It remains
to be seen if even farming out data files to a Grid of clusters can
scale.

Reconstruction algorithm currently used, such as PandoraPFA, will
require some profiling and optimization to potentially perform at the
expected rates. Moreover, there is a strong likelihood that
parallelizing at the scale of each event may be required to
efficiently exploit the growing CPU count in commodity cluster nodes.
Methods to serve individual events and return the results of the
process are needed.

This event-level parallelism marks a qualitative shift.  At this scale
the ``batch processing'' system must be aware, to some extent, of the
details of the ``batch job''.  Developing the services to support such
a processing paradigm should be best done in a way that can
nonetheless remain general purpose.

Novel reconstruction techniques are being developed that may take the
granularity to sub-event levels.  The need to support this scale of
parallel processing puts a stronger requirement on the need to develop
general systems to handle farming smaller units of computing.

Finally, it is recognized that a major shift is needed in the
generally monolithic development practices of
``physicists-programmers'' in order to exploit the high CPU-count
environments in modern computational clusters such as HPC.  Training
in general parallel processing techniques is needed and specifically
competency in exploiting GPU co-processors must be increased if
neutrino data processing will exploit modern large-scale computing
resources.

\noindent{\em Belle II Experiment:} During the 2020/2025 timescale,
Belle II is expected to be running at designed luminosity and
generating ~16Gbps of raw data. This will drive Belle II grid
computing to expand the existing workflow to include all elements of
the computing infrastructure such as CPU, networking, and storage
locality. Additionally, provenance and simulation/modeling can provide
guidance to the workflow scheduler.  The end goal is for each job
workflow to be properly matched to the available resources given the
current and forecasted conditions of the distributed system. This
would improve the overall processing efficiency and stability of the
distributed computing effort.  With the growing number of experiments
using the Grid resource allocation is a concern.

Automation of data-centric tasks will be required in order to handle
the anticipated large distributed data volume.  A common solution for
storage accounting, data mobility, data integrity, and data healing
would benefit the DOE-HEP community. Additionally, a common DOE-HEP
data availability solution should be developed to minimize cost in
developing redundant solutions.

New software development should focus on profiling/optimizing existing
software or writing new software to leverage the new architectures
available at HPC centers. Many techniques can potentially be applied
to improve the existing code. Belle II and other DOE-HEP experiments
would greatly benefit from training in code optimization, parallel
processing techniques and access to new hardware technologies.

\newpage

\noindent{\bf\scshape{1.9~Evolution of HEP Facilities: Compute Engines}}  

\medskip

\noindent{\bf Authors:} S. Fuess and P. Spentzouris

\medskip

\noindent{\bf{\em 1.9.1~Current Science Activities and Computational 
    Approaches}}

Computing for Theoretical HEP combines tasks at leadership class
machines with tasks on dedicated commodity clusters that are based at
HEP facilities. (At HEP facilities, the principal communities are
Lattice QCD and Accelerator Modeling.)  This computing is
program-driven, with users allocated resources via a proposal process.
The program and associated projects tend to be of long (months/years)
duration, with no specific time criticality of operation.  As
computing for theoretical HEP is already closely aligned with ASCR,
further discussion will focus on computing for experimental HEP.

Experimental HEP data processing is dominated by pleasingly-parallel
event-based simulation and analysis.  In HEP parlance, an event refers
to both an underlying physics process and the data associated with a
detector during a slice in time, often related to a beam spill or
colliding beam crossing.  The principal computing steps associated
with data processing -- event simulation, reconstruction, and analysis
-- are primarily performed with single-threaded, large memory,
independent instances of experiment-developed applications.
Multi-threaded variants of applications are beginning to appear, but
performance improvement has not yet dictated a mass migration to this
model.

Experimental HEP data processing is thus best matched to a High
Throughput Computing (HTC) model, where the processing aspect is
straightforward and much of the complication lies in moving data to
and from the processing units.

The typical HTC hardware architecture is a multi-processor, multi-core
x86 machine 
\begin{wrapfigure}[14]{l}{4in}
\centering \includegraphics[width=4in]{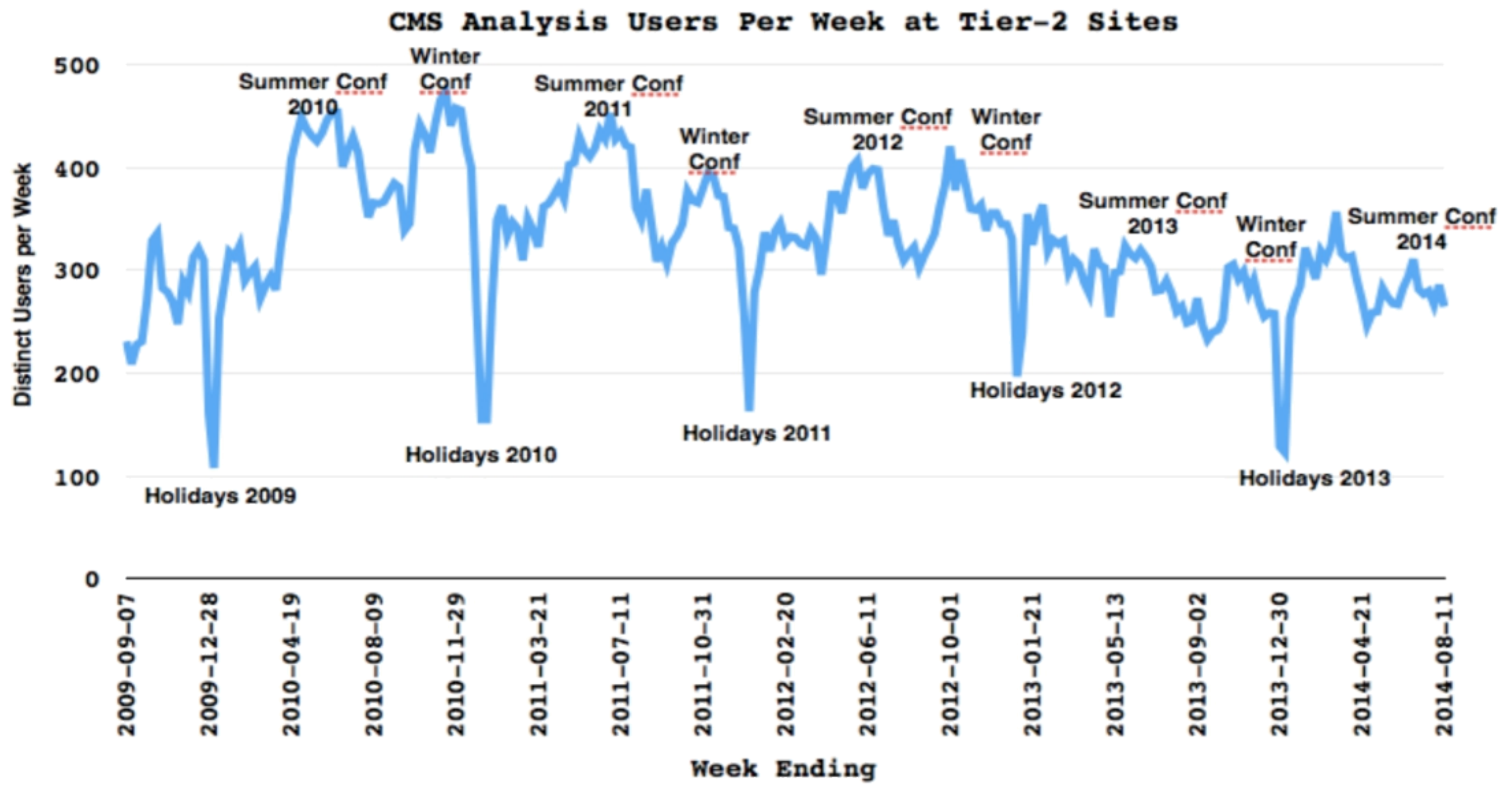}

\vspace{-0.2cm}

\caption{\small{\em Time variation of CMS user jobs.}}   
\label{cms_jobs}
\end{wrapfigure} 
\noindent
(currently capable of 20 to 64 applications or threads) with O(2GB)
memory per core, with a multi-Gbit or 10 Gbit network interface per
machine.  [Note: in the following we will use the term ``core'' to
represent an independent processing unit; processors capable of
multiple threads per core (hyperthreading) are considered as the
equivalent of multiple cores.]

The approximate size of the HEP computing infrastructure, in units of
cores as defined above, is:
\begin{itemize}
\item World-wide CMS Experiment: 100K cores, including the CMS Tier-1
  and Tier-2 facilities
\begin{itemize}
\item US CMS: 15K cores at Fermilab, 25K cores at Tier-2 and Tier-3
  sites
\end{itemize}
\item General purpose (non-CMS) at Fermilab: 12K cores
\item Similar (or larger) values for world-wide and US ATLAS, and for
  the ATLAS Tier-1 center (at BNL) and Tier-2 facilities
\end{itemize}

A significant portion of HEP computation occurs in bursts, associated
with new beam or detector operations, with re-processing efforts on
existing data utilizing newly developed or improved algorithms, or
with intensive preparations in advance of conferences or publication.
As an example, Figure~\ref{cms_jobs} illustrates the time variation of
CMS user jobs.

The HEP computing community is in the initial stages of exploring
commercial cloud resources for their processing intensive jobs.  This
effort has been enabled on several fronts -- with grants from cloud
service providers, with network peering arrangements with the
principle science networks (ESnet, Internet2), and general outreach
from the providers in terms of education and establishment of business
relationships.

HEP data processing jobs are efficiently distributed among the
processing resources, so that the average utilization (defined as a
processing job occupying an available core) exceeds 95\%.

A single HEP processing job is likely to be CPU limited, but the I/O
resource needs of the job are likely to highly impact the architecture
of the required computing systems.  Event and detector simulation jobs
require minimal input (but multiple simulation jobs are likely to
share the same input, for example beam parameters) but produce output
data files of size of order 1GB.  Reconstruction (of both simulated
and real data) and analysis jobs require unique input files (again of
order 1 GB each), possibly shared input files (for example calibration
information), and produce output data files of order 1 GByte.  Network
or storage I/O rates need to be sufficient to not dominate the total
processing time.

CMS computing needs are listed in Section~A1.5. It contains a table
from which one can calculate that a typical simulated event is of size
1.5 MB.  A simulated data file of 1 GB thus contains ~666 events.  The
table also contains information from which one can calculate that a
single simulated event requires approximately 50 seconds to
reconstruct on a typical current CPU; hence a 1 GB simulated data file
requires approximately 9 hours.  This roughly illustrates the
relationship between input and output data sizes and processing time.

\medskip

\noindent{\bf{\em 1.9.2 Evolution of Science Activities and
    Computational Approaches on the 2020/2025 Timescale}}

In this time period, experimental HEP will see:
\begin{itemize}
\item Continuing and increasing computational needs for LHC computing.
\item The emergence of DUNE as the central long-baseline neutrino
  experiment with extensive associated simulation needs.
\item Multiple smaller efforts associated with muon science and
  short-baseline neutrino physics.
\end{itemize}
Estimates of computational needs are given in the next section.

HEP computational techniques are expected to evolve to utilize
parallel and/or accelerator assisted software algorithms.  The
per-core memory footprint should scale accordingly.  In addition,
traditional HEP Facilities such a the Fermilab facility are expected
to evolve to incorporate a ``rental'' resource model that allows
``elasticity'' in provisioning of resources according to demand.  HPC
cycles, either programed (persistent) or opportunistic are an
appealing component of this model.  In order to make utilization of
such resources feasible we need to work with the HPC ASCR community to
develop the necessary software infrastructure that will allow us to
monitor available cycles, provide integrated workflow management
capabilities that will allow us to submit and monitor jobs in the HPC
machines, and apply an appropriate security layer. In addition, data
management will require significant networking 
\begin{wrapfigure}[13]{l}{3.5in}
\centering \includegraphics[width=3.5in]{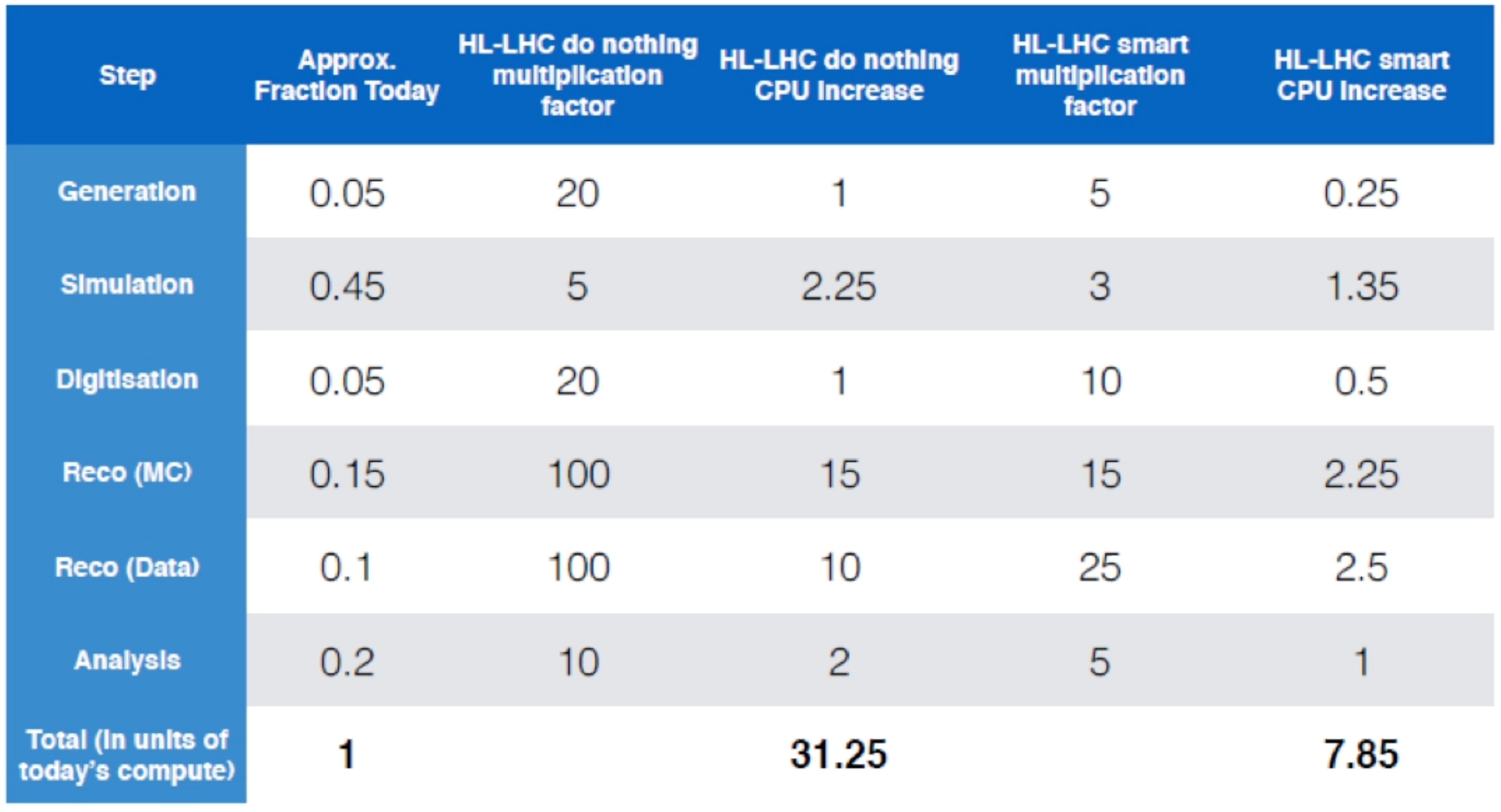}

\vspace{-0.2cm}

\caption{\small{\em Scale increase of LHC processing estimated for
    HL-LHC running.}}   
\label{cpu_lhc}
\end{wrapfigure} 
\noindent
capacity and monitoring capabilities to make this efficient.

A significant effort leading up to this time period will be devoted to
developing the cost models for utilizing the mixture of external cloud
provider resources, HPC resources, or dedicated HTC resources.  Each
resource type may be most suitable for specific phases and functions
of the HEP computational load.  The selection of the resource type
will be made opaque to the user submitting the job, with the decision
to be determined based upon aggregate usage patterns, resource
availability and cost, and criticality of the job.

\medskip

\noindent{\bf{\em 1.9.3 Compute, Data, and Services Needs on the
    2020/2025 Timescale}}

Figure~\ref{cpu_lhc}~\cite{scale} provides an estimate of the scale
change, relative to current
\begin{wrapfigure}[13]{l}{2.5in}
\centering \includegraphics[width=2.5in]{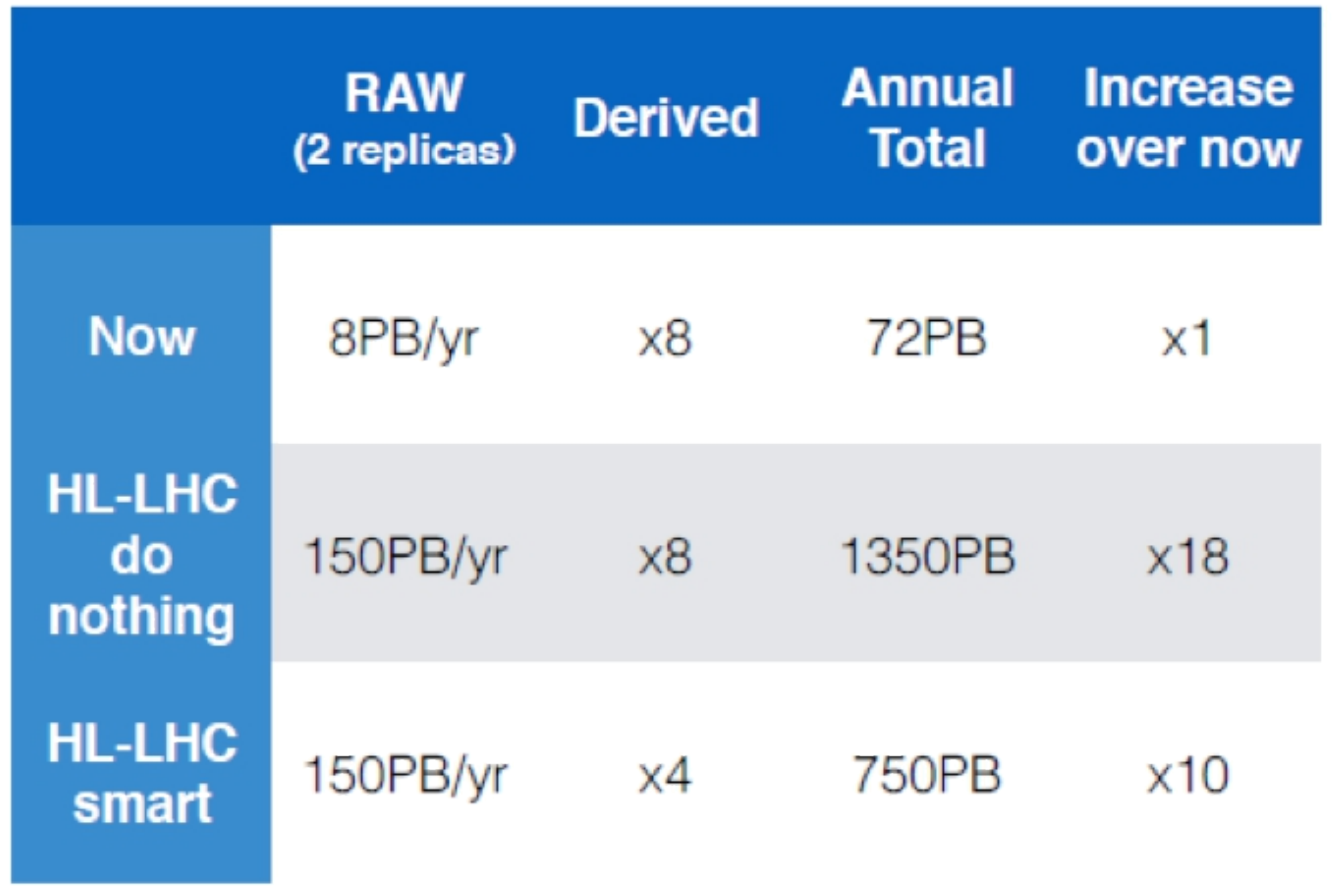}

\vspace{-0.2cm}

\caption{\small{\em Estimate of data sizes for HL-LHC.}}   
\label{data_lhc}
\end{wrapfigure} 
\noindent
needs, for the processing necessary to handle LHC data by the era of
the High Luminosity LHC (HL-LHC)~\cite{hllhc}, scheduled to begin
operation in approximately 2025~\cite{hllhc2}. The message of
Figure~\ref{cpu_lhc} is that even with optimization of algorithms to
handle more complex event data, HL-LHC alone will require nearly an
order of magnitude increase in processing capacity.  An order of
magnitude increase is consistent with hardware performance and
capacity improvements expected in the same time frame.

Ref.~\cite{scale} also provides an estimate of data produced in the HL-LHC
era, illustrated in Figure~\ref{data_lhc}.  An order of magnitude increase in data
is again consistent with expected hardware performance and capacity
improvements, with one implication that tape will likely be a required
storage component.

\newpage

\begin{center}
{\bf\scshape{Appendix 2: Case Studies}}
\end{center}

The white papers in the previous Appendix presented broad coverage of
each HEP focus area. The case studies in this Appendix are meant to
drill down to the level of an individual application (or set of
applications targeted to a single science goal). The case studies
focus at the level of individual computational codes (or code
frameworks) and make statements about their current status and usage
patterns. They also cover what work needs to be done to get ready for
next-generation computing architectures and what the specific
computing requirements are likely to be on the 2020/2025 timescale.

Although each case study represents a substantial unit of work, the
examples covered below can provide only a cursory glimpse of HEP
computational activity. They range from those requiring a relatively
modest computational resource (``mid-scale'') to others that run on
the largest supercomputers, or exploit very large-scale high
throughput computing. The aim of this Appendix is to provide a flavor
of a subset of HEP computing tasks and their future roadmaps. (The one
exception is the `Dark Energy Survey in Hindsight' case study that
focuses on lessons learnt and not on a future roadmap.)

\newpage

\noindent{\bf\scshape{2.1~Advanced Modeling of Accelerator Systems
    (ACE3P)}}   

\medskip

\noindent{\bf Authors:} L. Ge, K. Ko, O. Kononenko, Z. Li, C.-K. Ng,
L. Xiao

\medskip

\noindent{\bf{\em 2.1.1~Description of Research}}

\noindent{\bf Overview and Context:} Accelerator modeling using high
performance computing has provided the capability of high fidelity and
high accuracy simulation of accelerator structures and systems for the
design, optimization and analysis of accelerators. Running on DOE
state-of-the-art supercomputers, parallel electromagnetics computation
has enabled the design of accelerator cavities to machining tolerances
and the analysis of large-scale accelerator systems to ensure
accelerator operational reliability. The modeling effort has supported
many operational and planned accelerator projects within the DOE
accelerator complex and beyond such as LHC Upgrade and PIP-II in HEP,
CEBAF Upgrade, FRIB and e-RHIC in NP, as well as LCLS, LCLS-II in BES.

\medskip

\noindent{\bf Research Objectives for the Next Decade:} Accelerator
modeling will continue to solve challenging computational problems in
accelerator projects and accelerator science by developing
multi-physics capabilities and fast numerical algorithms to facilitate
optimized accelerator design, hence saving time and reducing
costs. Advances of enabling technologies in scalable linear algebra
solvers, mesh adaptation and dynamic load balancing that are specific
to large-scale finite element simulation are required to achieve the
scientific goals using emerging computer architectures.

\medskip

\noindent{\bf{\em 2.1.2~Computational and Data Strategies}}

\noindent{\bf Approach:} The computational problems include the
solution of a sparse linear system, whose convergence requires the use
of direct solvers with reduced memory footprint, and thus the
development of solvers scalable in memory and on multi-core
architectures is necessary. In addition, large datasets of particle
and field data generated in the time domain are transferred back to
local computers for visualization. With improvement of network
bandwidth, remote visualization on LCFs can mitigate the problem of
data transfer of increased sizes of datasets.

\medskip

\noindent{\bf Codes and Algorithms:} ACE3P is a comprehensive set of
parallel finite element electromagnetics codes with multi-physics
thermal and mechanical characteristics for simulation on unstructured
grids. The simulation workflow starts from preprocessing for model and
mesh generation on desktop computers, job execution on LCF, and then
postprocessing including visualization and data analysis locally or
remotely. ACE3P solves for eigenvalues and harmonic excitation
problems in the frequency domain, employs an implicit scheme and the
particle-in-cell method in the time domain, and uses Runge-Kutta
algorithms for particle tracking in electromagnetic fields.

\medskip

\noindent{\bf{\em 2.1.3 Current and Future HPC Needs}}

\noindent{\bf Computational Hours:} Currently ACE3P uses 2.5M CPU
hours on NERSC computers. A computational challenging problem will be
to model dark current effects in the entire linac of an accelerator
such as the superconducting linac in PIP-II and its upgrade to high
power operation. The problem size will be 20-30 times larger than that
of current simulation. It is anticipated that a growth of more than an
order of magnitude to 50M CPU hours is required in the next decade.

\medskip

\noindent{\bf Parallelism:} The codes use MPI for parallelism with the
average number of cores in the order of 5,000.  Parallelism on
multi-core nodes focuses on the development of hybrid linear solvers
that are scalable in memory. The average number of cores will increase
by an order of magnitude to 50,000 due to the increase in problem
size.

\medskip

\noindent{\bf Memory:} ACE3P simulation in frequency domain requires
large per core memory and hence benefits from compute nodes with large
memory. Currently, the simulation uses 64 GB of memory on a compute
node and the aggregate memory can reach up to 2 TB for electromagnetic
simulation. Future aggregate memory will increase by an order of
magnitude to 40 TB for multi-physics simulation.

\medskip

\noindent{\bf Scratch Data and I/O:} A typical run in the time domain
generates 1-2 TB of data including field and particle snapshots and
checkpoint files. A total of 50 TB scratch space is required for ACE3P
users to perform their simulations concurrently. The current I/O
bandwidth is estimated to be 20 GB/sec. Future requirements will
increase the size of output datasets to 20 TB and the I/O bandwidth to
80 GB/sec to maintain reasonable I/O percentage of the runtime, which
is about 20\%.

\medskip

\noindent{\bf Long-term and Shared Online Data:} Several of the
production runs are shared for the collaboration. It is estimated 5 TB
storage for long-term data is required, which will increase to 50 TB
in the next decades.

\medskip

\noindent{\bf Archival Data Storage:} About 50 production runs need to
be archived. The estimated current space is 100 TB and the future
storage will increase to 800 TB.

\medskip

\noindent{\bf Workflows:} The data generated from simulations on LCFs
are transferred back to local computing resources for analysis, and
hence maintaining and enhancing adequate data bandwidth from the
remote facility are essential to the scientific process. For the next
decade, the use of remote processing and visualization of data will
alleviate the demand for high bandwidth of data transfer.

\medskip

\noindent{\bf Many-Core and/or GPU Readiness:} ACE3P's current
parallel implementation uses MPI. The plan to build a hybrid
programing paradigm with OpenMP is under way, for example, for
particle tracking. In addition, ACE3P will benefit from improvement of
third-party linear algebra libraries on multi-core architectures.

\medskip 

\noindent{\bf Software Applications, Libraries, and Tools:} N/A

\medskip

\noindent{\bf HPC Services:} N/A

\medskip

\noindent{\bf Additional Needs:} N/A

\newpage

\centerline{\bf Requirements Summary Worksheet}

\begin{figure}
\centering \includegraphics[width=6.5in]{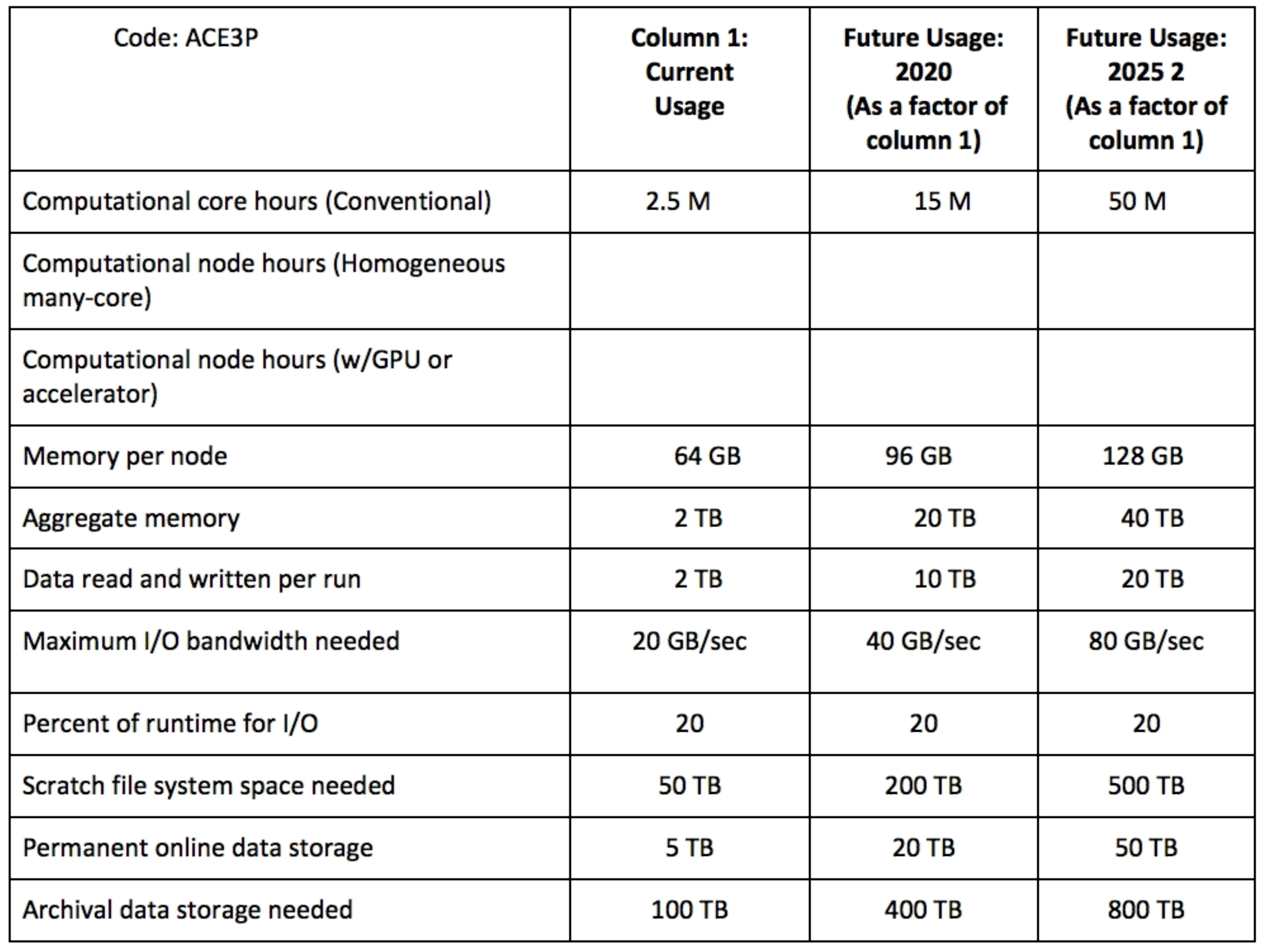}
\end{figure}

\newpage

\noindent{\bf\scshape{2.2~Luminosity Optimization in High Energy
    Colliders Using High Performance Computers (BeamBeam3D)}}

\medskip

\noindent{\bf Authors:} J. Qiang

\medskip

\noindent{\bf{\em 2.2.1~Description of Research}}

\noindent{\bf Overview and Context:} The event rate in a high energy
collider such as the LHC is directly proportional to the luminosity of
the colliding beams. The achievement of higher luminosity (integrated)
is critical for new scientific discovery. However, colliding beam
effects (also called beam-beam effects) are a limiting factor for the
final achievable luminosity in all high energy colliders including the
LHC. The research focus is on understanding and mitigating these
effects through self-consistent simulations and optimizing the
collider operation/design to achieve a higher luminosity through using
high performance computers.

\medskip

\noindent{\bf Research Objectives for the Next Decade:} The research
objectives for the next decade are to study the beam-beam effects and
compensation methods including long-range beam-beam effects (if
possible) for the LHC operation, the HL-LHC upgrade and other
future high energy colliders such as FCC to achieve optimal
luminosity.  This study will integrate multi-bunch simulation
including both the head-on beam-beam effects and the long-range
beam-beam effects, an important factor limiting luminosity lifetime,
with the machine parameter optimization. The computational and data
analysis/processing goals are to do multi-level parallel simulation
and data analysis to identify the best machine parameter settings for
the optimal luminosity.

\medskip

\noindent{\bf{\em 2.2.2~Computational and Data Strategies}}

\noindent{\bf Approach:} The computational problem is to solve
multiple coupled 6D Vlasov-Poisson equations (can be on the order of a
thousand) and to optimize the collider luminosity with respect to the
machine setting including beam-beam effects. The strategies used to
solve the equations are based on a multi-species self-consistent
particle-in-cell method with evolution-based parallel machine
parameter optimization.

\medskip

\noindent{\bf Codes and Algorithms:} The code used in this study is
BeamBeam3D. The algorithm is a parallel particle-in-cell method with
particle-field decomposition to achieve a perfect load balance. This
algorithm will be integrated with a parallel optimization algorithm
for machine parameter optimization.

\medskip

\noindent{\bf{\em 2.2.3 Current and Future HPC Needs}}

\noindent{\bf Computational Hours:} At present, we typically use about
one and half million core-hours on conventional cores.  We expect that
it will increase to 10 to 20 million core-hours through 2020 and 2025

\medskip

\noindent{\bf Parallelism:} Currently, we are using two-level
parallelization: one level for multi-group parallel parameter scans
and one level for parallel simulation. Our current plan to increase
the level of parallelization is to use multi-thread share memory
programming together with MPI, and to parallelize the multiple bunch
collisions at multiple interaction points.

\medskip

\noindent{\bf Memory:} Currently, we are using a few million
macroparticles for a single simulation run. Together with the parallel
parameter scan, this requires 1-100 GB memory of a parallel
computer. In future studies, we plan to increase this by one more
level of parallelism to account for multi-bunch long-range beam-beam
effects. In the LHC study, each beam has more than 2000 bunches. A
full optimization study including colliding beam effects with
potentially a larger number of macroparticles in the simulation will
require 200-2000 TB memory (DRAM).

\medskip

\noindent{\bf Scratch Data and I/O:} The on-line scratch storage
needed in our study is on the order of 10 GB  to 20 TB depending on the
applications. We hope that the runtime for I/O can be controlled
within 5\% of the total computing time.

\medskip

\noindent{\bf Long-term and Shared Online Data:} The online-long-term
storage that we need today is about 2TB. In 2020 and 2025, the storage
that we need is about 20 TB and 40 TB.

\medskip

\noindent{\bf Archival Data Storage:} Currently, we probably have
stored about 4 TB archival data. We will probably need 50 TB and 100
TB in 2020 and 2025.

\medskip

\noindent{\bf Workflows:} N/A

\medskip

\noindent{\bf Many-Core and/or GPU Readiness:} Our code is not ready
for this yet. We plan to add OpenMP into the code to exploit the
shared memory architecture inside a node and to use MPI for cross-node
communication.

\medskip 

\noindent{\bf Software Applications, Libraries, and Tools:} N/A

\medskip

\noindent{\bf HPC Services:} N/A

\medskip

\noindent{\bf Additional Needs:} N/A

\newpage

\centerline{\bf Requirements Summary Worksheet}

\begin{figure}
\centering \includegraphics[width=6.5in]{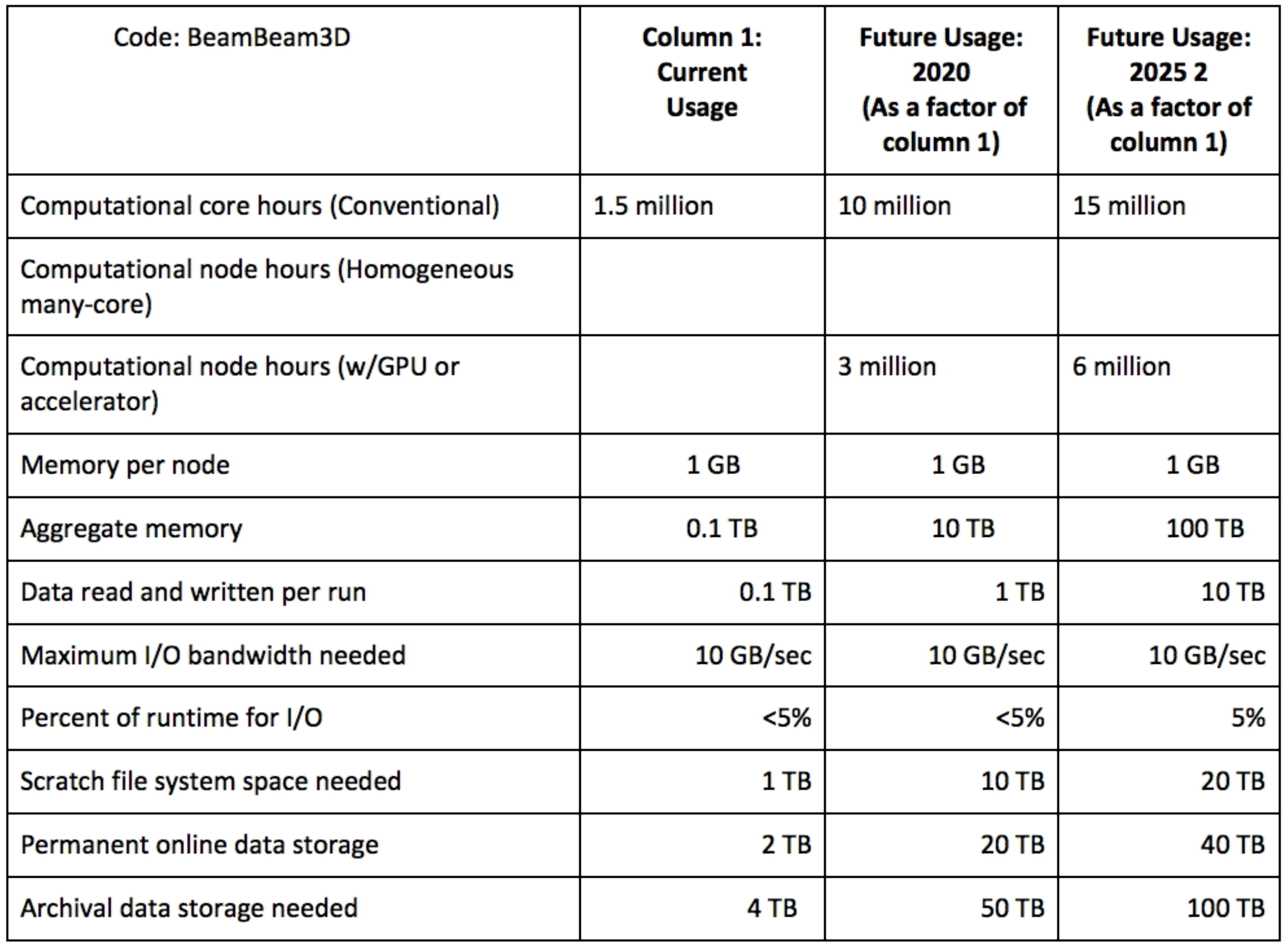}
\end{figure}

\newpage

\noindent{\bf\scshape{2.3~Computational Cosmology (HACC)}}  

\medskip

\noindent{\bf Authors:} H. Finkel, N. Frontiere, S. Habib, K. Heitmann, A. Pope

\medskip

\noindent{\bf{\em 2.3.1~Description of Research}}

\noindent{\bf Overview and Context:} High-fidelity cosmological
surveys require immense simulation capability to make maximal use of
information in, e.g., the spatial clustering of galaxies, to make
cosmological inferences. Structure formation simulations are integral
in survey planning, error characterization, and for calculating
observable signatures to which the data will be compared. Galaxies
form in highly over-dense regions but trace delicate structures that
span distances of hundreds of millions of light-years, demanding
dynamic ranges of roughly six order of magnitude in spatial resolution
and even more in mass resolution. Much like galaxy surveys, simulation
data products are rich and can be interrogated in various ways at
several levels of data reduction, requiring significant resources for
data archiving, access, and distribution.

\medskip

\noindent{\bf Research Objectives for the Next Decade:} Gravity is the
dominant force on large length scales; currently HACC (Hardware/Hybrid
Accelerated Cosmology Code) is gravity-only, and the effects of other
interactions involved in the process of galaxy formation are modeled
in post-processing using a variety of techniques. Next-generation
surveys require significantly higher mass resolution and inclusion of
baryons and feedback effects while maintaining survey-relevant volumes
and resolutions.

\medskip

\noindent{\bf{\em 2.3.2~Computational and Data Strategies}}

\noindent{\bf Approach:} HACC's N-body methods use tracer particles to
track the phase-space distribution of matter in an expanding
universe. Force-splitting methods are used to calculate long-range
forces using MPI distributed-memory methods and short-range forces
using shared-memory methods. Individual simulations must run longer
than individual job runtimes (and mean-time-to-failure) requiring high
performance IO for checkpoint/restart files. Particle methods should
scale well for the next generations of large-scale HPC systems.

\medskip

\noindent{\bf Codes and Algorithms:} The HACC framework runs on all
types of HPC systems, including power-efficient multi/many-core
systems and accelerated systems. Long-range forces are computed by a
high-order spectral particle-mesh method. Portable, custom-written MPI
code is used to perform a distributed memory FFT to solve for
long-range forces. Short-range force calculations are adapted to the
underlying architecture. Tree methods are used to generate particle
interaction lists on multi-core and many-core CPUs, while an OpenCL
code employs a simpler fixed spatial-scale data structure for
calculations on GPUs. Custom-written (non-collective) MPI-IO code with
internal checksums is used to checkpoint/restart and output. files. We
will add a new higher-order smoothed-particle hydrodynamics (CRK-SPH)
method to HACC in order to scale baryonic physics calculations to
future HPC systems.

\medskip

\noindent{\bf{\em 2.3.3 Current and Future HPC Needs}}

\noindent{\bf Computational Hours:} Currently HACC simulations run
under ALCC and INCITE awards at the scale of 200M
core-hours/year. This includes running the GPU version of the code
with the OLCF charging factor included in the above number. Future
usage is expected to go up by a factor of $\times 20$ in the
pre-exascale era, to about $\times 200$ at exascale.

\medskip

\noindent{\bf Parallelism:} Production runs regularly use 262,144 MPI
ranks and 8 threads per rank on 32,768 nodes (524,288 cores) of IBM
Blue Gene/Q systems for a total of $\sim$2.1 million
threads. Performance tests of the full code scaled up to $\sim$6.3
million threads on 98,304 nodes (1,572,864 cores) of IBM Blue Gene/Q
at greater than 90\% parallel efficiency and ~69\% of peak
performance. Tests of the MPI-parallel FFT have scaled up to $\sim$1
million MPI ranks for a 16,3843 grid. The production GPU version the
code regularly uses 16,384 nodes of Titan at one MPI rank per node and
achieves full occupancy on the NVIDIA K20 GPUs.

\medskip

\noindent{\bf Memory:} Memory (DRAM) requirements driven by science
cases for large-volume survey simulations already exceed
next-generation system capabilities (10+PB desired versus 1+PB
available). HACC strong-scales to 100MB/core, which is likely
sufficient for exascale architectures. The use of NVRAM as a local
memory extension is being investigated.

\medskip

\noindent{\bf Scratch Data and I/O:} The scratch storage requirements
track the total RAM available in the machine multiplied by the number
of snapshots. This can easily lead to $\sim$100+PB requirements, which
will be mitigated to some extent by using in situ analysis (already
included in HACC). I/O bandwidth requirements are $\sim$1TB/sec until
$\sim$2020, and $\sim$10TB/sec in the 2020-2025 timescale. Obviously,
we would like to minimize I/O overhead (we are thinking of how to use
NVRAM to reduce this overhead to a very small fraction, if not zero.)
We would like to have less than 10\% of our time be lost to I/O.

\medskip

\noindent{\bf Long-term and Shared Online Data:} We already need
$\sim$10PB of active (spinning disk) long-term storage (which we don't
have). This number can easily hit the Exabyte scale by
2020-2025. Currently, we use Globus transfer services for sharing data
between the LCFs and NERSC.

\medskip

\noindent{\bf Archival Data Storage:} Currently we have 4PB (soon to
be 6PB) in archival storage (HPSS). We expect this to scale out by a
factor of up to 1000 by 2025 (but everything will depend on data
access bandwidth).

\medskip

\noindent{\bf Workflows:} We use SMAASH, a simulation management and
analysis system, for workflow management. We will continue to develop
SMAASH to handle large-scale simulation campaigns.

\medskip

\noindent{\bf Many-Core and/or GPU Readiness:} See discussion above;
HACC is ready for next-generation many-core and GPU systems, with
early science projects on Cori, Summit, and Theta. We tune
data structures and algorithms for the short-range force calculations,
the major computational hot-spot, to make efficient use of the
available memory hierarchy and bandwidth on each architecture that we
use at scale.

\medskip 

\noindent{\bf Software Applications, Libraries, and Tools:} Currently
we place minimal reliance on (non-vendor supported) libraries because
of past experiences on HPC systems. We do not expect this to change in
the near future.

\medskip

\noindent{\bf HPC Services:} A federated authentication system coupled
to a well-tuned and supported data transfer system (e.g., Globus)
would help greatly with data sharing and publication.

\medskip

\noindent{\bf Additional Needs:} N/A

\newpage

\centerline{\bf Requirements Summary Worksheet}

\begin{figure}
\centering \includegraphics[width=6.5in]{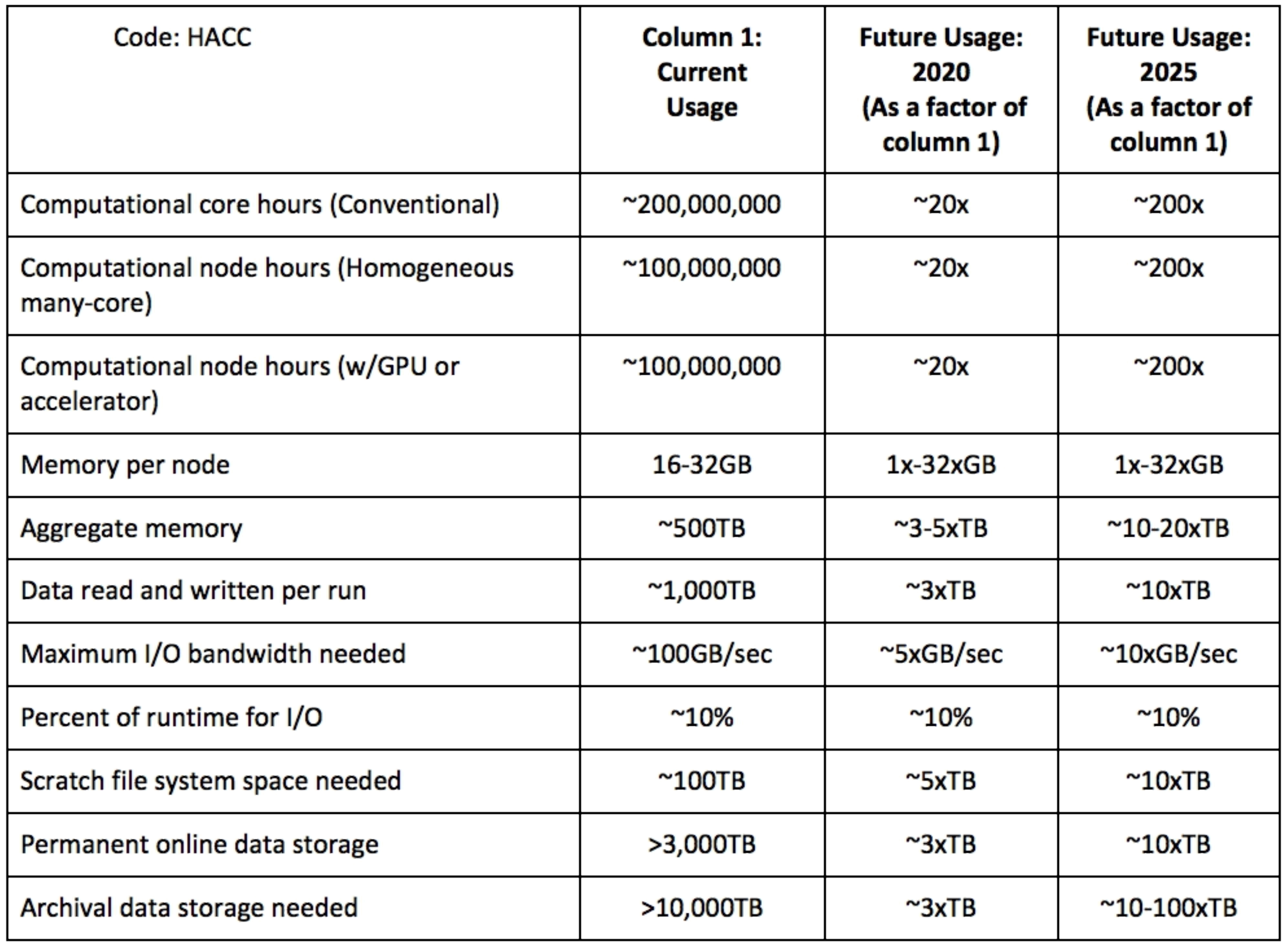}
\end{figure}

\newpage

\noindent{\bf\scshape{2.4~Cosmic Reionization (ART)}}  

\medskip

\noindent{\bf Authors:} N. Gnedin

\medskip

\noindent{\bf{\em 2.4.1~Description of Research}}

\noindent{\bf Overview and Context:} The primordial anisotropies in
the Cosmic Microwave Background (CMB) have been measured with great
precision. Modern constraints on the dark matter, dark energy, or
modifications of gravity are inconceivable without using these
measurements as one of the main data sets. Hence, maximizing the
achievable precision from the CMB data is crucial. As statistical
errors continue to decrease, the importance of systematic
uncertainties in CMB modeling increases. Of these, cosmic reionization
is by far the most important. The process of reionization leads to the
transition from mostly neutral to a highly ionized state for most of
the cosmic gas. This ionized gas serves as a semi-transparent screen
for the CMB, affecting the anisotropies in a well-understood
manner. Because observational constraints on reionization are limited,
theoretical modeling, including numerical simulations, plays a very
important role. The Cosmic Reionization On Computers (CROC) project
aims, over the course of several years, to model -- fully
self-consistently -- all the relevant physics, from radiative transfer
to gas dynamics and star formation, in simulation boxes up to 100
comoving Mpc and with spatial resolution reaching 100 pc in physical
units.

\medskip

\noindent{\bf Research Objectives for the Next Decade:} The next
decade will see three major observational advances: 1) The James Webb
Space Telescope (JWST, 2018 launch), will provide orders of magnitude
increase in the data sample sizes of primary reionization sources --
high redshift galaxies. 2) 30-meter class telescopes will increase
many-fold the number of known high-redshift quasars that serve as
lightposts against which the intergalactic gas is observed. 3) radio
observatories, e.g., the Hydrogen Epoch of Reionization Array (HERA)
or Square Kilometer Array (SKA) will map neutral hydrogen at large
scales during cosmic reionization. To match the flood of observational
data, higher precision simulations will be required. The ``ideal''
reionization simulation for matching the new data should reach 100 pc
spatial resolution and mass resolution of about 10$^6$ solar masses. For
a 100 Mpc box, such a simulation would use over 100 billion
particles/cells, and consume close to 500 million CPU-hours. At
present this is unfeasible, but would be a routine calculation some
time in the second half of the next decade.

\medskip

\noindent{\bf{\em 2.4.2~Computational and Data Strategies}}

\noindent{\bf Approach:} Cosmological reionization simulations model a
diverse range of physical processes, from gas dynamics and
gravitational collapse to cosmological radiative transfer and various
atomic processes in the primordial gas. In order to achieve the high
dynamic range needed for reionization simulations, we use the Adaptive
Mesh Refinement (AMR) approach, implemented in the Adaptive Refinement
Tree (ART) code. The large size of the existing and future simulations
will present a serious problem for data storage and analysis. Hence, a
substantial effort will have to go into developing efficient and
scalable algorithms for dynamic data compression and on-the-fly
analysis.

\medskip

\noindent{\bf Codes and Algorithms:} The ART code uses an N-body
method to solve the Vlasov equation for dark matter and a modern
Riemann solver for modeling gas dynamics. Gravity is solved on the
fully refined mesh with a multi-grid relaxation solver. Approximate
methods are used for modeling radiative transfer and star
formation. Atomic processes are solved exactly with a sub-stepping
technique and implicit solvers.

\medskip
\newpage

\noindent{\bf{\em 2.4.3 Current and Future HPC Needs}}

\noindent{\bf Computational Hours:} Currently we use between 50 and
100 million hours per year. ``Ideal'' reionization simulations mentioned
above will require of the order of 500 million hours each, and a
statistical ensemble of at least 5 of them will be required, so we are
talking about 2.5-3 billion CPU hours for the 2020-2025 period.

\medskip

\noindent{\bf Parallelism:} The current implementation of the ART code
scales to about 10,000 nodes and 20 cores/node for the largest
currently feasible simulation sizes. Going beyond this scaling will
not be possible with the current implementation both for numerical and
algorithmic reasons. We are currently starting work on the next
generation of the ART code that should scale on exascale platforms
(i.e., reach greater than million core scaling), with the expectation
that the new code will be production ready around 2020.

\medskip

\noindent{\bf Memory:} All of the current and future simulations are
CPU limited and the memory requirements are not critical -- i.e., they
are highly suitable for the future machines with low
memory-to-peak-performance ratio. The total memory requirements for
the ``ideal'' simulation described above will be of the order of 1
PB. The per-node memory requirement will depend on the particular
implementation of the next version of the code, but in no case should
be below 16 GB.

\medskip

\noindent{\bf Scratch Data and I/O:} Because of the persistent value
of these simulations, a sensible number of snapshots will need to be
stored. The exact storage requirement will depend on the degree of
compression available to us; something in the range of 10-30 PB seems
to be reasonable. The IO bandwidth will be crucial, however, and will
need to exceed the IO performance of the BlueGene/Q by at least a
factor of 100.

\medskip

\noindent{\bf Long-term and Shared Online Data:} At present we need
about 300TB of active data storage. In the period 2020-2025 this
requirement will grow by a factor of 10-20.

\medskip

\noindent{\bf Archival Data Storage:} At present we have about 1PB of
archival storage used for simulation outputs.  In the period 2020-2025
this requirement will grow by a factor of 10-20.

\medskip

\noindent{\bf Workflows:} We don't need any workflows and do not plan
to use any in the future. 

\medskip

\noindent{\bf Many-Core and/or GPU Readiness:} The current version of
the ART code is not ready for exascale and inhomogeneous
architectures. As we discussed above, work on the next generation,
exascale AMR code has already started, and the new code is expected to
be operational by about 2020.

\medskip 

\noindent{\bf Software Applications, Libraries, and Tools:} N/A

\medskip

\noindent{\bf HPC Services:} N/A

\medskip

\noindent{\bf Additional Needs:} N/A

\newpage

\centerline{\bf Requirements Summary Worksheet}

\begin{figure}
\centering \includegraphics[width=6.5in]{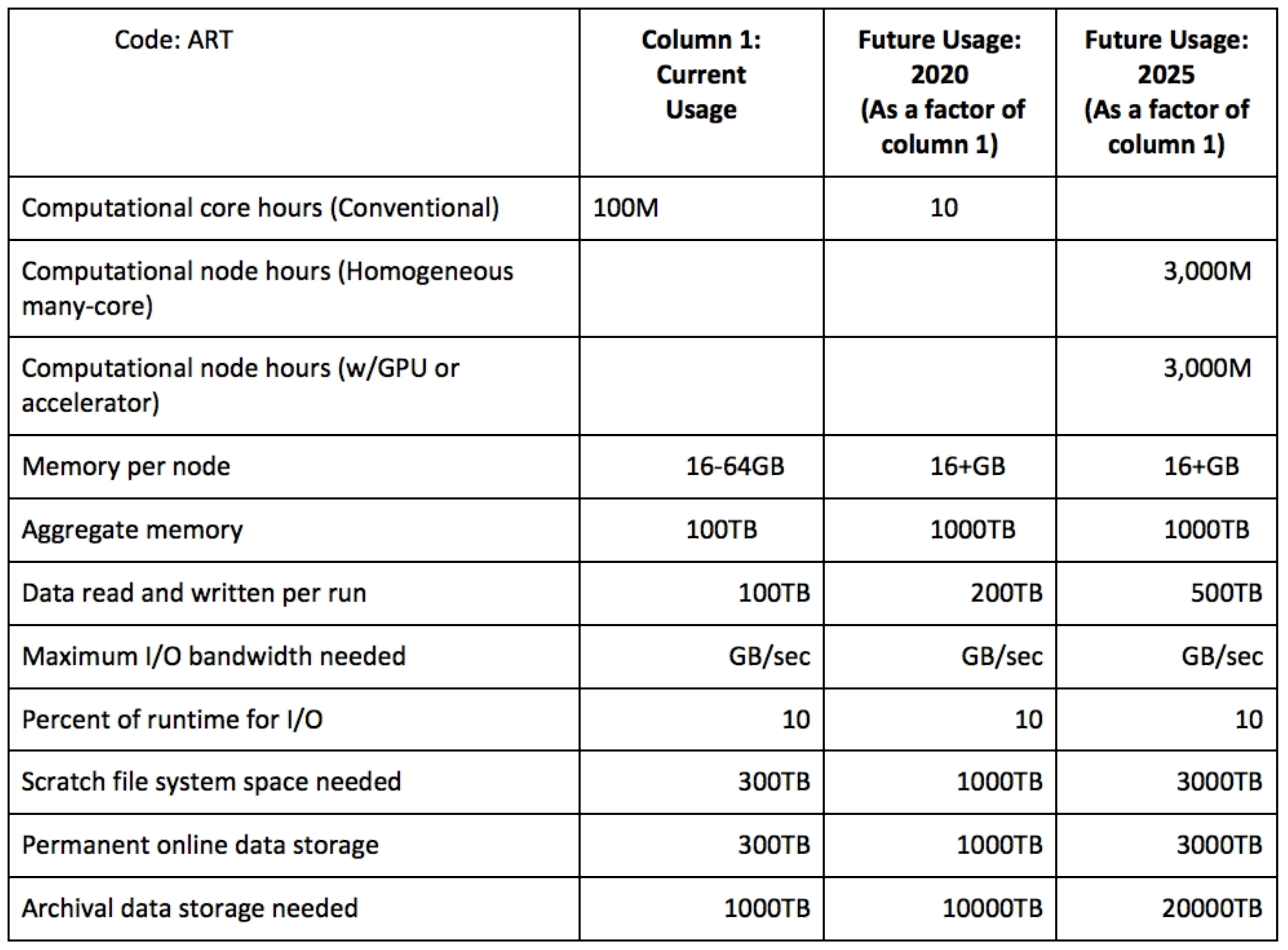}
\end{figure}

\newpage

\noindent{\bf\scshape{2.5~Dark Energy Survey in Hindsight}}  

\medskip

\noindent{\bf Authors:} D. Petravick

\medskip

\noindent{\bf{\em 2.5.1~Description of Research}}

\noindent{\bf Overview and Context:} 
In astronomy, the term Data Management refers to the creation (or
adoption) of production codes; the configuration management needed to
assemble the codes into a coherent set of pipelines; the creation of
the set of services needed for a working system -- including job
management, and management of a collection of data; the effort to
support production, assess the quality of the results, and provide all
other effort to build a systematically coherent set of measurements
supporting a multi-year observational program. DES is organized into a
collaboration. The overall system that is now in place is constructed
to maximize the collaboration's participation in production
activities.

The Dark Energy Survey itself consists of a wide area survey and a
survey that focuses on type 1a supernovae.  The primary set of
measurements supporting the wide field survey is provided by annual
data releases.  The primary set of measurements supporting the
supernova program is prompt processing of specially designated
supernova fields, which are observed on a six-day cadence. In addition
to providing measurements directly supporting the DES science
outcomes, DES Data Management (DESDM) supports DES observers with
prompt processing of all data, and identifying which of the raw data
meeting survey requirements, and identifying fields which must be
re-observed.

Nightly supernova processing produces measurements for the supernova
working group. The central group provides front-end processing
requiring re-use of the DES software stack and the use of
batch-oriented production frameworks. The central group also provides
a VM infrastructure to the supernova science working group for
essential back-end processing of the data.  Data products from this
stage are placed in the central relational database.

Annual weak lensing processing is done within the weak lensing
group. The collaboration has demonstrated a data ingest of shear
catalogs back into the central relational database.

All in all, production is an activity that is carried out in active
collaboration with the science working groups. Within the
collaboration, data is not hidden as it is computed -- all data
products, including nightly quality assurance (which catalogs all
objects albeit with crude calibrations), to the data accumulating for
an annual release are available to the collaboration (though not
necessarily supported by the central group).  Availability of this
data allows for prompt, unanticipated science results.

Additional processes, which were not envisioned in the original
proposal have emerged. After DESDM provides an annual release, a
Release Scientist further characterizes that release, providing
additional characterization, subsetting, and augmentation.  This is
ingested, and integrated into the master relational table views and
joins.

Lastly, work is underway for additional integration for the production
of value-added catalogs that are specialized value-added data products
of interest to one, or a few science working groups.  The LINEA group
in Brazil provides infrastructure for production of these data items,
and the LINEA provided infrastructure is being directly integrated into
the central relational database, which provides efficient access to
the measurements computed in the annual releases.

\medskip

\noindent{\bf Research Objectives for the Next Decade:} N/A

\medskip

\noindent{\bf{\em 2.5.2~Computational and Data Strategies}}

\noindent{\bf Approach:} Processed data are available as FITS images
and tables held in files or as relational tables accessed via SQL, as
apropos. The original technical concept for the DESDM was to provide a
quick and low-cost system based on:
\begin{enumerate}
\item Use of grid computing, both TeraGrid (now XSEDE) and the
  OpenScience Grid (OSG). 
\item Use of a body of community codes, in lieu of bespoke codes for
  important operations, including astrometry, object detection and
  classification, and image co-addition. 
\item A distributed file system, based on independent servers and a
  replica catalog.
\item A system of mutually mirrored relational databases supporting a
  system of primary, secondary and tertiary data and production sites.
\item A production infrastructure that produces supernova pipelines,
  annual release processing consisting of single epoch image
  processing, image co-addition, weak lensing processing, photometric
  redshifts, and other data products.
\end{enumerate}
As of 2015, the system in place has processed data from a science
verification year, and two full years of survey operations. As such we
are able to survey the current state of the system, with lessons
learned.

DES now primarily accepts bulk computing sites on the basis of their
supporting file systems and proximity to the central file store at
NCSA. DES experience is as follows:
\begin{enumerate}
\item DES uses the Fermigrid OSG site and NERSC for processing single
  exposures. Jobs pull their inputs directly from the central file
  store to local disk on an OSG worker node, which has sufficient
  attached disk.   Jobs push their output back to the central file
  store at NCSA before completing. The transport protocol is HTTP.
  The number of HTTP servers is scaled to support the needed
  bandwidth. 
\item DES uses computers with a central, global file system for image
  co-addition. Coadd processing involves gathering all images
  overlapping a 0.25 sq. degree area of sky for processing.  The
  primary resources have been the iForge Cluster at NCSA, which has a
  well-provisioned GPFS file system. 
\end{enumerate}

\medskip

\noindent{\bf Codes and Algorithms:} DES uses an approach based on the
use of community codes. This has driven the DES production framework
to adopt the assumption that all codes are ``hostile'', each with a
different interface, and some optimized for interactive use.
Experience at NCSA has indicated that many production systems for
emerging data-intensive science require production systems capable of
handling hostile codes.  Lessons learned include:
\begin{enumerate}
\item Dealing with many small files
\item Informing scientists how production was done as ``how the
  community codes were called'' instead of how ``wrappers to the
  community codes were called''
\item Budgeting effort to include acquiring thorough acumen about
  these codes and effective liaison to the code developers 
\end{enumerate}
That said, DES's software processes have adapted to these
circumstances. DES provides a strong example of code-reuse, and has
developed techniques to accept codes from its collaborators -- and
from the community -- that are reasonably stand-alone and
framework-independent.

\medskip
\newpage

\noindent{\bf{\em 2.5.3 Current and Future HPC Needs}}

\noindent{\bf Computational Hours:} At the $\sim 1$M core-hour annual
use. 

\medskip

\noindent{\bf Parallelism:} Independently parallel jobs.

\medskip

\noindent{\bf Memory:} Few GB/core.

\medskip

\noindent{\bf Scratch Data and I/O:} DES is like many emerging
production efforts in other science domains, where integrating
community codes results in many small files, which can be problematic
for the file system.

\medskip

\noindent{\bf Long-term and Shared Online Data:} DES attempted to
implement a file architecture similar to a data grid.  Files would be
held by primary, secondary, and tertiary authorized sites. The scope
of this project was greater than could be realized. Given the
few-petabyte scale of the persistent archive, a central file store,
based on GPFS at NCSA was substituted. This gave the production system
a hub-and-spoke data architecture. The central data store was
implemented in the NCSA storage condominium, with Fermilab tape
providing disaster recovery capabilities.

The original DES concept was to mirror the collaboration's relational
data to databases at various sites.  A straightforward implementation
is to replay database logs into slave databases.  However this
technique was found to have shortcomings. An example shortcoming was
that fast data ingest operations bypass the logging mechanism, dooming
the system to very slow data ingest.  Since DES used Oracle, DES
reverted to a large central Oracle RAC, which provides a good data
service to the collaboration.

Use of the relational technology is critical for DES calibration
activities, and for providing subsets of data. DES has been successful
in ingesting refinements to annual release, computing outside of the
main production effort and has ingested data produced outside the
production framework at NCSA into its relational schema, preserving
object identities.

\medskip

\noindent{\bf Archival Data Storage:} DES archival storage is at the
$\sim 1$~PB level.

\medskip

\noindent{\bf Workflows:} Described above.

\medskip

\noindent{\bf Many-Core and/or GPU Readiness:} Not used because the
community code base does not have many-core or GPU applications.

\medskip 

\noindent{\bf Software Applications, Libraries, and Tools:} N/A

\medskip

\noindent{\bf HPC Services:} Careful design of a processing
infrastructure would allow annual processing or reprocessing to be run
in backfill mode. Service level issues, such as job scheduling would
need to be addressed. DES currently mitigates job scheduling with
glide in techniques. 

\medskip

\noindent{\bf Additional Needs:} DES nightly processing requires
nightly availability.  Because the computational requirements for this
are relatively modest and dominated by availability and reliability
concerns, this is a more problematic use case for a conventional HPC
environment.

\newpage

\noindent{\bf\scshape{2.6~Lattice QCD (MILC/USQCD)}}  

\medskip

\noindent{\bf Authors:} D. Toussaint

\medskip

\noindent{\bf{\em 2.6.1~Description of Research}}

\noindent{\bf Overview and Context:} Lattice field theory calculations
are essential for interpretation of many experiments done in
high-energy and nuclear physics, both in the US and abroad (e.g.,
BaBar at SLAC, CLEO-c at Cornell, CDF and D0 at Fermilab, and Belle at
KEK, Japan).  New data is beginning to arrive from the LHCb experiment
at the LHC, and BESIII in Beijing. In many cases, lattice QCD
calculations of the effects of strong interactions on weak interaction
processes (weak matrix elements) are needed to realize the full return
on the large investments made in the experiments.  In all such cases,
it is the uncertainties in the lattice QCD calculations that limit the
precision of the standard model tests.  Our objective is to improve
the precision of the calculations so that they are no longer the
limiting factor.

\medskip

\noindent{\bf Research Objectives for the Next Decade:} During the
next decade we will continue our work on decays and interactions of
heavy quarks, seeking to match the accuracy of experiments scheduled
for the 2020 time frame.  Also, lattice computations of the hadronic
contributions to the magnetic moment of the muon will be essential to
understanding the results of the experiment planned at Fermilab.  In
this time frame we expect that computations of the
quark-line-disconnected contributions to masses and other properties
will be increasingly important.  Both the g-2 and the disconnected
computations will require high statistics, but not necessarily much
smaller lattice spacings than are currently used.  Thus we expect a
slowdown in the rate at which the lattice spacing has been pushed
down, and the size of our simulations increased, in favor of higher
statistics.  

\medskip

\noindent{\bf{\em 2.6.2~Computational and Data Strategies}}

\noindent{\bf Approach:} Lattice QCD is a theory of quarks and gluons
(gauge fields) defined on a four-dimensional space time grid.  We use
a Monte Carlo method to create gauge configurations in proportion to
their weight in the Feynman path integral that defines the theory.
Once a suitable ensemble of configurations is created and stored, it
can be used to study a wide variety of physical phenomena.  The
generation of configurations is a long stochastic simulation, so must
run at high speed.  However, with the stored configurations,
subsequent work can be done in parallel and speed of a single job is
not critical as long as there is sufficient capacity to run multiple
jobs.

\medskip

\noindent{\bf Codes and Algorithms:} The MILC collaboration has
developed its code over a period of 20 years and makes it freely
available to others.  It has evolved to match our physics goals and to
accommodate changes in computers.  Currently containing approximately
180,000 lines, it is used by several collaborations worldwide.  Our
code has made increasing use of a library of specialized data-parallel
linear algebra and I/O routines developed over the past several years
with support from the DOE's SciDAC program.  These packages were
developed for the benefit of the entire USQCD Collaboration, which
consists of nearly all of the physicists in the United States working
on the numerical study of lattice gauge theory.  

There are a number of algorithms used in the generation of gauge
fields.  The heart of the algorithm is a dynamical evolution similar
to molecular dynamics.  In order to calculate the forces driving the
evolution, a multimass conjugate gradient solver is required to deal
with the quarks.  This solver takes the majority of the time in the
calculation and it becomes increasingly important as the up and down
quark masses are reduced.  It has only recently become feasible to use
up and down quark masses as light as in Nature.

\medskip

\noindent{\bf{\em 2.6.3 Current and Future HPC Needs}}

\noindent{\bf Computational Hours:} 

\medskip

\noindent{\bf Parallelism:} Our codes currently use MPI at the large
scale, and have the option to use OpenMP for a second level of
parallelism.  We also can and do use GPUs to accelerate the most
computationally intensive parts of the calculation.

\medskip

\noindent{\bf Memory:} 

\medskip

\noindent{\bf Scratch Data and I/O:} Some new analysis methods require
temporary storage of a large number of eigenvectors, so short term (as
opposed to archival) storage may be more important.

\medskip

\noindent{\bf Long-term and Shared Online Data:} Because lattice QCD
is so computationally expensive, the groups that create the gauge
configurations have often made them publicly available.  These
configurations can be used for a wide variety of physics topics and by
sharing their configurations with other collaborations the scientific
impact can be maximized.  The NERSC Gauge Connection was a pioneering
service in support of sharing configurations world wide and remains an
important service of NERSC that relies on its storage facilities.
Thus, both high-end computing and storage are essential to our
research.

\medskip

\noindent{\bf Archival Data Storage:} In the range of 250TB by 2025. 

\medskip

\noindent{\bf Workflows:} N/A

\medskip

\noindent{\bf Many-Core and/or GPU Readiness:} Lattice QCD codes are
using many-core and GPU resources already. 

\medskip 

\noindent{\bf Software Applications, Libraries, and Tools:} N/A

\medskip

\noindent{\bf HPC Services:} N/A

\medskip

\noindent{\bf Additional Needs:} N/A

\newpage

\centerline{\bf Requirements Summary Worksheet}

\begin{figure}
\centering \includegraphics[width=6.5in]{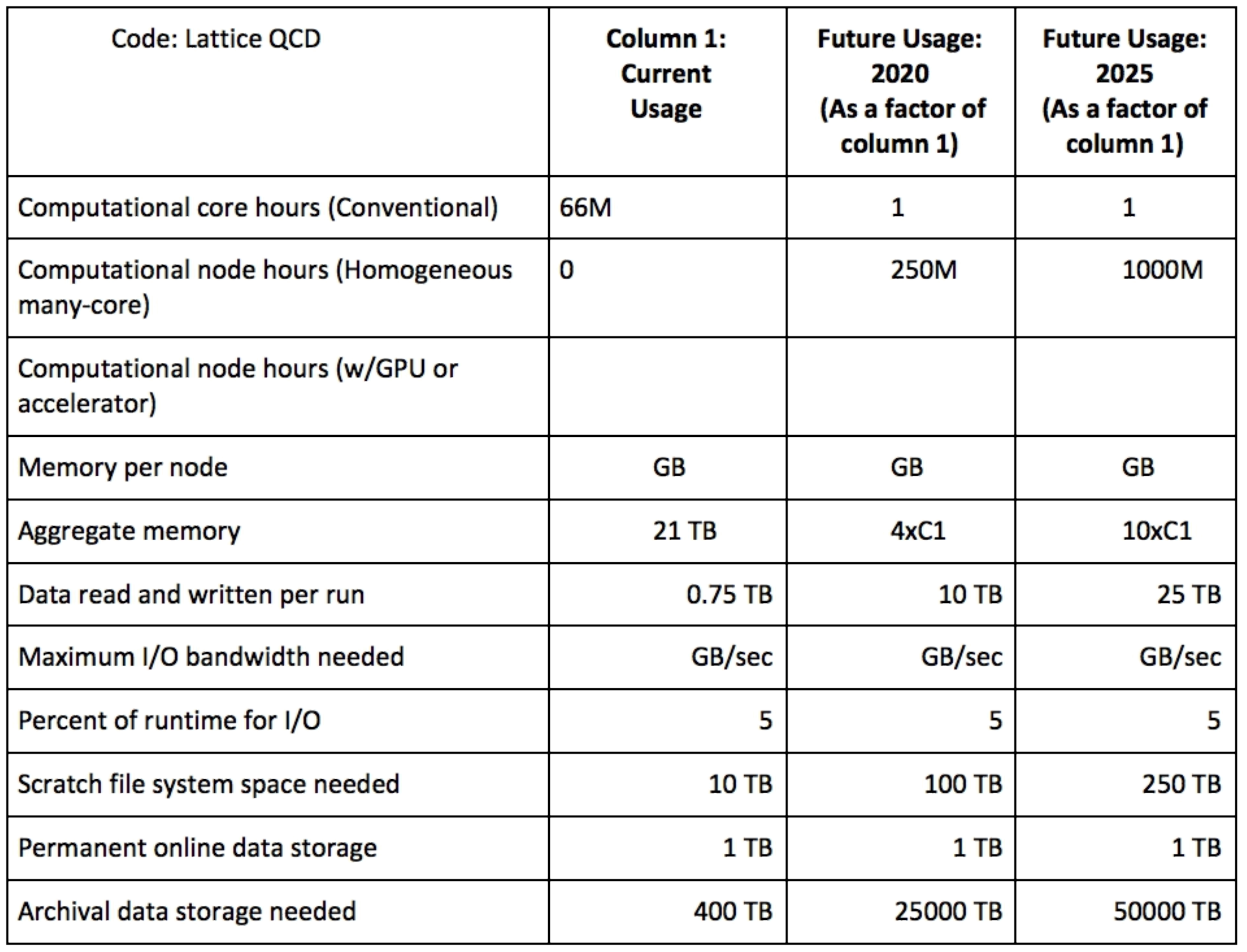}
\end{figure}

\newpage

\noindent{\bf\scshape{2.7~Event Generation (Sherpa)}}  

\medskip

\noindent{\bf Authors:} S. Hoeche

\medskip

\noindent{\bf{\em 2.7.1~Description of Research}}

\noindent{\bf Overview and Context:} High Energy Theory can be broadly
classified into two sub-fields: Precision calculations in the Standard
Model, and phenomenological analysis of new physics scenarios. Both
are connected by Monte-Carlo event generators, which are used to make
particle level predictions based on the former and used for the
latter.  The computation of SM reactions has reached a high degree of
automation, with next-to-leading order (NLO) QCD perturbation theory
being the standard means of obtaining cross sections at collider
experiments like the LHC. The results have been instrumental in
extracting properties of the Higgs boson, and they will continue to
play a dominant role in the future. Dedicated next-to-next-to leading
order (NNLO) QCD calculations exist for important reactions such as
Higgs-boson production, both at the inclusive and at the fully
differential level. They have been included in parton shower
Monte-Carlo event generators in order to make particle-level
predictions for high precision measurements in collider experiments.

\medskip

\noindent{\bf Research Objectives for the Next Decade:} We are rapidly
moving towards full automation of NLO electroweak
calculations. Tremendous progress has also been made in NNLO QCD
calculations, raising hopes for an eventual automation in the
future. The first complete three-loop result for inclusive Higgs-boson
production paves the way for more ultra-precise SM predictions at
hadron colliders. They will become mandatory as future data will have
reduced experimental errors and theory uncertainties begin to limit
our understanding of Nature. Predictions must eventually be made fully
differentially at the particle level in order to be maximally useful
for experiment, i.e. they must be interfaced to event generators.

\medskip

\noindent{\bf{\em 2.7.2~Computational and Data Strategies}}

\noindent{\bf Approach:} We use adaptive Monte-Carlo methods to
integrate over the high-dimensional phase space of multi-particle
production processes at hadron colliders. We also use numerical
methods to solve small sets of linear equations on a point-by-point
basis during the Monte-Carlo integration. We store phase-space
configurations in the form of Root Ntuples for later use by experiment
or theory.

\medskip

\noindent{\bf Codes and Algorithms:} We use C++ codes built from
scratch, with minimal dependencies on external libraries. We interface
Root for the storage of phase space configurations and LHAPDF for the
parameterization of structure functions.

\medskip

\noindent{\bf{\em 2.7.3 Current and Future HPC Needs}}

\noindent{\bf Computational Hours:} Typical NLO QCD calculations
currently require between 50k and 500k CPU-hours each. NNLO
differential calculations require between 50k and 1M CPU-hours
each. Parton-shower matched predictions require of the order of 100k
CPU hours, depending on the complexity of the final-state.

\medskip

\noindent{\bf Parallelism:} Sherpa has been parallelized using both
MPI and POSIX threads. First studies have been done on Xeon Phi, but
the code is not ready for production. Other NLO codes (MCFM) have
demonstrated scaling up to 10$^6$ threads using a hybrid MPI+OpenMP
approach.

\medskip

\noindent{\bf Memory:} High-multiplicity NLO calculations currently
set the limit. They require between 1 and 6 GB shared memory per node,
with the aggregate memory set by the number of nodes. At current
technology we expect this to be rising slowly over the next years.

\medskip

\noindent{\bf Scratch Data and I/O:} Typical NLO QCD calculations
currently require between 0.1 and 1.5 TB. The results can be
re-analyzed to obtain predictions for different SM parameters. NNLO
differential calculations are typically performed for a single set of
observables, but the possibility to store phase-space configurations
is investigated. We expect this to take 1 to 10 TB per calculation.

\medskip

\noindent{\bf Long-term and Shared Online Data:} None currently.

\medskip

\noindent{\bf Archival Data Storage:} None currently.

\medskip

\noindent{\bf Workflows:} Not standardized.

\medskip

\noindent{\bf Many-Core and/or GPU Readiness:} Sherpa is generally not
ready for accelerators. Automated NLO generators and general-purpose
MC event generators perform many different tasks, and they are
typically designed in a highly object oriented way. This makes it
difficult to outsource parts of the calculation in an efficient
manner.

\medskip 

\noindent{\bf Software Applications, Libraries, and Tools:} N/A

\medskip

\noindent{\bf HPC Services:} N/A

\medskip

\noindent{\bf Additional Needs:} N/A

\newpage

\centerline{\bf Requirements Summary Worksheet}

\begin{figure}
\centering \includegraphics[width=6.5in]{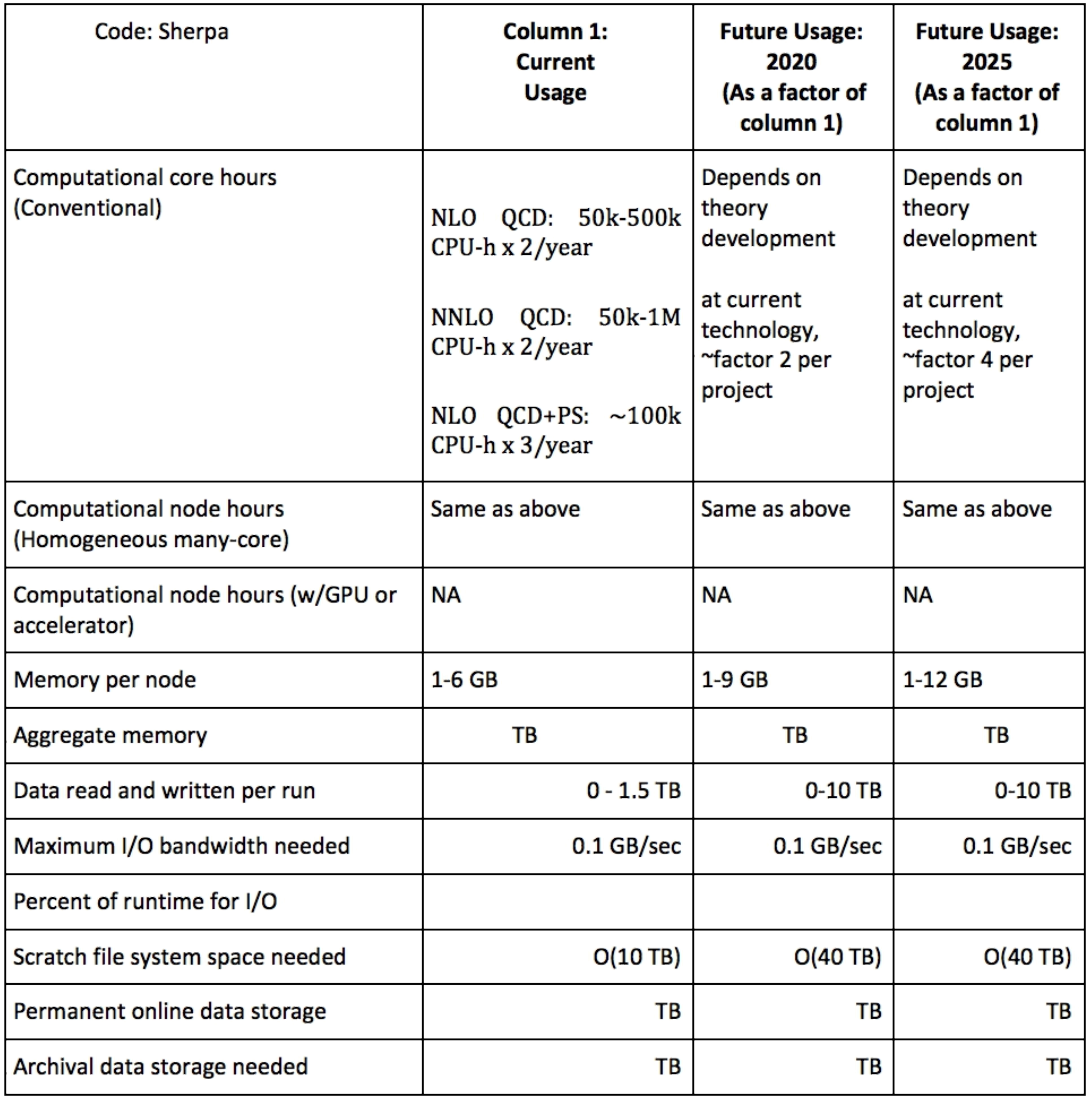}
\end{figure}

\newpage

\noindent{\bf\scshape{2.8~Energy Frontier Experiment (ATLAS)}}  

\medskip

\noindent{\bf Authors:} T. LeCompte

\medskip

\noindent{\bf{\em 2.8.1~Description of Research}}

\noindent{\bf Overview and Context:} We collide beams of particles
together and then measure the trajectories of the particles created in
this collision.  These are then compared to simulated events at
several levels -- e.g., comparing the response of the detectors to
particular particles with given energies or comparing the number of
events of a particular topology to expectations.  This is not a simple
process, and we are every bit as dependent on the simulation chain as
we are on the actual data chain.  The ATLAS experiment uses
approximately one billion CPU-hours per year to generate, simulated,
digitize, reconstruct and analyze simulated data.  Over 100 petabytes
of data (including replicas) are stored.  This is achieved primarily
via the LHC Computing Grid (LCG, or just ``Grid'') -- these are
networked clusters of commodity PCs.

\medskip

\noindent{\bf Research Objectives for the Next Decade:} The scientific
goals are described in the P5 report
(http://www.usparticlephysics.org/p5/) but in short the experiments
are looking for new particles and phenomena, and making precise
measurements of the Higgs boson properties.  To do this, there exists
a twenty-year plan of running the Large Hadron collider and ultimately
collecting 100$\times$ the data at nearly twice the energy.  The data
in hand will be tripled in roughly 2-3 years, tripled again in another
2-3 years, and then will grow linearly until 100$\times$ has been
collected, around 2035.  ATLAS would like to keep the ratio of
simulated to real events at least constant over the next decade,
although the likely funding scenarios make this impossible if we
restrict ourselves to LGC resources.

\medskip

\noindent{\bf{\em 2.8.2~Computational and Data Strategies}}

\noindent{\bf Approach:} We take advantage of event-level parallelism
to divide our tasks into 24-hour single PC jobs.  Input and output
from these jobs resides on Grid ``Storage Elements''.  We have just
recently been using High Performance Computers (HPCs) and have over
the last year offloaded about 6\% of the Grid production.

\medskip

\noindent{\bf Codes and Algorithms:} We use dozens of codes, often
with millions of lines of code.  These have in general never been
optimized for parallelism of any kind.

\medskip

\noindent{\bf{\em 2.8.3 Current and Future HPC Needs}}

\noindent{\bf Computational Hours:} 

\medskip

\noindent{\bf Parallelism:} We use event-level parallelism -- one
event per core.  We are investigating the use of accelerators; we have
seen some improvement with some codes running a single thread.
Whether this extends to an ensemble of cores sharing one or many
co-processors remains to be seen.

\medskip

\noindent{\bf Memory:} Our code was developed under the assumption
that 2 GB is available for every process.  In some cases, with
relatively minor code changes we can drastically reduce the amount of
memory -- more than an order of magnitude in one case.  It appears that
in some cases we would get better performance by using fewer cores to
repurpose some memory as a RAM disk, thereby achieving an overall gain
by speeding up the I/O.

\medskip

\noindent{\bf Scratch Data and I/O:} Everything we have run so far is
I/O limited.  This is driven not by the total I/O, but rather by the
overhead of having millions of ranks that can potentially read or
write data.  Because our Grid jobs are fast (hours) and our HPC jobs
are faster (20-30 minutes) we tend not to need much scratch area per
job.

\medskip

\noindent{\bf Long-term and Shared Online Data:} All input and output
must be available on the LCG.  HPC sites can store additional copies,
or if willing, act as Grid Storage Elements, but this must happen in
some way.

\medskip

\noindent{\bf Archival Data Storage:} 

\medskip

\noindent{\bf Workflows:} A job will be submitted to the production
manager program, assigned to an HPC, run there, and the output placed
on an SE, just as Grid jobs do.  This works today, with some room for
improvement.  The alternative, submitting hundreds of thousands of
jobs by hand, is impractical.

\medskip

\noindent{\bf Many-Core and/or GPU Readiness:} We are just now
exploring these.  The first step in transitioning is to use Cori,
Theta and even Titan as a testbed, to delineate the present
limitations.  Then we can address them in detail.

\medskip 

\noindent{\bf Software Applications, Libraries, and Tools:} N/A

\medskip

\noindent{\bf HPC Services:} Small, easy-to-use, machines like Carver
provide an important resource for development, testing, and training.
While the very largest machines get the most attention, I would hope
that these kinds of machines figure into the planning.

\medskip

\noindent{\bf Additional Needs:} N/A

\newpage

\centerline{\bf Requirements Summary Worksheet}

\begin{figure}
\centering \includegraphics[width=6.5in]{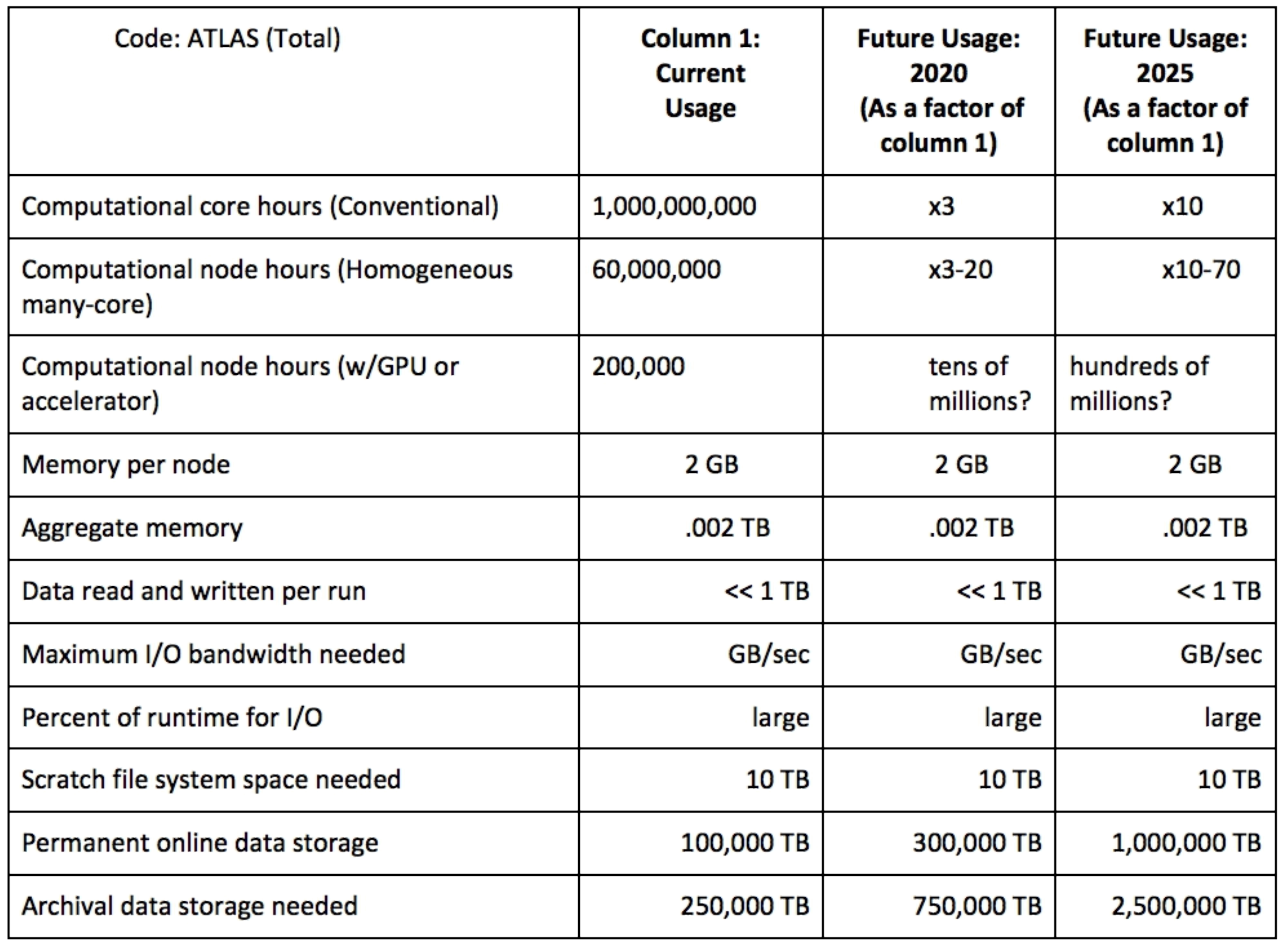}
\end{figure}

\newpage

\noindent{\bf\scshape{2.9~Cosmic Microwave Background (TOAST)}}  

\medskip

\noindent{\bf Authors:} J. Borrill

\medskip

\noindent{\bf{\em 2.9.1~Description of Research}}

\noindent{\bf Overview and Context:} The CMB carries the imprint of
the entire history of the Universe in the fluctuations of its
temperature and polarization on different angular scales. To detect
fainter signals in the CMB we have to gather larger datasets,
resulting in exponential data growth paralleling Moore's Law for the
last and next 20 years. Such growth constrains us to reduce CMB data
sets using no worse than log-linear analysis algorithms, and to follow
Moore's Law to massively parallel systems at state-of-the-art HPC
facilities.

\medskip

\noindent{\bf Research Objectives for the Next Decade:} The quest is
to measure the B-mode polarization signal, whose fluctuation spectrum
contains vital information on the energy scale of inflation and the
sum of neutrino masses. The program runs from a suite of 3rd
generation ground-based and balloon-borne projects over the next 5-10
years, to future missions such as CMB-S4 (DOE/NSF), the LiteBIRD
(JAXA/NASA), PIXIE (NASA), and COrE+ satellites (ESA), in the 2020's
and beyond. Our primary goals are (i) to increase the veracity of our
simulations subject to scaling constraints, and (ii) to improve the
efficiency of the simulation generation/reduction pipeline under the
twin pressures of exponential data growth and increasing complexity in
HPC architectures.

\medskip

\noindent{\bf{\em 2.9.2~Computational and Data Strategies}}

\noindent{\bf Approach:} Although exact solutions exist for the
analysis of a CMB data set, they depend on the full data covariance
matrix and scale as O(N$^3$). Exponential data growth has made this
approach intractable, so we employ approximate analysis algorithms and
Monte Carlo methods for debiasing and uncertainty quantification. The
dominant computational challenge is the generation and reduction of
O(10$^4$) realizations of the data. To minimize I/O costs we generate
each realization of the time domain data on-the-fly and reduce it to a
set of pixelized maps before any data touch disk. Since we are now
constrained by spinning disk storage, in future we may need to
pre-reduce the maps as well.

\medskip

\noindent{\bf Codes and Algorithms:} Our analysis proceeds in several steps: 
\begin{itemize}
\item real or simulated time-ordered data are reduced to pixelized
  maps at each frequency,  
\item CMB and foreground maps are separated based on different
  frequency dependencies,  
\item fluctuation power spectra are derived from the CMB map, and 
\item likelihoods of the cosmological parameters are derived from the
  power spectra. 
\end{itemize}
The first step dominates computation, particularly when simulations
are required, and is implemented by the Time-Ordered Astrophysics
Scalable Tools (TOAST) software, which includes time-ordered data
generation and reduction capabilities, and provides hooks and memory
management for experiment-specific tools; key algorithms are
pseudo-random number generation and preconditioned conjugate gradient
solvers, with underlying algorithms for Fourier and Spherical Harmonic
transforms and sparse matrix-vector multiplication.

\medskip

\noindent{\bf{\em 2.9.3 Current and Future HPC Needs}}

\noindent{\bf Computational Hours:} For almost 20 years CMB data
analysis has used about 1\% of the NERSC cycles annually, with
requirements growing in lockstep with NERSC's capacity and
capability. Provided we can maintain our computational efficiency on
the coming generations of systems, our needs will likely stay at
around the percent level.

\medskip

\noindent{\bf Parallelism:} Almost all of the workhorse codes are now
hybrid MPI/OpenMP; work is ongoing to upgrade or replace those which
are not. Analysis of the biggest data sets (Monte Carlo simulations)
is explicitly parallelized over realizations. This is managed as
separate jobs running simultaneously within a batch submission, but
efficiencies could be exploited using distinct MPI communicators to
run different realizations independently and concurrently.

\medskip

\noindent{\bf Memory:} Though CMB detector data require whole-data
reduction, there are subsets that localize secondary data (e.g.,
telescope pointing and detector properties). Current memory per core
generally supports our smallest data unit (a stationary interval of
samples from a single detector), and the growth in memory per node
mirrors that of the next (a stationary interval from all of the
detectors at one frequency). Explicit access to manage intermediate
data storage (e.g., burst buffers) would be useful.

\medskip

\noindent{\bf Scratch Data and I/O:} We are already severely hampered
by limited persistent (``project'') spinning disk space, and what we
do have exhibits I/O performance so limited that we routinely
pre-fetch entire data sets to fast but transient scratch disk spaces.
I/O performance itself is currently mostly limited by sub-optimal data
formats; this issue is being addressed.

\medskip

\noindent{\bf Long-term and Shared Online Data:} Our hope is to keep
data spinning on persistent disk for extended periods (years) and
bring analysis to the data. Currently the Planck collaboration
maintains 200TB of NERSC project space as a purchased service, but
such allocations (and larger) will need to be made routine for
next-generation experiments.

\medskip

\noindent{\bf Archival Data Storage:} Absent sufficient spinning disk
for active data sets, fast, reliable, flexible, automated archiving
and restoration tools are vital to project data management. Ideally
this would also be integrated into batch scheduling so that, for
example, archived data could be restored during a job's queuing time.

\medskip

\noindent{\bf Workflows:} CMB data analysis proceeds as a series of
data compressions, and the reduction from time samples to map pixels
marks the transition from tightly- to loosely-coupled analysis
steps. From the map domain onwards, subsequent steps are typically
carried out by sub-domain experts, and inspected in detail at each
step, so much of traditional workflow software is inappropriate. Time
domain processing is seeing a significant move towards (typically
Python) scripted workflows for rapid prototyping and modularity. We
will require the ability to load and execute such scripts efficiently
over large fractions of the biggest HPC systems.

\medskip

\noindent{\bf Many-Core and/or GPU Readiness:} It will be critical for
us to transition our core codes to many-core architectures. We expect
to dedicate a substantial effort to this end.

\medskip 

\noindent{\bf Software Applications, Libraries, and Tools:} We
typically maintain our own installation of many libraries and users
load both generic CMB and experiment-specific modules. The range of
systematic effects in CMB data from different experiments requires a
flexible analysis infrastructure for understanding and realistically
simulating each configuration's data. At present this is being
prototyped in Python, though running this efficiently in an HPC
environment is a predictable challenge.

\medskip

\noindent{\bf HPC Services:} CMB missions are decade-long projects
with large numbers of collaborators participating the group analysis
of exponentially growing data sets. We therefore need to be able to
rely on access to continuously state-of-the art HPC resources, for a
large number of users scattered around the world, for the duration of
the mission.

\newpage

\centerline{\bf Requirements Summary Worksheet}

\begin{figure}
\centering \includegraphics[width=6.5in]{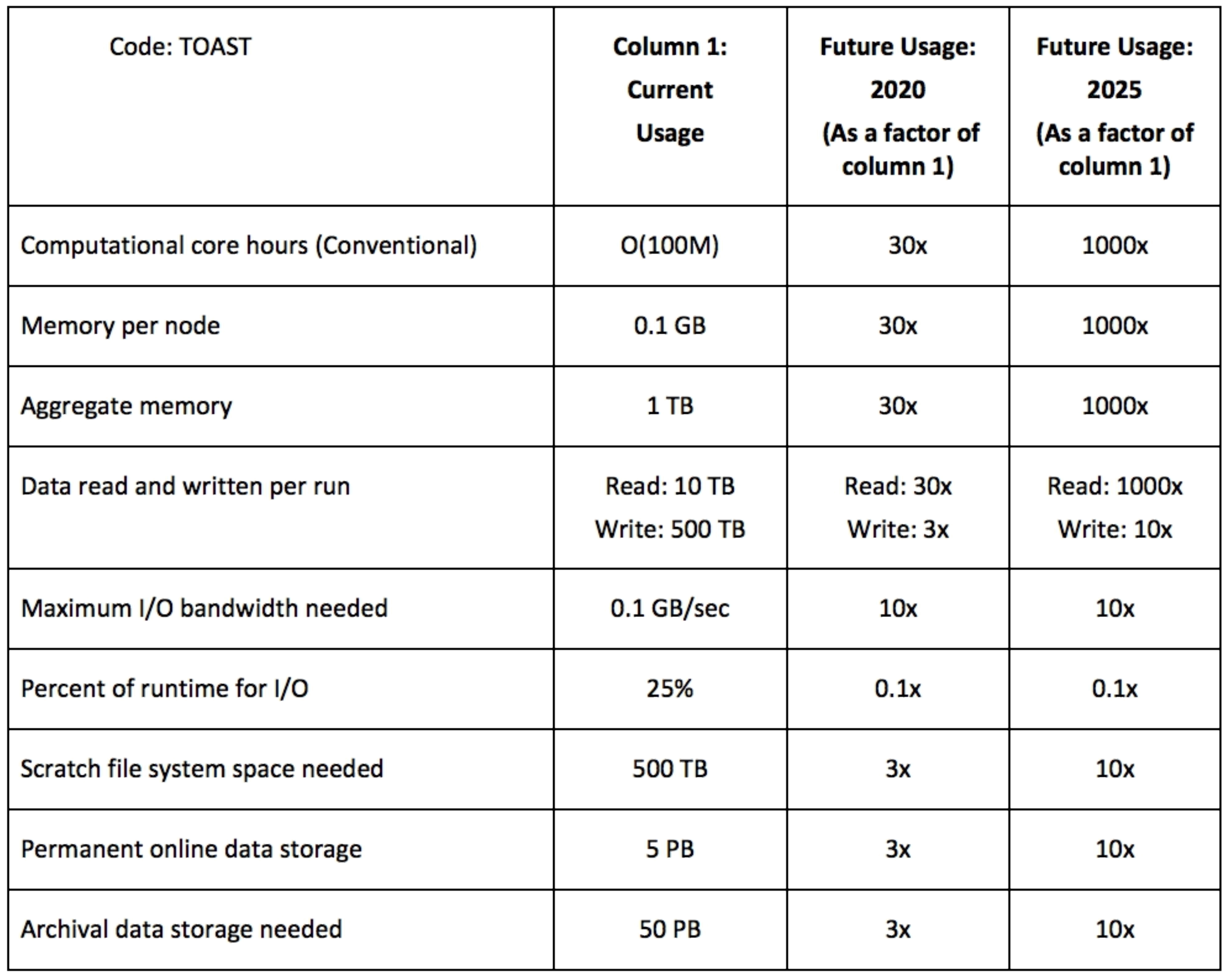}
\end{figure}

\newpage

\begin{center}
{\bf\scshape{Appendix 3:~Acronym Index}}
\end{center}

\noindent ACE3P --- Advanced Computational Electromagnetics 3-D Parallel\\
\noindent ALCF --- Argonne Leadership Computing Facility\\
\noindent AMR --- Adaptive Mesh Refinement\\
\noindent ART --- Adaptive Refinement Tree\\
\noindent ASCR --- Advanced Scientific Computing Research\\
\noindent AST --- Advanced Superconducting Test Accelerator\\
\noindent ATF --- Accelerator Test Facility\\
\noindent ATLAS --- A Toroidal LHC ApparatuS\\
\noindent AWA --- Argonne Wakefield Accelerator\\
\noindent BAO --- Baryon Acoustic Oscillations\\
\noindent BES --- Basic Energy Sciences\\ 
\noindent BELLA --- Berkeley Lab Laser Accelerator\\
\noindent Belle II --- B detector at SuperKEKB\\
\noindent BOSS/eBOSS --- (extended) Baryon Oscillation Spectroscopic Survey\\
\noindent CEBAF --- Continuous Electron Beam Accelerator Facility\\
\noindent CKM --- Cabibbo-Kobayashi-Maskawa\\
\noindent CORAL -- Collaboration of Oak Ridge, Argonne, Livermore\\
\noindent CMB --- Cosmic Microwave Background\\
\noindent CMS --- Compact Muon Solenoid\\
\noindent CP --- Charge Parity\\
\noindent CPU --- Central Processing Unit\\
\noindent CUDA --- Compute Unified Device Architecture\\
\noindent DES --- Dark Energy Survey\\
\noindent DESI --- Dark Energy Spectroscopic Instrument\\
\noindent DOE --- Department of Energy\\
\noindent DUNE --- Deep Underground Neutrino Experiment\\
\noindent DWA -- Dielectric Wakefield Accelerator\\
\noindent FACET --- Facility for Advanced Accelerator Experimental Tests\\
\noindent FRIB --- Facility for Rare Isotope Beams\\
\noindent FASTMath --- Frameworks, Algorithms, and Scalable
Technologies for Mathematics\\
\noindent FFT --- Fast Fourier Transform\\
\noindent GPU --- Graphics Processing Unit\\
\noindent HACC --- Hardware/Hybrid Accelerated Cosmology Codes\\
\noindent HAWC --- High-Altitude Water Cherenkov Observatory\\
\noindent HEP --- High Energy Physics\\
\noindent HEP-FCE --- High Energy Forum for Computational Excellence\\
\noindent HL-LHC --- High Luminosity Large Hadron Collider\\
\noindent HPC --- High Performance Computing\\
\noindent HTC --- High Throughput Computing\\
\noindent INCITE --- Innovative and Novel Computational Impact on
Theory and Experiment\\
\noindent IOTA --- Integrable Optics Test Accelerator\\
\noindent JLSE --- Joint Laboratory for System Evaluation\\
\noindent LArTPC --- Liquid Argon Time Projection Chamber\\
\noindent LCF --- Leadership Computing Facility\\
\noindent LCLS --- Linac Coherent Light Source\\
\noindent LHC --- Large Hadron Collider\\
\noindent LHCb -- Large Hadron Collider beauty experiment\\
\noindent LPA --- Laser-Plasma Accelerators\\
\noindent LSST --- Large Synoptic Survey Telescope\\
\noindent LSST-DESC --- LSST Dark Energy Science Collaboration\\
\noindent LZ --- Merger of LUX and ZEPLIN dark matter experiments\\
\noindent MILC --- MIMD Lattice Computation\\
\noindent MPI --- Message-Passing Interface\\
\noindent MSSM --- Minimal Supersymmetric Standard Model\\
\noindent NERSC --- National Energy Research Scientific Computing
Center\\
\noindent NESAP --- NERSC Exascale Science Applications Program\\
\noindent NP --- Nuclear Physics\\
\noindent NUMA --- Non-Uniform Memory Access\\
\noindent NVRAM --- Non-Volatile Random-Access Memory\\
\noindent OLCF --- Oak Ridge Leadership Computational Facility\\
\noindent OpenCL --- Open Computing Language\\
\noindent OpenMP --- Open Multi-Processing\\
\noindent OSG --- Open Science Grid\\
\noindent P5 --- Particle Physics Project Prioritization Panel\\
\noindent PGAS --- Partitioned Global Address Space\\
\noindent PIC -- Particle In Cell\\
\noindent PIP-II --- Proton Improvement Plan-II\\
\noindent PM --- Particle-Mesh\\
\noindent P$^3$M --- Particle-Particle Particle-Mesh\\
\noindent PPM --- Piecewise Parabolic Method\\
\noindent PWFA --- Plasma Wakefield Accelerators\\
\noindent QCD --- Quantum Chromodynamics\\
\noindent QUEST --- Quantification of Uncertainty in Extreme
Scale Computations\\
\noindent SKA -- Square Kilometer Array\\
\noindent SAM --- Semi-Analytic Modeling\\
\noindent SCET --- Soft-Collinear Effective Theory\\
\noindent SciDAC --- Scientific Discovery through Advanced Computing\\ 
\noindent SUPER --- Sustained Performance, Energy, and Resilience\\
\noindent SuperCDMS --- Super Cryogenic Dark Matter Search\\
\noindent SPH --- Smoothed Particle Hydrodynamics\\
\noindent T2K --- Tokai to Kamioka neutrino experiment\\
\noindent TOAST --- Time-Ordered Astrophysics Scalable Tools\\
\noindent UMER --- University of Maryland Electron Ring\\
\noindent UQ --- Uncertainty Quantification\\
\noindent WORM --- Write-Once, Read-Many\\
\noindent XSEDE --- Extreme Science and Engineering Discovery Environment\\ 

\newpage

\begin{center}
{\bf\scshape{Appendix 4: Acknowledgments}}
\end{center}

The authors acknowledge the support and help of many of our colleagues
who provided or helped collect important pieces of information
presented in the report. We record our indebtedness to the ASCR and
HEP attendees at the Exascale Requirements Review meeting held June
10-12, 2015 in Bethesda, MD. The agenda and presentations at the
meeting can be found at
\texttt{https://www.nersc.gov/science/hpc-requirements-reviews/exascale/HEP/}. We
have endeavored to encapsulate the conclusions of the discussions at
the meeting as completely as possible in the report. In this we were
greatly aided by a number of the attendees who made their detailed
notes of the discussions available to us. We also acknowledge support
from the Computational HEP program to the HEP Forum for Computational
Excellence (HEP-FCE) for help in coordinating the HEP component of the
work carried out for this report.

We are grateful to the responsible DOE program managers Lali
Chatterjee (HEP) and Carolyn Lauzon (ASCR) for their active interest
and help in organizing the activity that led to this report.


\newpage

\begin{center}
{\bf\scshape{5~References}}
\end{center}

\begin{thebibliography}{100}
\bibitem{p5} Building for Discovery: Strategic Plan for U.S. Particle
  Physics in the Global Context, Report of the Particle Physics
  Project Prioritization Panel (P5),\\
  \texttt{http://www.usparticlephysics.org/p5/}
\bibitem{snowmass} Snowmass 2013 Study Electronic Proceedings:\\
  \texttt{https://www-public.slac.stanford.edu/snowmass2013/Index.aspx};
  computationally relevant Snowmass documents are collected at
  \texttt{http://hepfce.org/documents/}
\bibitem{higgs_atlas} The ATLAS Collaboration, {\em Phys. Lett. B}
  {\bf 716}, 1, 2012
\bibitem{higgs_cms} The CMS Collaboration, {\em Phys. Lett. B} {\bf 716}, 30, 2012
\bibitem{neut} S.F.~King, et al., {\em New J. Phys.} {\bf 16}, 045018, 2014
\bibitem{weinberg_rev} D.~Weinberg, et al., {\em Phys. Rep.} {\bf
    530}, 87, 2013
\bibitem{dm_review} J.L.~Feng, {\em Annu. Rev. Astron. Astrophys.}
  {\bf 48}, 495, 2010
\bibitem{cms} CMS Experiment home page. \texttt{http://cms.web.cern.ch/}
\bibitem{atlas} ATLAS Experiment home page. \texttt{http://atlas.ch/}
\bibitem{craig_susy} N.~Craig, ``The State of Supersymmetry after Run
  I of the LHC,'' 2013, arXiv:1309.0528 [hep-ph]. Available at
  \texttt{http://arxiv.org/abs/1309.0528}.  
\bibitem{pip2_fnal} Proton Improvement Plan-II home page. \texttt{https://pip2.fnal.gov/}
\bibitem{dune_fnal} DUNE collaboration home page. \texttt{https://web.fnal.gov/collaboration/DUNE/SitePages/Home.aspx}
\bibitem{g-2} Muon g-2 Experiment home page. \texttt{http://muon-g-2.fnal.gov/}
\bibitem{m2e} Mu2e home page. \texttt{http://mu2e.fnal.gov/index.shtml}
\bibitem{belle2} Belle II Collaboration home page. \texttt{http://belle2.kek.jp/}
\bibitem{desi} M.~Levi, et al., ``The DESI Experiment, a white paper
  for Snowmass 2013,'' 2013, arXiv:1308.0847 [astro-ph.CO].
  Available at \texttt{http://arxiv.org/pdf/1308.0847.pdf}. 
\bibitem{lsst} P.A.~Abell, et al., ``LSST Science Book, Version 2.0,''
  2009, arXiv:0912.0201 [astro-ph.IM]. Available at
  \texttt{http://arxiv.org/abs/0912.0201}.  
\bibitem{cmbs4} K.N.~Abazajian, et al., ``Inflation Physics from the
  Cosmic Microwave Background and Large Scale Structure,''
  2013, arXiv:1309.5381 [astro-p.CO]. \\ Available at
  \texttt{http://arxiv.org/abs/1309.5381}.
\bibitem{ska} Square Kilometre Array Telescope home page. 
\texttt{https://www.skatelescope.org/}
\bibitem{lz} The LZ Dark Matter Experiment home page. \texttt{http://lz.lbl.gov/}
\bibitem{s_cdms} Super Cryogenic Dark Matter Search home page. 
\texttt{http://cdms.berkeley.edu/}
\bibitem{fermi_exp} Fermi Gamma-Ray Space Telescope home page.
\texttt{http://fgst.slac.stanford.edu/}
\bibitem{hawc_exp} The High-Altitude Water Cherenkov Gamma-Ray
  Observatory home page. \\\texttt{http://www.hawc-observatory.org/}
\bibitem{miniapp} M.A.~Heroux, et al., ``Improving Performance via
  Mini-application,'' Sandia Report SAND2009-5574 (2009). Available
  at \texttt{https://mantevo.org/MantevoOverview.pdf}.
\end{thebibliography}

\medskip

\begin{thebibliography}{99}
\bibitem{accsim} P. Spentzouris, E. Cormier-Michel, C. Joshi,
  J. Amundson, W. An, D.L. Bruhwiler, J.R. Cary, B. Cowan, V.K. Decyk,
  E. Esarey, et al., ``Working Group Report: Computing for Accelerator
  Science'', Oct 8, 2013. Conference: C13-07-29.2; arXiv:1310.2203
  [physics.acc-ph]
\bibitem{beams} J. Amundson, A. Macridin, P. Spentzouris, et al.,
  ``High Performance Computing Modeling Advances Accelerator Science
  for High Energy Physics'', IEEE Comput. Sci. Eng. 16, 32 (2014)
 \bibitem{mars} MARS: \texttt{www-ap.fnal.gov/MARS/}
 \bibitem{impact} IMPACT: \texttt{http://blast.lbl.gov/BLASTcodes\_Impact.html}
 \bibitem{synergia} Synergia:
   \texttt{https://web.fnal.gov/sites/synergia/SitePages/Synergia\%20Home.aspx} 
 \bibitem{beambeam} BEAMBEAM3D:
   \texttt{http://blast.lbl.gov/BLASTcodes\_BeamBeam3D.html} 
\bibitem{materials}
  \texttt{https://confluence.slac.stanford.edu/display/AdvComp/Materials+for+CW14} 
 \bibitem{adv1} R.A. Fonseca et al., Lecture Notes in Computer Science
   2331, 342-351 (2002); R.G. Hemker, ``Particle-in-Cell Modeling of
   Plasma-Based Accelerators in Two and Three Dimensions'', PhD
   Dissertation, UCLA (2000); arXiv:1503.00276 
\bibitem{adv2} C.K. Huang et al., ``QuickPIC: A highly efficient PIC code
  for modeling wakefield acceleration in plasmas'',
  J. Comp. Phys. 217, 658 (2006); W. An et al., ``An improved
  iteration loop for the 3D quasi-static PIC algorithm: QuickPIC'',
  J. Comp Phys. 250, 165 (2013). 
\bibitem{adv3} VSim: \texttt{https://www.txcorp.com/vsim/}
\bibitem{adv4} Warp: \texttt{http://blast.lbl.gov/BLASTcodes\_Warp.html},
\texttt{https://warp.lbl.gov}
\bibitem{P5} Report of the Particle Physics Project Prioritization
  Panel (P5), ``Building for Discovery: Strategic Plan for
  U.S. Particle Physics in the Global Context'', report of the
  Particle Physics Project Prioritization Panel (P5),
 \texttt{http://usparticlephysics.org/p5/}
\bibitem{bella} BELLA: \texttt{http://bella.lbl.gov/}
\bibitem{facet} FACET: \texttt{http://facet.slac.stanford.edu/}
\bibitem{awa} \texttt{http://www.hep.anl.gov/awa/}
\bibitem{pip2}
  \texttt{http://www.alcf.anl.gov/projects/accelerator-modeling-discovery} 
\bibitem{hl_lhc} J. Qiang, BeamBeam3D case study, this workshop
\bibitem{struct_req} C. Ng, ACE3P case study, this workshop
\bibitem{adv_exa} J.-L. Vay, ``Traditional HPC needs: particle
  accelerators'', this workshop 
\end{thebibliography}

\medskip

\begin{thebibliography}{99}
\bibitem{springel} V. Springel, Mon. Not. Roy. Astron. Soc. 364, 1105 (2005)  
\bibitem{habib} S. Habib, et al., arXiv:1410.2805, New Astronomy, in press
\bibitem{hot} M.S. Warren, arXiv:1310.4502, Proceedings SC'13
\bibitem{stadel} M.D. Dikaiakos and J. Stadel, ICS Conference Proceedings (1996) 
\bibitem{bryan} G.L. Bryan et al., Astrophys. J. Supp., 211, 19 (2014)
\bibitem{nyx} A. Almgren et al., Astrophys. J., 765, 39 (2013)
\bibitem{ramses} R. Teyssier, A\&A, 385, 337 (2002)
\bibitem{art} A.V. Kravtsov, A.A. Klypin, and Y. Hoffman, Astrophys. J. 571, 2 (2002) 
\bibitem{springel2} V. Springel, Mon. Not. Roy. Astron. Soc. 401, 791 (2010)
\bibitem{hopkins} P.F. Hopkins, Mon. Not. Roy. Astron. Soc. 450, 53 (2015)
\end{thebibliography}

\medskip

\begin{thebibliography}{99}
\bibitem{usqcd1} ``Lattice Gauge Theories at the Energy Frontier'', 
arXiv:1309.1206 [hep-lat] 
\bibitem{usqcd2} ``Lattice QCD at the Intensity Frontier'', 
\texttt{http://www.usqcd.org/documents/13flavor.pdf}
\bibitem{snow1} ``Report of the Snowmass 2013
Computing Frontier working group on Lattice Field Theory -- Lattice
field theory for the energy and intensity frontiers: Scientific goals
and computing needs''; arXiv:1310.6087
\bibitem{snow2} ``Planning the Future of Particle Physics
(Snowmass 2013): Chapter 9: Computing'', arXiv:1401.6117
\bibitem{nersc_req} ``Large Scale Computing and Storage Requirements 
for High Energy Physics: Target 2017'', 
\texttt{https://www.nersc.gov/assets/pubs\_presos/HEP2017FINAL-Nov13.pdf} 
\bibitem{ex_2008}``Challenges for Understanding the Quantum Universe
  and the Role of Computing at the Extreme Scale'',\\
  \texttt{http://science.energy.gov/$\sim$/media/ascr/pdf/program-documents/docs/Hep\_report.pdf}
\end{thebibliography}

\medskip

\begin{thebibliography}{99}
\bibitem{} S. Hoeche, L. Reina, M. Wobisch, et al., ``Working Group
  Report: Computing for Perturbative QCD'', arXiv:1309.3598 [hep-ph].
\end{thebibliography}

\medskip

\begin{thebibliography}{99}
\bibitem{higgs} CMS Collaboration, ``Observation of a new boson at a
  mass of 125 GeV with the CMS experiment at the LHC'',
  Phys. Lett. B716, 30 (2012), 10.1016/j.physletb.2012.08.021 
\bibitem{cms_exp} See
  \texttt{http://cms.web.cern.ch/org/physics-papers-timeline} for a
  timeline of CMS physics publications sorted by physics topic.
\bibitem{cmscomp} See for instance K. Bloom, ``CMS Use of a Data
  Federation'', J. Phys. Conf. Ser. 513 042005
  doi:10.1088/1742-6596/513/4/042005.
\bibitem{wlhccg} See \texttt{http://wlcg.web.cern.ch for more information}
\bibitem{giwms} I. Sfiligoi et al., ``The Pilot Way to Grid Resources
  Using glideinWMS'', 2009 WRI World Congress on Computer Science and
  Information Engineering, Vol. 2, pp. 428-432. doi:10.1109/CSIE.2009.950. 
\bibitem{run3} For details of the LHC long-range schedule, see\\
  \texttt{http://lhc-commissioning.web.cern.ch/lhc-commissioning/schedule/LHC-schedule-update.pdf}
\bibitem{upgrades} For many more details on the upgrades to the
  detectors and anticipated changes to the computing model, see CMS
  Phase-II Technical Proposal, in preparation, especially Chapter 8.
\bibitem{spdsheet} Planning spreadsheet: \texttt{http://cern.ch/go/T7LB}
\end{thebibliography}

\begin{thebibliography}{99}
\bibitem{sam} A. Norman, ``Data Handling best Practices'', 2015 FIFE
  Workshop,
  \texttt{https://indico.fnal.gov/getFile.py/access?contribId=35\&resId=0\\
\&materialId=slides\&confId=9737}
\bibitem{tape} INSIC's 2012-2022 International Magnetic Tape Storage
  Roadmap: \texttt{http://www.insic.org/news/2012Roadmap/12index.html}
\bibitem{info2} G. Stewart, CHEP 2015, ``Evolution of Computing and
  Software at LHC: from Run 2 to HL-LHC'',\\
  \texttt{http://indico.cern.ch/event/304944/session/15/contribution/551/material/slides/0.pdf}
\bibitem{info1} B. Panzer-Steindel, ``Technology, Market and Cost
  Trends'', 2015 WLCG Collaboration Workshop,\\
  \texttt{http://indico.cern.ch/event/345619/session/1/contribution/10/material/slides/1.pdf}
\end{thebibliography}

\medskip

\begin{thebibliography}{99}
\bibitem{scale}
  \texttt{https://indico.cern.ch/event/304944/session/15/contribution/551/material/slides/0.pdf}  
\bibitem{hllhc} HL-LHC: \texttt{http://hilumilhc.web.cern.ch/}
\bibitem{hllhc2}
  \texttt{http://indico.cern.ch/event/345619/session/1/contribution/23/material/slides/1.pdf}  
\end{thebibliography}
\end{document}